\documentclass[letterpaper,11pt]{article}
\pdfoutput=1 % if you are submitting a pdflatex (i.e. if you have
\usepackage{jheppub} 
\usepackage{graphicx}
\usepackage{amsmath}
\usepackage{amsfonts}
\usepackage{slashed}
\usepackage{amssymb}
\usepackage{hyperref}
	\hypersetup{colorlinks=true, linkcolor=blue, citecolor=red, urlcolor=cyan, linktoc=page}
\usepackage{xfrac}
\usepackage{empheq}

\newcommand{\mc}{\mathcal}
\renewcommand{\t}{\tilde}
\newcommand{\del}{\partial}
\newcommand{\Del}{\nabla}
\newcommand{\be}{\begin{eqnarray}}
\newcommand{\ee}{\end{eqnarray}}
\newcommand{\ep}{\epsilon}

\newcommand{\sla}{\slashed}
\newcommand{\p}{\partial}
\newcommand{\h}{\hat}
\newcommand{\w}{\wedge}
\newcommand{\sh}{\hat{\star}}
\newcommand{\st}{\tilde{\star}}
\newcommand{\til}{\tilde}

\newcommand{\wtn}{\tilde{\nabla}}
\newcommand{\un}{\underline}
\newcommand{\G}{G_3^{(0)}}
\newcommand{\bG}{\bar{G}_3^{(0)}}
\newcommand{\comment}[1]{}
\newcommand{\dnot}{\delta_\perp^6(y, Y_*)}
\newcommand{\sperp}{\star_\perp}
\newcommand{\0}{(0)}

\title{Dimensional Reduction for D3-brane Moduli}

\author[a]{Brad Cownden,}

\author[a,b]{Andrew R.~Frey,}

\author[c]{M.~C.~David Marsh,}

\author[d]{and Bret Underwood}

\affiliation[a]{Department of Physics \& Astronomy\\ University of Manitoba,
Winnipeg, Manitoba R3T 2N2, Canada}

\affiliation[b]{Department of Physics\\
University of Winnipeg, Winnipeg, Manitoba R3B 2E9, Canada}

\affiliation[c]{Department  of  Applied  Mathematics  and  Theoretical  
Physics\\ University  of  Cambridge, Cambridge, CB3 0WA, United Kingdom}

\affiliation[d]{Department of Physics \\ 
Pacific Lutheran University, Tacoma, WA 98447, USA}

\emailAdd{cowndenb@myumanitoba.ca}
\emailAdd{a.frey@uwinnipeg.ca}
\emailAdd{m.c.d.marsh@damtp.cam.ac.uk}
\emailAdd{bret.underwood@plu.edu}

\abstract{Warped string compactifications are central to many
attempts to stabilize moduli  and connect string theory with
cosmology and particle phenomenology.  We present a
first-principles derivation of the low-energy 4D effective theory
from dimensional reduction of a D3-brane in a warped Calabi-Yau
compactification of type IIB string theory with imaginary self-dual
3-form flux, including effects of D3-brane motion beyond the probe
approximation, and find the metric on the moduli space of brane
positions, the universal volume modulus, and axions descending from
the 4-form potential. As D3-branes may be considered as carrying
either electric or magnetic charges for the self-dual 5-form field
strength, we present calculations in both duality frames. Our
results are consistent with, but extend significantly, earlier
results on the low-energy effective theory arising from D3-branes in
string compactifications.  }


\begin{document}
\maketitle
\flushbottom

\section{Introduction}
\label{sec:intro}

In any theory with extra dimensions, a first step towards
understanding its dynamics is to construct a low energy, 4-dimensional
effective theory. In such an effective description, the dynamical 
degrees of freedom arise as particular fluctuations of
the fields present in the higher dimensions. As a classic example,
5-dimensional gravity on a circle gives rise at low energies to a
4-dimensional effective theory with gravity, a $U(1)$ gauge field, and
a scalar, all of which arise as fluctuations of the bulk 5-dimensional
metric.
In string theory, the low energy limit of compactification also leads
to a 4-dimensional effective theory in which the 4-dimensional degrees
of freedom arise from fluctuations of the original 10-dimensional
fields. These can include bulk fields such as the metric and $p$-form
gauge fields, as well as localized sources like branes. The details of
the effective theory depend on the details of the compact
space and other fields of the higher-dimensional background, and
much work has gone into deriving these effective theories 
(see \cite{Grana:2003ek,hep-th/0403067,Grimm:2004ua,hep-th/0509003} for
example). Typically, these effective theories are derived by dimensionally 
reducing the higher dimensional action for some particular ansatz of the higher
dimensional fields (specifically a zero mode of the appropriate differential
operator on the extra dimensions). However, as we emphasize, 
care must be taken that the ansatz chosen
is a consistent solution to the higher-dimensional equations of motion.

One set of essential ingredients in string compactifications are the dynamics 
of localized sources such as D-branes.
D-branes arise in string compactifications as sources of Standard-Model-like 
fields \cite{Maharana:2012tu}, 
supersymmetry-breaking uplifiting \cite{Kachru:2003aw},
and sources of cosmic inflation (as in \cite{Kachru:2003sx}; see also 
the reviews \cite{Burgess:2013sla,Baumann:2014nda}).
The 4-dimensional effective theory for D-branes is commonly obtained by 
dimensional reduction of 
the localized Dirac-Born-Infeld and Chern-Simons actions in the probe 
approximation. However, the probe approximation cannot address 
several related conceptual problems.

\begin{figure}[t]
\centering \includegraphics[width=.9\textwidth]{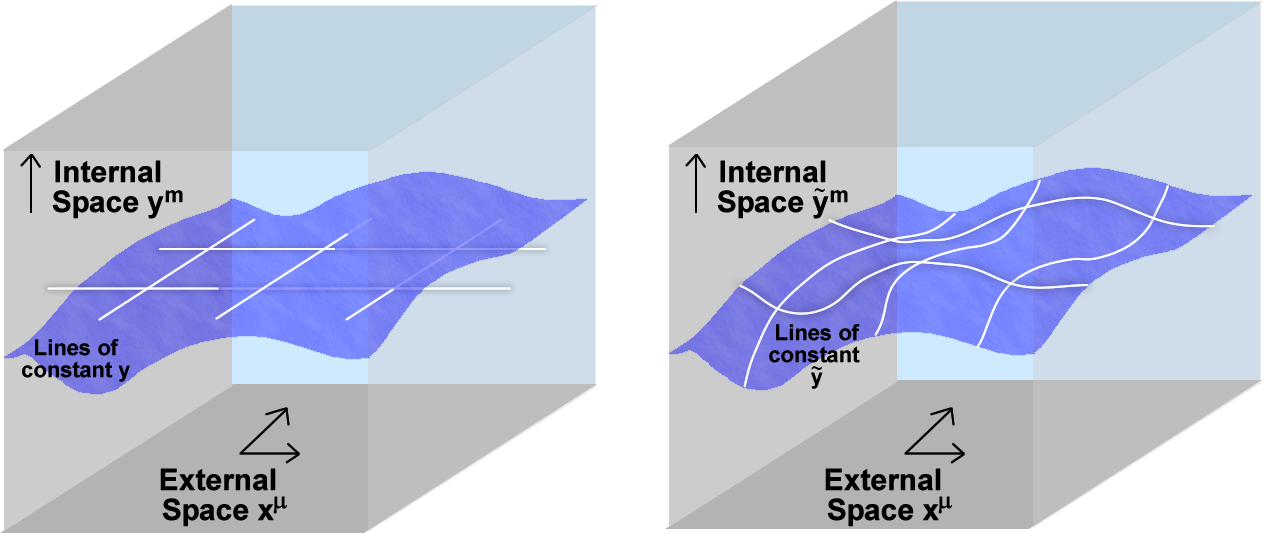}
\label{fig:DBraneFluct}
\caption{A spacetime-dependent fluctuation of a brane can be described by 
its transverse coordinates (left). However,
it is also possible to redefine the coordinates in a spacetime-dependent way, 
so the brane fluctuation is ``gauged away'' (right). In this
case, the brane fluctuation is ``eaten'' by the metric.}
\end{figure}

As an example, consider a D$p$-brane 
in an internal space described by the coordinates $\{y^m\}$, 
so that the brane spans 4-dimensional
spacetime $\{x^\mu\}$ embedded with coordinates $Y$.
Let us write the D-brane embedding as a constant reference value plus a 
spacetime-dependent fluctuation $Y = Y_0 + \delta Y(x)$.
In the probe approximation, the D-brane degrees of freedom are described by 
the spacetime-dependent transverse coordinates 
$\partial_\mu Y = \partial_\mu \delta Y(x)$ and are independent fluctuations 
in 10-dimensions.
However, when the brane is coupled to gravity it is always possible to make 
a spacetime-dependent coordinate redefinition (diffeomorphism) of the transverse
coordinates $y \rightarrow \t y, Y \rightarrow \t Y$ so hypersurfaces of 
constant $\t y$ coincide with the worldvolume of the D-brane, 
as in Figure \ref{fig:DBraneFluct}.
The brane embedding coordinates no longer encode the D-brane degrees of 
freedom, since they are now, by definition, spacetime-independent 
$\partial_\mu \t Y = 0$.
The D-brane degrees of freedom have been ``eaten'' by the metric, 
and the effective theory for the D-brane degrees of freedom now arises
not from the localized sources but from the dynamics of the metric.
It is interesting to note that this lack of diffeomorphism invariance of the 
D-brane transverse coordinates implies that the true diffeomorphism
invariant degrees of freedom describing the motion of D-branes are 
a combination of the transverse coordinates and the metric -- a combination
of open and closed string sectors.

One might think that the simplest resolution is just to fix the gauge
such that all of the degrees of freedom are found in the transverse
coordinates and not in the metric. 
Unfortunately, the linearized higher dimensional equations of motion (EOM) do not
allow one to choose a gauge with vanishing metric fluctuations.
As an example, consider a scalar field in a 4-dimensional homogeneous and
isotropic cosmological background. Fluctuations of the scalar field about a 
time-dependent background $\phi = \phi_0(t) + \delta\phi(t,x)$ can be removed 
by an appropriate time redefinition $t\rightarrow \tilde t(t,x)$ so
that the true gauge-invariant degree of freedom is a combination of scalar 
field and metric fluctuations \cite{Mukhanov:1990me}.
It is not possible to work in a fixed gauge in which the fluctuation only 
appears in the scalar field, setting the metric
fluctuations to zero, because there is a contribution to the off-diagonal 
Einstein equations of the form
\be
0 = G_{0i} - \kappa_4^2 T_{0i} = -\frac{1}{2}\kappa_4^2\, \dot \phi_0\, 
\partial_i \delta \phi\, .
\ee
These off-diagonal Einstein equations act as non-dynamical constraints on 
the fluctuations, and must be solved for a consistent dynamical description 
of the scalar perturbations, even at linear order. As is well known, the
correct ansatz for scalar cosmological perturbation theory includes 
simultaneous fluctuations of the metric and scalar field,
allowing one to consistently solve the constraint equations.
The metric fluctuations then play an important role in determining the 
equation of motion for the dynamics of the scalar degree of freedom.

Similarly, there is no consistent gauge in which the metric does not 
contain any of the degrees of freedom for the D-brane. In particular,
an ansatz with a static metric but spacetime-dependent D-brane fails in
two ways even at linear order.  First, the change in brane position 
backreacts on the 10-dimensional metric (specifically the warp factor).
In addition, there is a contribution 
to the off-diagonal 10-dimensional Einstein equations through the
energy-momentum tensor of the form
\be
0 = G_{\mu m} - \kappa_{10}^2 T_{\mu m} = - \kappa_{10}^2 T_3 \partial_\mu Y_m(x) 
\delta^6(y,Y)\, .
\ee
These constraints on the D-brane motion arise from the required gauge-fixing
described above, much like the Gauss law constraint of electromagnetism 
or the Hamiltonian and momentum constraints of gravitation.
As a result, the probe-brane effective theory based on the D-brane 
degrees of freedom residing entirely in the localized transverse coordinates is
incomplete.  The appropriate 10-dimensional ansatz will require 
including parts of the D-brane degrees of freedom in both the metric and the 
transverse coordinates, as we will see.  

One concern with moving beyond the 
probe approximation is that inserting the backreacted brane solution into 
the effective action could cause a ``self-energy'' problem in which the 
effective action diverges. We explicitly show in section 
\ref{sec:actionelectric} that when one \emph{carefully} performs a 
dimensional reduction which solves the constraint equations the resulting 
effective action does not contain any such divergent self-energy terms.

In this paper, as a test case and for concreteness, we will focus on the 
dynamics of D3-branes in the type IIB backgrounds of the form given in
\cite{hep-th/9605053,Dasgupta:1999ss,Greene:2000gh,hep-th/0105097}
(commonly called GKP compactifications).
In these backgrounds, 
the positions of D3-branes (along with metric K\"ahler moduli and various
axions) are moduli, and the effective theory is described by a 4D supergravity 
(possibly with spontaneous supersymmetry breaking).  As a result, 
the effective theory arises from a K\"ahler potential (see for example
\cite{hep-th/0208123,Kachru:2003sx,Grana:2003ek}). 
One challenge for the construction of an effective theory in GKP backgrounds
is that the metric becomes a warped product between the internal and external 
spaces, complicating the identification of the degrees of freedom.
The supersymmetry of the background along with the fact that scaling the
warp factor can be removed with a 10D diffeomorphism has allowed 
\cite{Koerber:2007xk,Martucci:2009sf,Martucci:2014ska,Martucci:2016pzt} 
to derive many aspects
of the effective theory without a direct dimensional reduction.

However, to work with generic warped compactifications without so much
structure, it is necessary to develop techniques for dimensional reduction
in warped metrics. For example, \cite{Underwood:2010pm} showed that 
the dilaton and volume modulus are identified under diffeomorphisms for
warped backgrounds with a dilaton profile.  
Fortunately, work on warped effective field theory (see 
\cite{hep-th/0201029,hep-th/0507158,0805.3700,Shiu:2008ry,0810.5768,Underwood:2010pm,1308.0323}) 
has led to some useful formalisms \cite{0805.3700,Underwood:2010pm} 
for constructing and analyzing fluctuations of these 10-dimensional
backgrounds and their effective theories.  In \cite{0810.5768}, the effective 
theory for the universal volume modulus and its associated axion was derived
for GKP compactifications, 
while \cite{1308.0323} improved and extended this analysis to include the 
rest of the axion sector.\footnote{A recent analysis in 
\cite{Martucci:2016pzt} shows that the supersymmetric formalism 
\cite{Koerber:2007xk,Martucci:2009sf,Martucci:2014ska} in fact agrees with
direct dimensional reduction \cite{1308.0323} in detail for GKP 
compactifications.}
Our first-principles approach is to choose an ansatz in a fixed gauge
(for diffeomorphisms and the supergravity form fields) and ensure that it
solves all the constraints of the 10D theory even when the 4D fluctuation
is off shell.

An initial attempt \cite{0905.4463} to derive the effective theory for
D3-branes in these backgrounds was unable to solve the constraint
equations coming from the higher-dimensional EOM and
was unable to derive the terms (including mixing terms) necessary to
construct the K\"ahler potential and effective theory. We will use the
formalism and techniques developed in
\cite{Underwood:2010pm,1308.0323} to obtain an ansatz which does solve
all of the higher dimensional EOM and to derive all
the necessary terms in the effective theory.  It is not essential that
the D3-brane we study in this paper is localized in a strongly warped
region: the effects discussed above are inherent to the D-brane's
interaction with the background, and thus will be present regardless
of the strength of the warping at the location of the D3-brane.

In fact, we will present two calculations of this effective action.
The self-duality of the type IIB five-form causes the action to vanish
if all the components of the 10D covariant tensor are included.  Following
\cite{hep-th/0105097}, we discard half the components of the five-form
in the dimensional reduction of the action.  If we keep components with
legs mostly along the external dimensions, the D3-brane carries an 
electric charge for the five-form; keeping the complementary set of 
components leads the D3-brane to carry magnetic monopole charge.  In the
former case, axions of the 4D theory arise as 2-form degrees of freedom,
and cross terms between the brane position and axion descend from the 
brane's Wess-Zumino action.  Meanwhile, in the 
latter case, we solve the nontrivial Bianchi identity by finding a 
field redefinition that gives the 5-form field strength
an explicit dependence on the brane position, and cross terms in the kinetic
action arise directly through backreaction of the brane on the field strength.  
This procedure is related to Dirac's original proposal for a Lagrangian
describing the coupling of magnetic monopoles and Maxwell fields
\cite{Dirac:1948um}; the relationship of our procedure to Dirac's theory
and its generalization to branes will appear in a forthcoming paper 
\cite{cownden}.

The plan of our paper is as follows.  
In section \ref{sec:background}, we review the background compactification,
the procedure for dimensional reduction, and the EOM of
type IIB supergravity with D3-branes.  We then give the dimensional reduction
in the version of the theory in which the D3-brane carries electric charge
for the five-form in \ref{sec:electric}, followed by the analogous 
calculation for the magnetically charged D3-brane in section \ref{sec:magnetic}.
In these two sections, we provide a short summary at the end of each
subsection encapsulating the important results as an aid to the reader.
We close with a discussion; conventions and formalism are found in the
appendices.

%%%%%%%%%%%%%%%%%%%%%%%%%%%%
%%%%%%%%%%%%%%%%%%%%%%%%%%%%
%%%%%%%%%%%%%%%%%%%%%%%%%%%%

\section{Background and Fluctuations}
\label{sec:background}

%%%%%%%%%%%%%%%%%%%%%%%%%%%%

\subsection{Flux Compactifications}% and D3-branes}
\label{sec:review}

To set our conventions, we work in the bosonic sector of
type IIB string theory as described by the low-energy supergravity (SUGRA) 
\be
\label{SIIB}
S_{\mathrm{IIB}} &=& \frac{1}{2\kappa_{10}^2} \int d^{10} x \sqrt{-g} 
\left( \mathcal R - \frac{\p_M \tau \p^M \bar\tau}{2 \, (\mathrm{Im} \, \tau)^2}
\right) - \frac{1}{2 \kappa^2_{10}} \int \left[ 
\frac{G_3 \w \star \bar G_3}{12 \, \mathrm{Im} \, \tau} + 
\frac{1}{4} \til F_5 \w \star \til F_5 \right. \nonumber \\
&& \left. + \frac{i}{4 \, \mathrm{Im} \, \tau} C_4 \w G_3 \w \bar G_3 \right] 
+ S_{\mathrm{loc}} \, ,
\ee
where $\mc R$ is the ten-dimensional (10D) Ricci scalar, and $\kappa_{10}$ is the 10D Newton's 
constant. We have defined the axio-dilaton $\tau = C_0 + ie^{-\phi}$,
combined the 3-forms $F_3 = d C_2$ and $H_3 = dB_2$ into the complex 3-form 
$G_3 = F_3 - \tau H_3$, and defined the 
five-form field strength as $\t F_5 = dC_4 - C_2 \w H_3$, 
which is constrained to 
be self-dual: $\t F_5 = \star \t F_5$. $S_{\mathrm{loc}}$ is the
action for all the local objects, including D3-branes.

The background fields of GKP compactifications take the form
\be
d s^2_{10}&=& e^{2A^{(0)}(y)} \h\eta_{\mu \nu} dx^\mu  dx^\nu + e^{-2A^{(0)}(y)} 
\t g_{mn} dy^m dy^n \, , \nonumber \\
\t F_5^{(0)} &=& \hat \epsilon \w \t d e^{4A^{\0}} + \t \star \t d e^{-4A^{\0}}\, ,
\ \  \t \star G_3^{(0)} = i G_3^{(0)}\, ,
\label{back metric}
\ee
where $\{x^\mu\}$ spans 4-dimensional (4D) spacetime and $\{y^m\}$ 
spans the internal dimensions.
The unwarped metrics $\h\eta_{\mu\nu}$, $\t g_{mn}$ are respectively
Minkowski and Calabi-Yau (CY).
We denote objects constructed with respect to $\h\eta_{\mu\nu}$ with a hat
and those with respect to $\t g_{mn}$ with a tilde.
The three-form field strength is harmonic, and 
the background warp factor, $A^{\0}(y)$, obeys a Poisson equation 
\be
\label{warp with sources}
\wtn^2 e^{-4A^{\0}} = - \frac{g_s}{2} \left| G^{(0)}_3 \right|^{\tilde 2} 
- 2 T_3 \kappa^2_{10} \til \delta^6 (y,Y^{\0})-\cdots \, ,
\ee
where $\cdots$ are other local sources including more D3-branes (which 
also contribute delta function sources), D7-branes, and
negative charge and tension orientifold O3- and O7-planes. 
Although only a single D3-brane (located at $Y$) is considered here and 
throughout, the extension to multiple, non-interacting branes of this type 
is trivial. Furthermore, we work in the orientifold limit
where four D7-branes are coincident with each O7-plane for simplicity, although
we expect our results to generalize straightforwardly to F-theory.

A nontrivial $G_3^{(0)}$ also stabilizes moduli of the compactification,
generically including all the complex structure moduli of $\t g_{mn}$ and
the axiodilaton $\tau$.  (See \cite{1308.0323} for a more complete review of 
the effects of the flux on moduli stabilization.)  As a result, we
assume that the complex structure and $\tau$ are constant.  In addition,
the background is supersymmetric at the classical level 
when $G_3^{\0}$ is $(2,1)$ in the CY
complex coordinates;\footnote{And primitive, which is automatically
satisfied on a generic CY.} if the flux breaks supersymmetry, it does so
spontaneously \cite{hep-th/0201029}, and the effective theory can be
described by 4D SUGRA.
% (nonperturbative effects such as gaugino condensation
%can restore supersymmetry, though additional supersymmetry-breaking
%contributions from anti-branes are often considered as part of a de Sitter
%compactification \cite{Kachru:2003aw}). 

With $\tau$ assumed to be constant, we can write $G_3=dA_2$ in terms of a 
complex potential $A_2=B_2-\tau C_2$.  In this case, it is common to redefine 
$C_4$ to set $\t F_5=dC_4+(ig_s/4)(A_2\w \bar G_3-\bar A_2\w G_3)$.  We 
will use this definition for $\t F_5$ henceforth.

%%%%%%%%%%%%%%%%%%%%%%%%%%%%

\subsection{Dimensional Reduction and the Kinetic Action}
\label{sec:quadratic}

To determine the effective 4D theory, we must first decompose the 10D fields
into orthogonal modes, each of which corresponds to a 4D degree of freedom.
In a product space compactification, these modes are simply the eigenfunctions
of some second order differential operator on the compact space (for example,
the Laplacian for a scalar or the Hodge-de Rham operator for a form field). 
However, the constraints described in the introduction 
complicate matters somewhat
for warped compactifications.  We are particularly interested in clasically 
massless moduli, so we consider only those modes that satisfy the massless 
Klein-Gordon equation in the external $x^\mu$ directions.  The following
sections will describe the structure of these massless
modes; here, we will outline the procedure of dimensionally reducing the
action once the linearized modes are known, following \cite{1308.0323}.

To find the two-derivative kinetic action in terms of a metric $G_{ab}(\phi)$ on
moduli space, we need find only the action to quadratic order in fluctuations
around the background --- as long as we expand the background around an
arbitrary point $\phi_0$ in moduli space, we recover the full dependence of
the metric on the moduli.  While our primary consideration is D3-brane
position moduli, we are interested in the metric on the K\"ahler moduli
space of the CY.  
We therefore also consider the universal volume modulus and axions
descending from the 4-form potential (which are partners of the K\"ahler
metric moduli).  Because the 4D effective theory is a SUGRA, the moduli
space is naturally described in terms of holomorphic coordinates, and the 
metric is defined in terms of a K\"ahler potential.

This quadratic action can be written in terms of the first-order 
fluctuations and the linearized EOM, as demonstrated in 
\cite{1308.0323} (and used implicitly in \cite{Shiu:2008ry}).  Specifically,
for type IIB SUGRA, we can write
\be
S &=& \frac{1}{4\kappa_{10}^2} \int d^{10}x\, \sqrt{-g}\, \delta g^{MN}
\delta E_{MN} +\frac{1}{4\kappa_{10}^2}\int\left(\delta C_4\w\delta E_6+
\frac{g_s}{2}\delta A_2\w\delta\bar E_8 
+\frac{g_s}{2}\delta\bar A_2 \w\delta E_8 \right)\nonumber\\
&&+\frac{T_3}{2} \int d^{10}x\, \sqrt{-g}\,\int d^4\xi\,
\sqrt{-\gamma}\, %\delta^{10}(x,X)\,
\delta X^{\sla M}\delta E_{\sla M}\ ,\label{quadaction1}
\ee
where $\delta E_{mn}$ is the linearized Einstein equation, $\delta E_6$
the linearized EOM for the 4-form, $\delta E_8$ for
the 2-form potential, and $\delta E_{\sla M}$ for the
D3-brane position. 
The EOM for the
world-volume metric $\gamma_{ab}$ (as described below) 
are higher-order (and constraints), so they do not contribute 
(though $\gamma_{ab}$ should be evaluated on the solution).
Note that we have labeled the brane embedding coordinate
$X^{\sla M}(\xi)$ with a slashed index to indicate that tensors at that
point do not contract with tensors at a general point $x^M$ in spacetime,
and this term is integrated over the brane worldvolume coordinates as
well as spacetime.  Furthermore, these are the EOM as they directly arise from
the variation of the action; in particular, the equations of type IIB
SUGRA are often reorganized to remove terms proportional to $E_6$
from $E_8$.  

There are also two subtleties to note. First, there is not
actually a covariant action in 10D that respects the 
self-duality of the 5-form field strength,\footnote{For the usual fields of
IIB SUGRA; Sen \cite{Sen:2015nph} has recently (during preparation of this
manuscript) described a covariant action with different field content.} 
so (\ref{quadaction1}) includes
only some of the components of $C_4$ and the corresponding EOM.  
We describe our approach to this subtlety in \S\ref{sec:SUGRAwD3}
below.  Second, \cite{1308.0323} arrives at the action (\ref{quadaction1})
after an integration by parts; in some cases, there are terms in the action
which are quadratic in fluctuations but contribute no terms to the linearized
EOM (for example, the axionic coupling $\theta F^2$ where
the field strength is first order).  These terms must be added separately
to the quadratic action, as we will see in section \ref{sec:actionelectric}.

The action (\ref{quadaction1}) should also be understood in a fixed gauge
(for diffeomorphisms, form gauge transformations, and worldvolume
reparameterization).  We choose a gauge in which the underlying CY metric
$\t g_{mn}$ of the internal directions does not fluctuate (and form potential
gauge as described below); for D3-branes with internal positions $Y^{\sla m}$ 
that vary slowly over the external spacetime, we find it convenient to use
a static gauge $\xi^a = \delta^a_{\sla \mu} X^{\sla\mu}(\xi)$ 
so that fluctuations of  $X^{\sla\mu}$ are gauged away (with 
$|\del_a Y^{\sla m}|$ assumed to be small).

%%%%%%%%%%%%%%%%%%%%%%%%%%%%

\subsection{Equations of Motion with D3-brane Sources}
\label{sec:SUGRAwD3}

In this section, we summarize the 10-dimensional EOM for 
type IIB supergravity with D3-brane sources.  The bulk SUGRA
equations are well-known \cite{Schwarz:1983qr}, but the brane sources are
less familiar.  We therefore simply state the bulk terms but provide a brief
derivation of the D3 sources. As in section \ref{sec:quadratic}, we 
use slashed indices at the D3-brane position; the parallel propagator 
$\Lambda_M^{\sla M}(x,X)$ (or its inverse) switches index type for fields 
evaluated at coincidence (i.e. when multiplied by a delta function).

The variation of the action
with respect to the metric, $E_{MN}$, is
\be
\label{EFE}
E_{MN} = R_{MN} - \frac{1}{2} g_{MN} \mc R - (T^5_{MN} +T^3_{MN}+ T^{D3}_{MN} + T^{loc}_{MN})\, ,
\ee
so that setting $E_{MN} = 0$ gives the 10-dimensional Einstein equations.
The contributions to the energy-momentum from the 5-form flux (given 5-form
self-duality) and background 3-form are
\be
\label{T5}
T^5_{MN} = \frac{1}{4 \cdot 4!} \til F_{MPQRS} \til F_N{}^{PQRS} \, ,\quad
T^3_{MN}=\frac{g_s}{4} \left(G_{(M}{}^{PQ} \bar G_{N)PQ}-g_{MN} |G|^2\right)\ .
\ee
The D3-brane coupling to the metric enters in the Dirac-Born-Infeld (DBI)
terms in the brane action; if we ignore couplings to the 2-form potentials
and world-volume gauge field (justified below),
we can replace the usual DBI action for the brane with a 
classically equivalent Polyakov-like form
\be
\label{SDBI}
S_{DBI} = -\frac{T_3}{2}  \int d^{10}x \sqrt{-g} \int d^4 \xi \sqrt{-\gamma} \, 
\left[\gamma^{ab} P(g)_{ab}-2\right] \delta^{10} (x,X(\xi)) \, ,
\ee
where $\gamma_{ab}$ is an independent worldvolume metric which equals
the induced metric $P(g)_{ab}= g_{\sla M\sla N}\p_a X^{\sla M} \p_b X^{\sla N}$ 
on shell. Regarding delta functions in curved spacetime, we refer the reader to appendix~\ref{sec:bitensors}.  The resulting energy-momentum from D3-branes is then
\be
\label{TD3}
T^{D3}_{MN} \equiv - \frac{2\kappa_{10}^2}{\sqrt{-g}}
\frac{\delta S_{DBI}}{\delta g^{MN}} = 
- \kappa_{10}^2 T_3 \int d^4 \xi \sqrt{-\gamma} \, \gamma^{ab} \Lambda_M^{\sla M}
\Lambda_N^{\sla N}g_{\sla M\sla P} g_{\sla N\sla Q} \p_a X^{\sla P} \p_b X^{\sla Q} 
\delta^{10}(x,X(\xi)) \, .\hspace{.2in}
\ee
A similar expression leads to the energy-momentum $T_{MN}^{loc}$ 
from localized sources other than the mobile D3-brane of interest.
For notational convenience, we will often make the parallel propagators 
implicit, writing for example $g_{M\sla P}=\Lambda_M^{\sla M}g_{\sla M\sla P}$.

Self-duality of $\til F_5$ raises some complications for dimensional reduction
at the level of the action.  Specifically, evaluating (\ref{SIIB}) 
on a self-dual 5-form leads to a vanishing kinetic term, whether on or
off shell.  The prescription we follow is to replace (\ref{SIIB}) by
a non-covariant action, retaining only half the components of $\til F_5$ 
and doubling the coefficient of the $\til F_5 \w \star \til F_5$ term in the
action.  We will make two distinct choices for the sets of components to keep:
the ``electric'' set with 4 or 3 legs on $x^\mu$ and 1 or 2 respectively on
$y^m$, and the ``magnetic'' set with 0 or 1 leg on $x^\mu$ and 5 or 4 on
$y^m$ (an equal number of components with 2 on $x^\mu$ and 3 on $y^m$ fall
in each set, but these all vanish for the moduli we consider).  The
corresponding components of $C_4$ for the electric set couple electrically to
the D3-brane, whereas the D3-brane is a magnetic source for the magnetic
set.\footnote{Technically, there is a WZ action coupling $C_4$ to the 
brane, but it appears at higher order in spacetime derivatives than we 
consider.}
It is worth noting that the components in each set are Hodge dual to
the components in the other set.

We begin by considering the theory for the electric components.  In this
case, the Bianchi identity is trivial ($d\til F_5=0$), and the source 
for $d\star\til F_5$ arises through the D3-brane Wess-Zumino (WZ) action
(again ignoring couplings to the 2-form potentials for now)
\be
S_{WZ} &=& \mu_3 \int d^{10} x \sqrt{-g} \int_\xi P(C_4) \delta^{10} (x, X(\xi))
\nonumber\\
&=&\mu_3 \int d^{4} \xi\sqrt{-\gamma} \int_{10} C_4 \w \star \, 
\epsilon_\| \, \delta^{10}(x,X(\xi))\, , \label{WZ}
\ee
where $\epsilon_\|$ is the push-forward of the antisymmetric world-volume
tensor. We define the push-forward as 
\be\label{epparallel}
\epsilon_\|^{\sla M\sla N\sla P\sla Q} \equiv \epsilon^{abcd}\p_aX^{\sla M}\p_bX^{\sla N}
\p_cX^{\sla P}\p_dX^{\sla Q} \, ,
\ee
and take parallel propagators to be implicit in the 10D Hodge star.
As we will see below, D3-branes which are mutually BPS with the background
have charge equal to tension $\mu_3=T_3$.
The resulting 5-form EOM, including contributions from the 3-form, 
is therefore
\be
E_6 = d\star \tilde{F}_5 -\frac{ig_s}{2} G_3\w\bar G_3
+ 2\kappa_{10}^2 T_3 \int d^4 \xi \sqrt{-\gamma} \, \star 
\epsilon_\| \, \delta^{10}(x,X(\xi)) \, ,
\label{E6}\ee
and vanishes on shell.\footnote{There is one subtlety with factors of 2; 
when varying the covariant action \eqref{SIIB}, we replace $\mu_3\to\mu_3/2$ 
in $S_{\mathrm{loc}}$ since the WZ term includes magnetic and electric couplings.
This is equivalent to keeping only the electric components of $\t F_5$
and doubling the coefficient of the $\t F_5 \star\t F_5$ term as in
\cite{hep-th/0105097}.}  
Since the 3-form has at most one leg in the
external spacetime, $G_3$ contributes only to the EOM and not the Bianchi
identity.

The EOM and Bianchi identity for the magnetic components of 
$\til F_5$ are simply given by exchanging 
$\til F_5\leftrightarrow\star\til F_5$ in the corresponding equations for
the electric components.  Therefore, the EOM is
$E_6=d\star\til F_5$, while the Bianchi identity becomes
\be
\label{BianchiF5}
0 = d \t F_5 -\frac{ig_s}{2} G_3\w\bar G_3
+ 2\kappa_{10}^2 T_3 \int d^4 \xi \sqrt{-\gamma} \, \star \epsilon_\| 
\, \delta^{10}(x,X(\xi)) \, .
\ee
We will consider both (\ref{E6}, \ref{BianchiF5}) in the static gauge later.
As discussed in \cite{hep-th/0201029,0810.5768,1308.0323},
the nontrivial Bianchi identity (\ref{BianchiF5}) for the magnetic
components means that perturbations of $C_4$ as defined in (\ref{SIIB})
and below are not globally defined on the CY manifold.  Those
references studied the Bianchi identity in the absence of D3-branes and
found a field redefinition with a globally-defined 4-form potential.
In terms of the new $C_4$ (in the absence of axions
descending from $A_2$), perturbations of $\t F_5$ are
$\delta\t F_5=d\delta C_4+(ig_s/2)(\delta A_2 \bG-\delta\bar A_2 \G)$.
We will review this field redefinition and demonstrate the additional
redefinition needed for the D3-brane contribution to the Bianchi identity
in section \ref{sec:magnetic}.

The 3-form is somewhat simpler; the Bianchi identity is trivial $dG_3=0$
for constant axio-dilaton $\tau$, and the EOM is
\be\label{E8} E_8=d\star G_3+iG_3\w\left(\t F_5+\star\t F_5\right) 
+\frac i2 A_2\w E_6+\cdots \, ,
\ee
with either electric or magnetic components for $\t F_5$.  The coefficients
of the last two terms are given in keeping with our prescription to double 
the $\t F_5$ kinetic term.  The dots represent brane source terms, 
which we ignore as discussed below.
Since we do not consider axions that descend from $A_2$ or their backgrounds, 
the last term given explicitly in (\ref{E8}) will not contribute. 

Finally, we need to determine the 10-dimensional D3-brane EOM. 
Due to the different couplings to the electric and magnetic components of 
the 5-form, these equations take different forms depending on which version
of the theory we consider.  Here we present the more familiar
electric version and discuss the magnetic version in section
\ref{sec:magnetic} below.
Using the normalization of (\ref{quadaction1}), variation of 
(\ref{SDBI}, \ref{WZ}) gives\footnote{The Euler-Lagrange equations also
apparently contain terms proportional to $\del_{\sla M} \delta^{10}(x,X)$;
however, in the variation of the action, these terms vanish upon converting
the $X^{\sla M}$ derivative to an $x^M$ derivative and integrating by parts.
We therefore do not consider them to be part of the EOM.\label{deltaderivEOM}}
\be
E_{\sla M} &=& \left\{\Del_a\left[\left(g_{\sla M\sla N}\del^a X^{\sla N}
+\frac 16 \frac{\mu_3}{T_3}\epsilon^{abcd}C_{\sla M\sla N\sla P\sla Q}
\del_b X^{\sla N}\del_c X^{\sla P}\del_d X^{\sla Q}\right)\right]\right.
\label{ED3_1}\\
&&\left.
-\left[\frac 12 \del_{\sla M}g_{\sla N\sla P}\del_a X^{\sla N}\del^a X^{\sla P}
+\frac{1}{4!}\frac{\mu_3}{T_3} \epsilon^{abcd}\del_{\sla M}
C_{\sla N\sla P\sla Q\sla R}\del_a X^{\sla N}\del_b X^{\sla P}\del_c X^{\sla Q}
\del_d X^{\sla R}\right]\right\}\delta^{10}(x,X)\ .\nonumber
\ee
This EOM allows us to set our sign convention for the D3-brane
charge (since $|\mu_3|=T_3$).  
Consider a static D3-brane $\del_a Y^{\sla m}=0$ in static gauge
in the background described above.  The $\sla M=\sla \mu$ equation becomes
$\Del_a\left[\left(1-\mu_3/T_3\right)\delta^{10}(x,X)\right]=0$, while the
$\sla M=\sla m$ equation becomes 
$-4\del_{\sla m}A\left(1-\mu_3/T_3\right)\delta^{10}(x,X)=0$.  Both of these
vanish for the choice $\mu_3=T_3$.

We now justify ignoring the 2-form potential couplings in both the DBI and
WZ actions, despite the fact that they appear in the background with
nontrivial $\G$.  The main point is that the background potentials have
completely internal legs and are pulled back to the world volume by two
powers of the small derivatives $\del_a Y^{\sla m}$.  Furthermore, in
a perturbative expansion of the DBI action, the pulled-back potential $P(B_2)$ 
enters either at second-order, or contracted with the world-volume field
strength.  These terms are 3rd and 4th order in fluctuations, so we ignore 
them.  The WZ terms containing $P(C_2)$ and $P(B_2)$ are similarly 3rd and
4th order and can also be ignored.  Finally, we set the world-volume
field strength to zero, since we do not consider vector degrees of freedom.

%%%%%%%%%%%%%%%%%%%%%%%%%%%%
%%%%%%%%%%%%%%%%%%%%%%%%%%%%
%%%%%%%%%%%%%%%%%%%%%%%%%%%%

\section{Electric D3-brane Couplings}
\label{sec:electric}

As we discussed in section \ref{sec:SUGRAwD3} above, because of
the 10D self-duality of $\t F_5$, a D3-brane can act as either an 
electric or magnetic
source for $C_4$. In this section, we consider the ``electric'' 
choice as defined above, namely, the choice to keep components of $\t F_5$
with legs mostly along the external spacetime.  
With this choice, 
a D3-brane has an electric coupling to $C_4$ through the WZ terms in its
action.

We begin by first presenting our ansatz for fluctuations in the
D3-brane position. As discussed in the introduction, an ansatz for
dimensional reduction must satisfy the constraint equations arising
from the 10D EOM (\ref{EFE}, \ref{E6}, \ref{ED3_1}). We will show how
our ansatz --- which involves  the presence of D3-brane degrees of
freedom not only as transverse coordinates but also in the  metric and
4-form gauge potential --- solves these constraint equations and is
thus the first known consistent ansatz for the dimensional reduction
of transverse D3-brane degrees of freedom.  In order to compute the
full effective action of the D3-brane, we need to understand how it
couples to the massless moduli, such as the volume modulus and $C_4$
axions.  Dimensional reduction of the volume modulus and axions in GKP
compactifications has been studied previously in
\cite{0810.5768,1308.0323}; however, those articles worked with the
other (magnetic) choice for components of $\t F_5$, which leads to a
slightly different ansatz for the linearized fluctuations.   Thus, we
briefly discuss the electric form of linearized fluctuations of these
moduli.  We will then see that these moduli can all be described by a
common ``electric'' ansatz, which we use to carry out the dimensional
reduction of the action to a 4D effective theory.

%%%%%%%%%%%%%%%%%%%%%%%%%%%%

\subsection{D3-brane Fluctuations in Electric Formalism}
\label{sec:ElectricD3}

%%%
% D3 Ansatz
%%%

As discussed in section \ref{sec:quadratic}, we will consider a
D3-brane with embedding coordinates $X^{\sla M} = \left\{ X^{\sla
  \mu}, Y^{\sla m}\right\}$ with  world volume parameterization in
static gauge $\xi^a = \delta^a_{\sla \mu} X^{\sla \mu}$ (so that
fluctuations in $X^{\sla \mu}$ are gauged away) and take small,
slowly varying fluctuations of the transverse coordinates of the
D3-brane $Y^{\sla m}(x) = Y^{\0\sla m} + \delta Y^{\sla m}(x)$.

As discussed in the introduction, diffeomorphisms and the constraint equations 
couple the D3-brane with the metric and 4-form $C_4$,
forcing us to go beyond the probe limit for the D3-brane.
Thus, fluctuations of the D3-brane transverse position appear in the metric at
linear order through the ansatz
\be
\label{metricansatzD3}
ds^2 = e^{2\Omega(x)}e^{2A(x,y)} \h\eta_{\mu\nu} dx^\mu dx^\nu 
+ 2e^{2\Omega(x)}e^{2A(x,y)} \p_\mu B_m^Y(x,y) dx^\mu dy^m + e^{-2A(x,y)} 
\til g_{mn} dy^m dy^n\, , \hspace{.3in}
\ee
which uses the same structure as 
\cite{0810.5768,1308.0323} for the volume modulus and axions.  
This is similar to the background
(\ref{back metric}); the additions are spacetime-dependence in the warp factor
$A(x,y)$, a ``compensator'' term containing
a 1-form $B_m^Y(x,y)$ needed to solve the constraints (and which vanishes
for spacetime-independent fluctuations), and a Weyl factor $\Omega(x)$
defined by 
\be\label{weyl}
e^{-2\Omega(x)}\equiv  \frac{1}{\til V}\int d^6y\,\sqrt{\til g}e^{-4A(x,y)} 
\quad \text{with} \:\: \til V\equiv \int d^6y\,\sqrt{\til g}\ .\ee
The Weyl factor is required to diagonalize the 4D graviton 
and warped volume fluctuations.
We work in a diffeomorphism gauge in which the CY metric $\til g_{mn}$
is fixed to its background form.

Fluctuations of the D3-brane position also appear in the electric 
components of the 4-form potential $C_4$ through the Weyl factor, warp 
factor, and compensator
\be
\label{C4electricD3} 
C_4= e^{4\Omega}e^{4A} \h\epsilon + e^{4\Omega}e^{4A}\h\star\h dB_1^Y\ 
\ee
with corresponding 5-form 
\be
\t F_5&=&e^{4\Omega}\h\epsilon\w\t de^{4A}+d\left(e^{4\Omega}e^{4A}\h\star\h dB_1^Y
\right)\nonumber\\
&&+\left[\t\star\t d e^{-4A}+\star d\left(e^{4\Omega}e^{4A}\h\star\h dB_1^Y
\right)-e^{2\Omega}e^{-4A} \t\star\left(\h dB_1^Y\w\t de^{4A}\right)\right]\ .
\label{F5ansatzD3}
\ee
The terms of \eqref{F5ansatzD3} in square brackets are required in the 
10D description of the field strength for self-duality (these are the 
magnetic components of $\t F_5$).

%%%
% Solving Constraints for D3
%%%

The constraints follow from inserting our ansatz 
(\ref{metricansatzD3},\ref{weyl},\ref{C4electricD3},\ref{F5ansatzD3}) into the 
10D EOM (\ref{EFE}), (\ref{E6}), and (\ref{ED3_1}).  We will see that
the constraints have specific solutions for $A(x,y)$, $B_m^Y(x,y)$, and 
$\Omega(x)$.

First, consider the constraints coming from the non-dynamical terms in 
the Einstein equations (see appendix~\ref{sec:ElectricEOM}). In particular, 
the $(\mu \nu)$ component yields
\be
\label{coulomb}
\wtn^2 e^{-4A(x,y)} &=& - \frac{g_s}{2} \left| G^{(0)}_3 \right|^{\til 2} 
- 2 \kappa^2_{10} T_3\til \delta^6 (y,Y)-\cdots \ ,
\ee
as well as
\be
\label{Bdiv}
\til\Del^{\til n} B_n^Y &=& e^{-2\Omega}\delta e^{-4A}-e^{-4A}\delta e^{-2\Omega}\ .
\ee
The first of these, (\ref{coulomb}), promotes the background Poisson
equation for the warp factor (\ref{warp with sources}) to include first-order 
contributions from the D3-brane ``instantaneously'' (i.e., separately at each
point $x^\mu$ in spacetime):
\begin{empheq}[box=\fbox]{align}
e^{-4A^{\0}(y)}\rightarrow e^{-4A(x,y)} &= 2 \kappa_{10}^2 T_3\ \t G(y,Y^{\0} 
+ \delta Y) + \cdots \nonumber \\
 &= 2 \kappa_{10}^2 T_3\ \t G(y,Y^{\0}) + 2 \kappa_{10}^2 T_3\ \delta Y^{\sla m} 
\partial_{\sla m} \t G(y,Y) + \cdots \nonumber\\
&= e^{-4A^{\0}(y)} + 2 \kappa_{10}^2 T_3\ \delta Y^{\sla m} \partial_{\sla m} 
\t G(y,Y) \ , \label{warpD3promote}
\end{empheq}
where the $\cdots$ represent the other contributions to the warp
factor due to fluxes and other localized sources (which are smooth at
$Y$) and $\t G(y,Y)$ is
the biscalar Green's function on the internal CY.\footnote{See 
appendix~\ref{sec:bitensors} for definitions and properties of bitensor
Green's functions.} 
In this sense, our diffeomorphism gauge
choice is analogous to the Coulomb gauge describing electromagnetic 
radiation.
Equation (\ref{warpD3promote}) also shows that the Weyl factor $\Omega(x)$ is 
independent of the D3-brane moduli at linear order, since
\be
\delta_Ye^{-2\Omega} =\frac{2\kappa_{10}^2T_3}{\t V}
\delta Y^{\sla m} \int d^6y\sqrt{\t g}\, \del_{\sla m}\t G(y,Y)
=-\frac{2\kappa_{10}^2T_3}{\t V}\delta Y^{\sla m} \int d^6y\sqrt{\t g}\,
\t\Del_n \t G^n_{\sla n}=0\ , \hspace{.2in}
\label{YOmega}
\ee
via the relation (\ref{Gs}) between scalar and tensor Green's functions.
Nevertheless, we will keep the Weyl factor in our calculations since it is 
important for defining the 4D Einstein frame.

Next, the mixed component of the Einstein equation gives a non-trivial 
constraint
\be
-\frac 12 \del_\mu\del_m e^{-4A} +\frac 12 e^{2\Omega}
\del_\mu\til\Del^{\til n}(\til dB_1^Y)_{mn}+\kappa_{10}^2
T_3\ \t g_{m\sla n} \del_\mu Y^{\sla n}\ \t\delta^6(y,Y) = 0
\label{mumD3constraint}
\ee
as well as a copy of (\ref{coulomb}) multiplied by $\partial_\mu B^Y_m$.
Since $\t\Del^{\t n}(\t dB_1)_{mn} = -\t\Del^2B_m+\t\Del_m\t\Del^{\t n}B_n$
on the Ricci flat CY, (\ref{Bdiv}) and (\ref{mumD3constraint}) yield
\be\label{BYGreen}
\til\Del^{2} B_m^Y = 2\kappa_{10}^2T_3 e^{-2\Omega} \t Y_m \t\delta^6(y,Y) \ ,
\textnormal{ where } \t Y_m = \t g_{m\sla n}\delta Y^{\sla n}\ .\ee 
The compensator is given by the bivector Green's function
\be
\boxed{B_m^Y(x,y) = -2\kappa_{10}^2T_3 e^{-2\Omega}\t g_{mn}
\delta Y^{\sla p}\t G^n_{\sla p}(y,Y)\, .} \label{BD3Solve}
\ee
The Green's function identity \eqref{Gs} implies that $B_1^Y$ automatically
satisfies \eqref{Bdiv}.

The EOM for the 5-form flux (\ref{E6}), evaluated for this ansatz in 
(\ref{app:E6}) in detail, also contributes a constraint equation.
The source term for (\ref{E6}) in static gauge is
\be
\label{staticD3charge}
\int d^4 \xi \sqrt{-\gamma} \, \star \epsilon_\| \, \delta^{10}(x,X(\xi))
=\star \epsilon_\| \, \delta^{6}(y,Y(x))=-\left(\t\epsilon+\h d\t\star\t Y_1
\right)\t\delta^6(y,Y)\ ,
\ee
and the constraints are another copy of (\ref{coulomb}) and
\be
\hat d \left[ \t\star\t de^{-4A} +e^{2\Omega}\t d\t\star\t dB_1
-2\kappa_{10}^2T_3\t\star \t Y_1\t\delta^6(y,Y) \right] = 0 \, .
\label{F5offdiag}
\ee
Equation \eqref{F5offdiag} is the 6-dimensional dual of 
(\ref{mumD3constraint}) and is thus automatically satisfied.

The D3-brane EOM (\ref{ED3_1}) contributes no new constraints; the 
$\sla M = \sla \mu$ component vanishes identically 
(see (\ref{app:ED3mu})) as in the background, 
while the $\sla M = \sla m$ component (shown below in 
\eqref{D3EOMdynamical}) contributes to the dynamical EOM only.

\subsubsection{Summary}
We have shown that the ansatz 
(\ref{metricansatzD3},\ref{weyl},\ref{C4electricD3},\ref{F5ansatzD3}) 
solves the constraints required to describe the motion of a D3-brane in
a GKP background \emph{beyond the probe limit}.  To first order in the 
fluctuation of the brane position, the warp factor, metric and $\t F_5$
compensator, and Weyl factor are (repeating our earlier results)
\begin{align}
\Aboxed{\,e^{-4 A(x,y)} &= e^{-4A^{\0}(y)} + 
2\kappa_{10}^2 T_3\ \delta Y^{\sla m} \partial_{\sla m} \t G(y,Y)\, ,}
\tag{\ref{warpD3promote}}\\
\Aboxed{\,B_m^Y(x,y) &= -2 \kappa_{10}^2  T_3 e^{-2\Omega} \t g_{mn}\ 
\delta Y^{\sla p}\t G_{\sla p}^{n} (y,Y)\, ,\ \ \ \ \ }
\tag{\ref{BD3Solve}} \\
\Aboxed{\,\ \ 
e^{-2\Omega(x)} &= \frac{1}{\t V} \int d^6y \sqrt{\t g}\ e^{-4A^{\0}(y)}
=e^{-2\Omega^{\0}}\, .\ \ \ } \label{WeylD3}
\end{align}
To our knowledge, this is the first ansatz in the literature
that describes the backreaction of the D3-brane on the field strength and
geometry and can be used to perform a consistent dimensional reduction.

\comment{
However, in order to determine the effective metric on the K\"ahler moduli
space, which includes the D3-brane, we need to also consider other K\"ahler moduli, e.g. the volume modulus and $C_4$ axions.
In the next section, we will review the ans\"atze and solutions for these moduli. }

%%%%%%%%%%%%%%%%%%%%%%%%%%%%

\subsection{K\"ahler Moduli in Electric Components}
\label{sec:electricKahler}

Having found an ansatz for the SUGRA fields for a moving D3-brane 
that solves all constraints, we can determine the dynamical EOM and
dimensionally reduce the quadratic action \eqref{quadaction1}.  However,
the full structure of the effective action is apparent only when we include
the complete moduli space.  In this paper, we consider the universal volume 
modulus $c$, 2-form axions $b_2^I$ descending from $C_4$, 
and brane positions $Y^{\sla m}$. Each of the axions is associated with a 
harmonic 2-form $\omega_2^I$ on the unwarped CY manifold (where 
$I=1,\cdots h^{1,1}$ runs over a basis).  In principle, the holomorphic 
moduli of the 4D SUGRA should include all the metric K\"ahler moduli as 
partners of the $b_2^I$; so far, the constraints have only been solved for
the volume modulus.  Some compactifications have additional axionic moduli
that descend from $A_2$.  Solutions to the constraints have been presented
in \cite{1308.0323}, and they can be added to our analysis in a 
straightforward manner.

%%%
%% Other Moduli Ansatz
%%%

Here we present a brief description of the
linearized fluctuations corresponding to all these moduli and the constraint
equations they must solve; more details are presented in appendix
\ref{sec:ElectricEOM}.

These moduli can all be described to linear order by a common metric
\be
\label{metricansatz}
ds^2 &=& e^{2\Omega}e^{2A} \h\eta_{\mu\nu} dx^\mu dx^\nu + 2e^{2A}e^{2\Omega} 
\p_\mu B_m dx^\mu dy^m + e^{-2A} \til g_{mn} dy^m dy^n 
\ee
and electric $C_4$ components\footnote{In principle, there can be an 
additional compensator term $-\h d b_2 K_1$
associated with each axion, but we show in the appendix that $K_1=0$.}
\be
\label{C4ansatz}
C_4 &=& e^{4\Omega} e^{4A} \hat \epsilon + e^{4\Omega} e^{4A} \hat \star 
\hat d B_1 + b_2^I \wedge \omega_2^I\ .% -\h db_2^I\w K_1^I\ .
\ee
The total compensator field is written as the sum from each moduli sector, so
\be 
\label{B1 tot}
\h d B_1(x,y)\equiv -\h dc(x)\wedge \t d K(y)
+e^{-4\Omega}\h\star\h db_2^I(x)\wedge B_1^{b,I}(y)+\h dB_1^Y(x,y)\ .
\ee
Note that we have used a notational shorthand since $\h d^2B_1\neq 0$ 
(that is, $\h dB_1$ is not actually an exterior derivative in spacetime)
except when the axions are on shell.
It is worth noting that there is a gauge in which the volume modulus 
compensator field, $K(y)$, appears in the 
$(\mu\nu)$ component of the metric; however, this is not possible for the 
other moduli, since perturbations
of $b_2^I$ and $Y^{\sla m}$ directly source $T_{\mu m}$.
The $C_4$ axions in the presence of background $G_3^{(0)}$ flux also source 
a compensator for the 2-form potential 
\be\label{A2ansatz}
\delta A_2 = - e^{-2\Omega} \hat \star \hat d b_2^I \w \Lambda_1^I\, .
\ee

The field strengths for our ansatz are 
\be
\t F_5&=&e^{4\Omega}\h\epsilon\w\t de^{4A}+d\left(e^{4\Omega}e^{4A}\h\star\h dB_1
\right)+\h d b_2^I\w \omega_2^I \nonumber \\
&&+\left[\t\star\t d e^{-4A}-e^{2\Omega}e^{-4A} \t\star\left(
\h dB_1\w\t de^{4A}\right) +\star d\left(e^{4\Omega}e^{4A}\h\star\h dB_1\right) 
+ e^{-2\Omega}\h\star\h d b_2^I\w e^{-4A}\t\star \omega_2^I\right]
\, ,\hspace{.5in}\label{F5ansatz2} \\
G_3 &=& G_3^{(0)} - e^{-2\Omega} \hat d \hat \star \hat d b_2^I \w \Lambda_1^I 
+ e^{-2\Omega} \hat \star \hat d b_2^I \w \t d \Lambda_1^I\, .
\label{G3ansatz}
\ee
As in the previous subsection, the terms in square brackets on the second
line of \eqref{F5ansatz2} are the magnetic components dual to the electric
components on the first line; the magnetic components subsume the
Chern-Simons terms $(ig_s/2)(\delta A_2 \bG-\delta\bar A_2 \G)$.
This form for $\t F_5$ differs from the form presented in 
\cite{0810.5768,1308.0323} by terms proportional to the 4D dynamical EOM 
$e^{4\Omega}e^{4A}\h d\h\star \h dB_1+e^{-4A}\h\star\h d\h\star\h d\t\star B_1$.
Therefore, both versions of the field strength represent the same on-shell
degrees of freedom with slightly different 4D field definitions, and 
the constraints for the volume modulus and axions remain unchanged.  We 
summarize them below.

%%%
% Solving Constraints for Other Moduli
%%%

The constraints arising from the external components of the Einstein 
equation are, as above,
\be 
\wtn^2 e^{-4A(x,y)} &=& - \frac{g_s}{2} \left| G^{(0)}_3 \right|^{\til 2} 
- 2 \kappa^2_{10} T_3\til \delta^6 (y,Y)-\cdots \ ,\label{coulomb2}\\ 
\til\Del^{\til n} B_n &=& e^{-2\Omega}\delta e^{-4A}-e^{-4A}\delta e^{-2\Omega}\ .
\label{Bdiv2}
\ee
Since the axions do not appear as sources for the warp factor in 
(\ref{coulomb2}), the warp and Weyl factors are independent of these 
degrees of freedom. However, the volume modulus appears as a 
spacetime-dependent shift of the warp factor \cite{0810.5768}
\be
\boxed{e^{-4A(x,y)} = e^{-4A^{\0}(y)}+c(x)+2\kappa_{10}^2T_3 \delta Y^{\sla m}(x)
\del_{\sla m}\t G(y,Y)\, .}
\label{warpshift1}
\ee
The Weyl factor is $e^{-2\Omega} = e^{-2\Omega^{\0}}+c(x)$ including the volume
modulus.  In addition to (\ref{Bdiv}), \eqref{Bdiv2} gives a Poisson equation
for the volume modulus compensator
\be
\t \nabla^2 K(y) = e^{-4A^{\0}(y)} - e^{-2\Omega^{\0}}
\label{KPoisson}
\ee
and $\t \nabla^{\t n} B_n^{b,I} = 0$.

The mixed component of $E_{MN}$ gives a nontrivial constraint
for all of the degrees of freedom, namely
\be
-\frac 12 e^{4A}\del_\mu\del_m e^{-4A} +\frac 12 e^{2\Omega}e^{4A} 
\del_\mu\til\Del^{\til n}(\til dB_1)_{mn}+\kappa_{10}^2
T_3e^{4A}\t g_{m\sla n} \del_\mu Y^{\sla n} 
\t\delta^6(y,Y)\nonumber\\
-2e^{-2\Omega}e^{4A}(\hat\star\hat db_2^I)_\mu\left[e^{-4A}
\left(\omega_2^I\right)_{mn}\del^{\t n}A
-\frac{ig_s}{8}\left(\t\star \left(\t d\Lambda_1\w\bG\right)_m
-\textnormal{c.c.} \right)\right]
&=&0 \ ,\hspace{30pt}\label{offdiagconstraint}
\ee
as well as a copy of equation (\ref{coulomb2}) multiplied by $\del_\mu B_m$.
We have already seen how this equation is satisfied for fluctuations in 
D3-brane position; for the volume modulus, we note that 
$\partial_\mu \partial_m e^{-4A} = 0$ and that its compensator is exact, so
the constraint is automatically satisfied.
%Moreover, since the 2-form axions produce no fluctuation in the warp factor 
%and $B_m^{b,I}$ is divergenceless,
For the 2-form axions, (\ref{offdiagconstraint}) becomes
\be
\label{axeinsteinconstraint}
\t\Del^{\t 2} B_m^{b,I} = 4e^{-4A}\left(\omega_2^I\right)_{mn}
\del^{\t n}A
-\frac{ig_s}{2}\t\star\left(\t d\Lambda_1^I\w\bG-\textnormal{c.c.}\right)\ ,
\ee
which we take to be the defining Poisson equation for the axion compensators,
as in \cite{1308.0323}.

The constraints coming from the 5-form EOM (\ref{app:E6}) 
again include a copy of (\ref{coulomb2}), as well as
$\t\star$ of (\ref{offdiagconstraint}).
For the axions, we can further rewrite this constraint as
\be \label{axF5constraint}
\t d \left[e^{-4A} \t \star \omega_2^I -\t \star \t d B_1^{b,I}+ 
\frac{i g_s}{2} \left(\Lambda_1^I \w G_3^{(0)} - 
\textnormal{c.c.}\right)\right] = 0\, ,\ee
which implies that
\be
\label{gamma4def}
e^{-4A} \t \star \omega_2^I + \frac{i g_s}{2} \left(\Lambda_1^I \w G_3^{(0)} - 
\textnormal{c.c.}\right) -\t \star \t d B_1^{b,I} = \gamma_4^I + \t d K_3^I\, ,
\ee
where $\gamma_4^I$ is harmonic and $K_3$ is given by a Poisson-like equation 
(whose precise form will be unimportant to us).

Finally, the constraint from the 3-form EOM (\ref{E8}) 
as given in (\ref{app:E8}) receives no contribution from the volume or
D3 position moduli.  For the axions, it is
\be \label{G3constraint} 
\t d\t\star\t d\Lambda_1 +i\omega_2^I \w\G =0\ .\ee
The contribution $\Lambda_1^I$ for
each axion takes the same form as in \cite{1308.0323}.

\subsubsection{Summary} 
We have shown that a generic ``electric'' ansatz, given by 
(\ref{metricansatz},\ref{C4ansatz},\ref{B1 tot},\ref{F5ansatz2}), 
can be used to describe the 
volume modulus, $C_4$ axions, and D3-brane position. The warp factor, 
Weyl factor, and compensators for each modulus can be determined by 
(\ref{warpshift1},\ref{KPoisson},\ref{axeinsteinconstraint},\ref{G3constraint}) 
--- along with corresponding expressions from section 
\ref{sec:ElectricD3} --- and are shown to satisfy all of the 10D constraint 
equations.  It is important to note that the constraints require fluctuations
in multiple 10D fields for each of moduli separately.

%%%%%%%%%%%%%%%%%%%%%%%%%%%%

\subsection{4D Effective Action in Electric Formalism}
\label{sec:actionelectric}

As discussed in section \ref{sec:quadratic}, 
the quadratic effective action is obtained by multiplying the fluctuations 
of the 10D fields with the first-order parts of the 10D dynamical EOM; 
in this case, there is an additional contribution to the quadratic action 
described below that takes a topological form and does not appear in the EOM.
Integration over the compact manifold projects onto the massless sector.

\subsubsection{Contributions to Effective Action}\label{sec:contribElectric}

As described above, the quadratic action is given by 
\begin{align}
S &= \frac{1}{4\kappa_{10}^2} \int d^{10}x\, \sqrt{-g}\, \delta g^{MN}
\delta E_{MN} +\frac{1}{4\kappa_{10}^2}\int\left(\delta C_4\w\delta E_6+
\frac{g_s}{2}\delta A_2\w\delta\bar E_8 
+\frac{g_s}{2}\delta\bar A_2 \w\delta E_8 \right)\nonumber\\
&+\frac{T_3}{2} \int d^{10}x\, \sqrt{-g}\,\int d^4\xi\,
\sqrt{-\gamma}\, %\delta^{10}(x,X)\,
\delta X^{\sla M}\delta E_{\sla M}\ ,\tag{\ref{quadaction1}}
\end{align}
where $\delta g_{MN},\delta C_4,\delta A_2$ are the first-order parts of
(\ref{metricansatz},\ref{C4ansatz},\ref{A2ansatz}).
However, as noted in section \ref{sec:quadratic}, there can generally be
``topological'' terms in the quadratic action, such as the instanton density
of 4D Yang-Mills theory, that do not appear in the linearized EOM.  We 
identify a contribution of this type below.

%%%%%
% Effective Action - D3 part
%%%%%

We begin with the contribution from the D3-brane sector.
The dynamical part of the D3-brane EOM is
\be
\delta E_{\sla m} &=& e^{-4A}e^{-4\Omega}\left( e^{2\Omega}\t g_{\sla m\sla n} 
\hat \partial^{2} Y^{\sla n} +e^{4A}e^{4\Omega}\h\del^2 B_{\sla m}
-e^{4A}e^{4\Omega}\h\del^2B_{\sla m}\right)\delta^{10}(x,X)\nonumber\\
&=& e^{-4A} e^{-2\Omega} \t g_{\sla m\sla n} \hat \partial^{2} Y^{\sla n} 
\delta^{10}(x,X)\ .\label{D3EOMdynamical}
\ee
Note that the contribution to the EOM from the WZ action has cancelled with
a term proportional to $g_{\mu m}$ in the DBI action in the first line of
\eqref{D3EOMdynamical}.  We obtain the contribution
\be
\label{D3electricEffAction2}
S_{eff}^{D3} &=& \frac{T_3}{2} \int d^4 x \int d^6y\, \sqrt{\t g}\, e^{2\Omega} 
\t g_{\sla m\sla n} \delta Y^{\sla m}\hat \partial^{2} Y^{\sla n}\, 
\t \delta^6(y,Y)\ .
\ee
There is no ``self-energy'' problem; the singular fields
sourced by the D3-brane cancel out of the effective action.
Previous attempts at constructing the effective action for
D3-branes (such as \cite{hep-th/0208123,0905.4463}) have only 
included this contribution from the DBI action on a fixed warped background.
However, as we have emphasized, there are, in principle, 
additional contributions to the effective action
arising from the presence of moduli dependence in the metric and flux sectors.

%%%%%
% Effective Action - Gravity part
%%%%%

For example, the dynamical Einstein equations, given in 
(\ref{app:Emunu},\ref{app:Emum},\ref{app:Emn}), are
\be
\delta E_{\mu\nu} &=& -2 e^{4A} e^{2\Omega} \hat \eta_{\mu\nu} (
\del^{\t \ell} A) \hat \partial^{2} B_\ell\, , \\
\delta E_{\mu m} &=& 0\, , \\
\delta E_{mn} &=& \hat \partial^{2}\left[ \t \nabla_{(m} B_{n)} - \t g_{mn} 
\t \nabla^{\t \ell} B_{\ell}\right]
+e^{-4A} e^{-2\Omega} \t g_{mn} \hat \partial^{2} \left(3\delta \Omega - 
2 \delta A\right)\, .
\ee
Inserting these into the expression for the gravity contribution to the effective action, we obtain
\be
S_{eff}^R 
&=& \frac{1}{4\kappa_{10}^2} \int d^4x \int d^6y \sqrt{\t g}
\, e^{4\Omega}  \left[ 16(\delta A + \delta \Omega)  (\del^{\t \ell} A) 
\hat \partial^{2} B_\ell - 10 \delta A \hat \partial^{2} 
\t \nabla^{\t \ell} B_\ell \right. \nonumber \\
&& \left. + 12 e^{-4A} e^{-2\Omega} \delta A \hat \partial^{2} \left(
3 \delta \Omega - 2 \delta A\right)\right]\, .
\label{RelectricEffAction}
\ee
Notice that (\ref{RelectricEffAction})
implies that there are contributions
to the effective action for D3-brane fluctuations from both the compensators
and the fluctuations in the warp and Weyl factors.
These include complicated mixings with the volume
modulus and $C_4$ axions and depend on the details of
the solutions for the compensators from 
(\ref{KPoisson}) and (\ref{axeinsteinconstraint}).

%%%%%
% Effective Action - 5-form part
%%%%%

Next, we need to include the contribution to the effective action from the
flux sectors. 
The relevant\footnote{\label{higher}There is an additional term of the form 
$e^{-4A} \t \star \hat d \hat \partial^{2} B_1$, but
it contributes only to higher-derivative terms.  We expect these to be 
modified by other corrections, such as threshold corrections.  They are also
ambiguous because they change under field redefinitions of the form
$u\to u+f(u)\h\del^2 u$.  For both these reasons, we do not consider them.} 
dynamic contributions to the 5-form EOM are
\be
\delta E_6 = \t d \left(e^{-4A} \t \star \hat \partial^{2} B_1\right) + 
e^{-2\Omega} \hat d \hat \star \hat d b_2^I \wedge \left(\gamma_4^I+\t dK_3^I
\right)\ .\label{E6eom}
\ee
The 5-form contribution to the effective action is then
\be
S_{eff}^5 &=& \frac{1}{4\kappa_{10}^2} \int d^4 x \int d^6y \sqrt{\t g}\, 
e^{4\Omega}\left[4 (\delta A + \delta \Omega) \left( \h \partial^{2} 
\t \nabla^{\t \ell} B_\ell - 4 \del^{\t \ell} A \hat \partial^{2} B_\ell\right) 
\right] \nonumber \\
&& -\frac{1}{4\kappa_{10}^2}\int  e^{-2\Omega} \hat d b_2^I\w \hat \star 
\hat d b_2^J\int \omega_2^I\w\gamma_4^J\, .
\label{F5EffAction}
\ee
We expand $\gamma_4^J=(C^{-1})^{JK}\t\star\omega_2^K$, so 
$\int \omega_2^I\gamma_4^J=3\t V(C^{-1})^{IJ}$ using the normalization of 
our basis forms.  Following \cite{1308.0323}, 
\be
\label{Cinvdef}\boxed{
(C^{-1})^{IJ}=\frac{1}{3\t V}\left\{\int e^{-4A}\omega_2^I\w\t\star\omega_2^J
+\frac{ig_s}{2}\int \omega_2^I\w\left(\Lambda_1^J\w\bG-\bar\Lambda_1^J\w\G
\right)\right\}\ ,}\ee
which is the identity when the CY
metric is \textit{formal}, meaning that the wedge product of harmonic forms
is always harmonic.
Meanwhile, the 3-form sector does not contribute to the quadratic 
action\footnote{Except at higher derivatives, which we ignore for the
reasons stated in footnote \ref{higher}.} because none of the components of
$\delta E_8$ have the correct legs to wedge nontrivially with $\delta A_2$.

%%%%%
% Effective Action - Axion-D3 part
%%%%%

Finally, there is an additional contribution to the effective action that is 
not captured by equation (\ref{quadaction1}). Specifically, the WZ action 
(\ref{WZ}) contains a term that is quadratic in field fluctautions but does 
not give a contribution to the linearized EOM:\footnote{The extra factor of
$1/2$ in \eqref{ax-D3} compared to the WZ form \eqref{WZ} follows from
careful comparison of combinatorial factors in a general 4-form versus 
a 2-form wedged with two 1-forms.}
\be\label{ax-D3}
S_{eff}^{ax-D3} =  \frac{T_3}{2} \int_{D3} b_2^I \w \h d Y^{\sla m}\w 
\h d Y^{\sla n}\omega^I_{\sla m \sla n} \, .
\ee
This is analogous to the axion-photon coupling $\theta F^2$, which contributes
to the EOM only at quadratic order in the fields but can nonetheless 
contribute to the action at quadratic order in fluctuations if $\theta$ has
a background value.  In infinite Minkowski space, a constant background
$b_2$ is gauge trivial, but there are nontrivial Wilson lines if the 
spatial dimensions are compactified on a large torus, for example, so we
must include this term.  The terms in the action with background $b_2$
are topological in the sense that they are total derivatives, also like
the axion-photon coupling.

%%%%%
% Effective Action - Combined
%%%%%

Adding all the contributions yields a series of cancellations.  
Specifically, all terms proportional to $\del^{\t\ell}A\h\del^2 B_\ell$ 
cancel.  Then the constraint equation for the compensator (\ref{Bdiv2}) 
allows us to simplify the remaining terms involving $\delta A,\delta\Omega$
to the form
\be 
S_{eff}^{A-\Omega} &=&\frac{8}{4\kappa_{10}^2} \int d^4x\int d^6y \sqrt{\t g}\, 
e^{-4A}e^{2\Omega}
\left(\delta A-\delta\Omega\right)\h\del^2\delta\Omega \\
&=&\frac{1}{4\kappa_{10}^2} \int d^4x\int d^6y \sqrt{\t g}\,\left[
e^{4\Omega}\left(c+2\kappa_{10}^2T_3\delta Y^{\sla m}\del_{\sla m}\t G(y;Y)
\right) \h\del^2 c +2e^{6\Omega}e^{-4A} c\h\del^2 c\right]\ .\nonumber
\ee
Since $\del_{\sla m}\t G(y;Y)=-\t\Del_n\t G^n_{\sla m}(y;Y)$, the brane-volume
cross-term integrates to zero.  With the definition of the Weyl factor,
the remaining terms combine, yielding
\begin{empheq}[box=\fbox]{align}
S_{eff} &= S^{D3}_{eff}+S^R_{eff}+S^5_{eff}+S^{ax-D3}_{eff}\nonumber\\
&=- \frac{3\til V}{4\kappa_{10}^2} \int d^4 x \, e^{4\Omega} 
\p^{\h \mu} c(x) \p_\mu c(x) - \frac{3\til V}{4\kappa_{10}^2} 
\int e^{-2\Omega} \left(C^{-1}\right)^{IJ} \hat d b_2^I\w 
\hat \star \hat d b_2^J \nonumber\\
& - \frac{T_3}{2} \int d^4 x \, e^{2\Omega}\, \t g_{\sla m\sla n}(Y) 
\partial^{\h \mu} Y^{\sla m} \partial_\mu Y^{\sla n} +
\frac{T_3}{2} \int_{D3} b_2^I \wedge \hat d Y^{\sla m} \w
\hat d Y^{\sla n} \omega^I_{\sla m \sla n}\, .\label{altogether}
\end{empheq}
This is the key result of this section: all of the complicated
structures from the ansatz, including compensators and Green's functions, 
end up cancelling non-trivially in the final effective action.
This remarkable set of cancellations illustrates the necessity for a 
consistent solution
of the 10D constraint equations; the constraints, and their solutions, 
are essential for simplifying the effective action.
Note that the \emph{only} contribution to the effective action by the warp 
factor is through the axion field space metric (\ref{Cinvdef}), 
as was determined in \cite{1308.0323}.

%%%%%%%%%%%%%%%%%%%%%%%%%%%%

\subsubsection{Scalar Axions and Kinetic Action}\label{ccs and dual}

Since the 4D effective theory is a supergravity, ultimately the quantity of
interest is the K\"ahler potential, so we should write the kinetic action
in terms of holomorphic scalar coordinates.  Here, we dualize the
2-form axions into scalars and find that using complex coordinates 
$y^m=(z^i, \bar z^{\bar \imath})$ on the CY provides some simpification.

The CY coordinates appear explicitly in the action \eqref{altogether} 
through the brane positions in the combinations 
$\t g_{\sla m\sla n}\del^{\h\mu}Y^{\sla m}\del_\mu Y^{\sla n}$ and
$\omega^I_{\sla m\sla n}\h dY^{\sla m}\h dY^{\sla n}$.  The first is trivially
re-written in complex coordinates as 
$2\t g_{i\bar\jmath}\del^{\h\mu}Z^i\del_\mu \bar Z^{\bar\jmath}$ (for legibility,
we drop the slash on quantities evaluated at the brane position in complex
coordinates).  However, we can re-organize the second further; by the
$\del\bar\del$ lemma, the harmonic 2-forms are locally 
$\omega^I_2=i\del\bar\del k^I(z,\bar z)$ in terms of potentials $k^I$ (when
$\omega_2$ is the complex structure form $\t J_2$, $k$ is the K\"ahler 
potential of the CY), and 
\be \omega^I_{\sla m\sla n} \h d Y^{\sla m}\h dY^{\sla n} =
i\del_i k^I_{\bar\jmath} \h d Z^i\h d\bar Z^{\bar\jmath}-i \bar\del_{\bar\imath}k^I_j
\h d \bar Z^{\bar\imath}\h d Z^j \ ,\ \ 
k^I_i = \del_i k^I\ ,\ \ k^I_{\bar\imath}=\bar\del_{\bar\imath}k^I \ .\ee
Furthermore, since $k$ is evaluated at the brane position, we have
$\h d k^I_{\bar\jmath} = \p_i k^I_{\bar \jmath} \, \h d Z^i + 
\p_{\bar\imath} k^I_{\bar\jmath} \, \h d Z^{\bar\jmath}$; since partial derivatives
commute, $\h d k^I_{\bar\jmath}\h d \bar Z^{\bar\jmath} =
\p_i k^I_{\bar \jmath} \, \h d Z^i\h d \bar Z^{\bar\jmath}$.

Now the kinetic action for the axions can be written as 
\be
\label{act}
S_{axion} &=& -\frac{3\t V}{4\kappa_{10}^2}\int e^{-2\Omega}(C^{-1})^{I J} 
\h d b_2^I \w \sh \h d b_2^J +i\frac{T_3}{2} 
\int b_2^I\w \left(\h d k^I_{\bar\imath}
\w\h d \bar Z^{\bar\imath}-\h d k^I_i\w\h d Z^i\right)\nonumber \\
&=&-\frac{3\t V}{4\kappa_{10}^2}\int\left[e^{-2\Omega}(C^{-1})^{I J} 
\h d b_2^I \w \sh \h d b_2^J- i \gamma \h d b_2^I \w \kappa_1^I \, , \right]
\ee
where we have defined 
$\kappa_1^I \equiv k_j^I \h d Z^j - k_{\bar\jmath}^I \h d Z^{\bar\jmath}$ and 
$\gamma=2\kappa_{10}^2T_3/3\t V$.
To define the scalar axion, we re-write the action \eqref{act} in terms of 
the field strength  $h^I_3 \equiv \h d b_2^I$ and enforce the Bianchi identity
$\h dh_3^I=0$ by introducing a Lagrange multiplier $b_0^I$, so
\be
\label{lagrange action}
S_{axion} = - \frac{3\t V}{4\kappa_{10}^2}\int \left( e^{-2\Omega}(C^{-1})^{IJ}
h^I_3 \w \sh h^J_3 - i\gamma h^I_3 \w \kappa^I_1 - 2 b_0^I \h d h^I_3 \right) 
\, . \ee
The classical EOM implies
\be
h_3^I =e^{2\Omega}C^{IJ} \sh\left(\h d b_0^J + i\frac \gamma 2 \kappa_1^J 
\right)\, .
\ee
Substituting this into \eqref{lagrange action} and simplifying gives the 
action in terms of the scalar axions as
\be
S_{axion} = - \frac{3\t V}{4\kappa_{10}^2} \int e^{2\Omega} C^{IJ}\left(
\h d b_0^I +i\frac\gamma 2 \kappa_1^I\right)\w \sh \left(\h d b_0^J + i
\frac\gamma 2 \kappa_1^J\right)\ .
\ee

With the brane coordinates written in terms of complex variables and the 
axions dualized to the conventional scalars, the effective action
\eqref{altogether} becomes 
\begin{empheq}[box=\fbox]{align}
S_{eff} &=  - \frac{3\til V}{4\kappa_{10}^2} \int d^4 x \left[ e^{4\Omega} 
\p^{\h\mu} c \p_\mu c +2\gamma e^{2\Omega}\, \t g_{i\bar\jmath}(Z,\bar Z) 
\partial^{\h \mu}Z^i \partial_\mu \bar Z^{\bar\jmath}+
e^{2\Omega}C^{IJ}\right.\nonumber \\
&\ \ \ \left. \times\left(\del^{\h\mu}b^I+i\frac\gamma 2
k^I_i\del^{\h\mu}Z^i-i\frac\gamma 2 k^I_{\bar\imath}\del^{\h\mu}\bar Z^{\bar\imath}
\right)\left(\del_{\mu}b^J+i\frac\gamma 2
k^J_j\del_{\mu}Z^j-i\frac\gamma 2 k^J_{\bar\jmath}\del_{\mu}\bar Z^{\bar\jmath}
\right)\right]\ .\label{ElectricEffActionTot}
\end{empheq}
This effective action for the volume modulus, $C_4$ axions, and D3-brane
positions is the primary result of this paper.  Compared to the action
found in \cite{Grana:2003ek}, which does not account for the effects of the 
warp factor or flux in kinetic terms, our result is similar, but 
we find that a nontrivial warp factor and flux appear through the 
Weyl factor (which corresponds to shifting the expectation value of the
volume modulus) and the metric $C^{IJ}$, as in \cite{0810.5768,1308.0323}.
In fact, our result matches that 
derived in \cite{Martucci:2014ska} using methods from 4D conformal 
SUGRA and corrected by flux dependence on metric moduli 
\cite{Martucci:2016pzt}; the contribution from the flux background 
to $C^{IJ}$ already appeared in direct dimensional reduction in 
\cite{1308.0323}.  

In the case that the CY metric $\t g_{mn}$ is formal or that we restrict 
to the universal axion ($\omega_2=\t J_2$, the almost complex structure),
$C^{IJ}=e^{2\Omega}\delta^{IJ}$.  With a single axion $b$, then the effective 
action (\ref{ElectricEffActionTot}) follows from the K\"ahler potential
\be
\mathcal{K} = -3 \ln\left[-i(\rho-\bar \rho)-\gamma k(Z,\bar Z)\right]
\ee
and holomorphic coordinate 
\be \rho = b_0 + i \left(e^{-2\Omega}+\frac{\gamma}{2} k(Z,\bar Z)\right)\ ,\ee
as we show in appendix \ref{sec:kahler}.
This takes the form proposed in \cite{hep-th/0208123,Kachru:2003sx} 
for D3-branes
in warped compactifications when restricted to a single K\"ahler modulus.

\subsubsection{Summary}
Starting with the ansatz of the previous subsection for the 10D SUGRA fields,
we have performed a consistent dimensional reduction beyond the probe 
limit of the effective action of a mobile D3-brane, the volume modulus, and
4-form axions in a warped GKP background.  A number of critical cancellations
occur because the 10D fields satisfy their constraint equations.

The effective action is 
\begin{empheq}[box=\fbox]{align}
S_{eff} &= - \frac{3\til V}{4\kappa_{10}^2} \int d^4 x \left[ e^{4\Omega} 
\p^{\h\mu} c \p_\mu c +\gamma e^{2\Omega}\, \t g_{i\bar\jmath}(Z,\bar Z) 
\partial^{\h \mu}Z^i \partial_\mu \bar Z^{\bar\jmath}+ e^{2\Omega}C^{IJ}\right. 
\nonumber \\
& \ \ \ \left. \times \left(\del^{\h\mu}b^I+i\frac\gamma 2
k^I_i\del^{\h\mu}Z^i-i\frac\gamma 2 k^I_{\bar\imath}\del^{\h\mu}\bar Z^{\bar\imath}
\right)\left(\del_{\mu}b^J+i\frac\gamma 2
k^J_j\del_{\mu}Z^j-i\frac\gamma 2 k^J_{\bar\jmath}\del_{\mu}\bar Z^{\bar\jmath}
\right)\right]\ \tag{\ref{ElectricEffActionTot}},
\end{empheq}
where the flux and warp factor appear through the metric
\begin{align}
\Aboxed{
(C^{-1})^{IJ}&=\frac{1}{3\t V}\left\{\int e^{-4A}\omega_2^I\w\t\star\omega_2^J
+\frac{ig_s}{2}\int \omega_2^I\w\left(\Lambda_1^J\w\bG-\bar\Lambda_1^J\w\G
\right)\right\}\ . \tag{\ref{Cinvdef}} }
\end{align}
There is no known explicit form for the corresponding K\"ahler potential,
though it reduces to the DeWolfe-Giddings \cite{hep-th/0208123,Kachru:2003sx} 
form when the CY has only a single K\"ahler modulus.

%%%%%%%%%%%%%%%%%%%%%
%%%%%%%%%%%%%%%%%%%%%
%%%%%%%%%%%%%%%%%%%%%
%%%%%%%%%%%%%%%%%%%%%

\section{Magnetic D3-brane Couplings}
\label{sec:magnetic}

We can also carry out the dimensional reduction in the version of type
IIB SUGRA in which we keep the mostly internal components of $\t F_5$.
In this version of the theory, a D3-brane does not couple to $C_4$ in
the  action,\footnote{Except at higher order in spacetime derivatives 
than we  consider.} but rather through a nontrivial Bianchi
identity.   To describe the SUGRA in this way
will require a new, but equivalent, expression
for the field strength ansatz which differs from
\eqref{F5ansatzD3} by terms that are proportional to equations of
motion.  Since it is identical on shell, this ansatz describes the same 
4D degrees of freedom as \eqref{F5ansatzD3}; off shell, the 4D effective 
action will differ only in higher-derivative terms, which are ambiguous
as they can be changed by a 4D field redefinition.  We start this section
by demonstrating that the new ansatz satisfies the same constraints as
required in the electric description of the D3-brane.  

At the same time, we will notice that the fluctuation in $\t F_5$ 
decomposes into an exact term (the contribution of a globally-defined 4-form
potential) and a delta-function supported term with an explicit dependence 
on the brane position.  This motivates us to divide $\t F_5$ 
in the presence of magnetic charges into an exact piece and 
terms that depend explicitly on brane position; the field redefinition
yields a 4-form potential without a Dirac string singularity, 
much as a field redefinition can be used to create a 4-form potential that
is invariant under 2-form gauge transformations needed to describe a 
background 3-form field strength.  We discuss this field
redefinition in section \ref{sec:bianchi}.  

A puzzle that arises when treating the D3-brane as a magnetic charge is how
the no-force condition in our background arises, since there is no WZ
coupling between the brane and $C_4$.  As it turns out, the field redefinition
described above solves this puzzle, since the explicit dependence of 
$\t F_5$ on the brane position modifies the D3-brane EOM.  In section
\ref{sec:magneticEOM}, we find the modified equation of motion for the brane
and demonstrate that a static D3-brane feels no force in backgrounds 
that are mutually BPS with the brane (including GKP backgrounds).

Finally, after reviewing the results of \cite{0810.5768,1308.0323} in a
``magnetic'' description, we present a unified ansatz for the volume
modulus, $C_4$ axions, and brane position at linear order. Using this
ansatz, we solve the constraints and integrate the quadratic action 
over the internal manifold to find the 4D effective action
for all moduli.

%%%%%%%%%%%%%%%%%%%%%%%%%%%%

\subsection{D3-brane Fluctuations in Magnetic Formalism}
\label{sec:magansatz}

Clearly, to represent the same 4D degree of freedom, the self-dual $\t F_5$
must be the same on shell whether we choose to describe IIB SUGRA using
the electric or magnetic components.  However, the magnetic components of
$\t F_5$ in the ansatz \eqref{F5ansatzD3} (which we used for the 
electrically-coupled D3-brane) are not easily described in terms of a 
4-form potential with the magnetic set of components.
Fortunately, it is possible to describe the same on-shell solution of the 10D 
theory by adding terms proportional to the 4D dynamical EOM to $\t F_5$; 
some of these combine with other magnetic components to take the form $dC_4$,
as we will see below.  Specifically, we take
\be
\t F^{mag}_5 &=& \t F^{elec}_5 - e^{4A}e^{4\Omega} \h d \sh \h d B^Y_1 - 
e^{-4A} \sh \h d \sh \h d \st B^Y_1 \nonumber\\
&=&\st \t d e^{-4A}- e^{2\Omega} \h d(\st \t d B^Y_1) +
\left[ e^{4\Omega} \h\epsilon \w \t d e^{4A}- 
e^{4\Omega}\sh \h d \t d (e^{4A} B^Y_1)\right]\ ,\label{F5mag}
\ee
where $\t F^{elec}_5$ is given by \eqref{F5ansatzD3}.   In the second line, 
the first two terms are the magnetic components of $\t F^{mag}_5$, while 
the terms in brackets are the electric components.
Throughout the remainder 
of this section, we consider only $\t F^{mag}_5$, and therefore we suppress
the superscript.  Our ansatz for the metric remains the same.

Since it differs from the ansatz in the electric formalism only by 
terms that are second order in spacetime derivatives, 
this modified ansatz for $\t F_5$
leaves the constraint equations unchanged, so it is still a valid ansatz for 
dimensional reduction.  The additional terms in \eqref{F5mag} 
lead to higher-derivative terms in the dynamical EOM,
which in fact vanish on shell and do not affect our analysis.
We verify that the constraints are unchanged by explicit calculation in
appendix \ref{sec:MagneticEOM}.

We can now address our claim that the magnetic components of this ansatz
are simply written in terms of a 4-form potential.  The key is the 
constraint \eqref{F5offdiag}, which can be written as
\be
\t\star\t d\delta e^{-4A} +e^{2\Omega}\t d\t\star\t dB_1
-2\kappa_{10}^2T_3\t\star \t Y_1\t\delta^6(y,Y) = 0 \, .
\ee
As a result, the fluctuation in the magnetic components is
\begin{empheq}[box=\fbox]{align}
\delta \t F_5 &= \st \t d \delta e^{-4A} - e^{2\Omega} \st \t d \h d B^Y_1 
\nonumber \\
&=  -d\left( e^{2\Omega} \st \t d B^Y_1 \right) +2\kappa_{10}^2 
T_3 \st \t Y_1 \t\delta^6(y,Y)\, .\label{first-order cont}
\end{empheq}
In other words, $\delta \t F_5 = d\delta C'_4 +S_5$, where $C'_4$ is a 
globally-defined potential (ie, has no Dirac string singularity and 
can therefore be defined with a single gauge patch)
and $S_5$ is the explicit dependence on the brane position 
required to solve the Bianchi identity.  
This is a (non-local) field redefinition of the potential which apparently
gives the 5-form an explicit dependence on the brane position.
We will explore this field redefinition in more detail in the following
subsection, including an analogy to the well-known Chern-Simons terms 
involving $A_2$ and $G_3$ in $\t F_5$.

\subsubsection{Summary}
The ansatz for 10D fields is somewhat modified in order to write the magnetic
components of $\t F_5$ in terms of a 4-form potential.  The metric and 5-form
are given by
\begin{align}
ds^2 &= e^{2\Omega}e^{2A} \h\eta_{\mu\nu} dx^\mu dx^\nu 
+ 2e^{2\Omega}e^{2A} \p_\mu B_m^Y dx^\mu dy^m + e^{-2A} 
\til g_{mn} dy^m dy^n\, , \hspace{.3in}\tag{\ref{metricansatzD3}}\\
\t F_5 &=\st \t d e^{-4A}- e^{2\Omega} \h d(\st \t d B^Y_1) +
\left[ e^{4\Omega} \h\epsilon \w \t d e^{4A}- e^{4\Omega}\sh \h d \t d (e^{4A} B^Y_1)
\right]\ ,\tag{\ref{F5mag}}
\end{align}
where the magnetic components are the first two terms.  We have found that 
the magnetic components of the field strength can be written as
$\t F_5 =S_5+dC'_4$, where $S_5$ contains explicit dependence on the brane
position and is described by a potential with a Dirac string singularity.
The redefined potential $C'_4$ is globally defined and has fluctuation
$\delta C'_4= -e^{2\Omega} \st\t dB_1^Y$.

This ansatz satisfies the same constraints as the ansatz we proposed in the
electric formalism.  Therefore, we still have $e^{2\Omega}=e^{2\Omega_{\0}}$,
\begin{align}
e^{-4 A(x,y)} &= e^{-4A^{\0}(y)} + 
2\kappa_{10}^2 T_3\ \delta Y^{\sla m} \partial_{\sla m} \t G(y,Y)\, , 
\ \textnormal{and}\tag{\ref{warpD3promote}}\\
B_m^Y(x,y) &= -2 \kappa_{10}^2  T_3 e^{-2\Omega} \t g_{mn}\ 
\delta Y^{\sla p}\t G_{\sla p}^{n} (y,Y)\, .
\tag{\ref{BD3Solve}}
\end{align}

\subsection{Non-Trivial Bianchi Identities, Field Redefintions, and EOM}
\label{sec:bianchi}

We recall from equation \eqref{BianchiF5} that the Bianchi identity for the
magnetic components of $\t F_5$ is 
\be
d \t F_5 =\frac{ig_s}{2} G_3\w\bar G_3
- 2\kappa_{10}^2 T_3 \sum_{D3/O3} \int d^4 \xi \sqrt{-\gamma} \, 
\star \epsilon_\| \, \delta^{10}(x,X(\xi)) \, ,
\ee
which has both distributed ($G_3$) and local sources.  The meaning of the 
distributed sources is well-understood --- the gauge-invariant field
strength contains both an exact term and Chern-Simons terms involving both
the potential and field strength for another SUGRA degree of freedom.
As a result, the 4-form potential $C_4$ (even at first order in perturbations) 
has a nontrivial gauge patching
in a background $G_3$; \cite{hep-th/0201029,0810.5768,1308.0323} 
demonstrated that this gauge transformation can be removed from perturbations
of $C_4$ by a simple field redefinition, so the perturbation in $\t F_5$
decomposes into an exact term and a (somewhat altered) Chern-Simons term. 

Similarly, the presence of a 
local magnetic source implies that $C_4$ must be defined on at least 
two patches glued together with a nontrivial gauge transformation (to remove
the Dirac-string-like singularity).  We show here that, as in the 
case of distributed sources, there is a field redefinition of the potential
that allows the perturbation in the field strength to be written as 
$dC'_4$ plus an analog of Chern-Simons terms with delta-function support.  
Our approach is to make a formal expansion of the source terms around an 
arbitrary fixed point; the nontrivial gauge patching can then be relegated
to a background potential that creates the zeroth order term in the
expansion, while the spacetime-dependent terms in the field strength are
separated into an explicit dependence on the brane position and the 
exterior derivative of an exact potential.
This explicit dependence of the field strength on the D3-brane position
in turn modifies the EOM for the brane's motion.  These techniques are
similar to Dirac's original proposal for magnetic monopoles in 4D Maxwell 
theory \cite{Dirac:1948um}; the relationship of our work to Dirac's and
the extension of Dirac's formalism to general branes is the subject of
an upcoming companion paper by two of us \cite{cownden}.

The key point in both cases is two-fold: the original $C_4$ is not suitable
for dimensional reduction because it is not globally defined (and cannot
be integrated over the CY in the usual way, for example) and is not
an entirely independent degree of freedom because its nontrivial gauge patching
depends on the values of other fields.  The field redefinitions we 
discuss below resolve both of these difficulties.

We begin with a brief review of the field redefinition in the case of a 
background $G_3$, largely following \cite{1308.0323}.  We then demonstrate
how to re-write $\t F_5$ with explicit dependence on the brane position 
in a generic background, working in static gauge.
The cases of 
distributed and local magnetic sources are independent, so we discuss them
separately.  We close with a discussion of the modified D3-brane EOM in section
\ref{sec:magneticEOM}.

\subsubsection{Field Redefinition in Background 3-Form}
\label{sec:backgroundG3}

Ignoring local sources, the 5-form Bianchi identity can be written as
$d \t F_5 =(ig_s/2) G_3\w\bar G_3 = (ig_s/4)d(A_2\w\bar G_3-
\bar A_2\w G_3)$.\footnote{When the axio-dilaton is constant; this 
discussion must be modified somewhat in a general F theory background.}
This, of course, leads to the well-known expression
$\t F_5 = dC_4 +(ig_s/4)(A_2\w\bar G_3-\bar A_2\w G_3)$ with Chern-Simons
terms acounting for the nontrivial Bianchi identity.  The appearance of the
2-form potentials requires $C_4$ to vary nontrivially under gauge 
transformations of $A_2$.  This is the usual definition of the 4-form 
potential in IIB SUGRA (one of the two common definitions, to be precise).

However, if $G_3$ has a harmonic background value $G_3^{(0)}$ (as in GKP 
compactifications), $A_2$ is defined only on coordinate patches, so $C_4$, 
including its fluctuation, 
must also be defined only in patches.  Defining $G_3=G_3^{(0)}+\delta G_3$
with $\delta G_3=d\delta A_2$ exact, we have
\be\label{bianchi_harmonicG3}
d \t F_5 =\frac{ig_s}{2}\left[ G_3^{(0)}\w\bar G_3^{(0)}+
\delta G_3\w\bar G_3^{(0)}+G_3^{(0)}\w\delta\bar G_3 +\delta G_3\w\delta\bar G_3
\right]\ .\ee
Similarly splitting $\t F_5=\t F_5^{(0)}+\delta \t F_5$, where 
$\t F_5^{(0)}$ satisfies the Bianchi identity for $G_3=G_3^{(0)}$, 
\be 
d \delta \t F_5 = \frac{ig_s}{2}d\left[\delta A_2\w\bG -\delta\bar A_2\w\G
\right]+\frac{ig_s}{4}d\left[\delta A_2\w\delta\bar G_3 -\delta\bar A_2\w
\delta G_3\right]\ .
\ee
This suggests writing
\be \delta \t F_5 = d\delta C'_4+
\frac{ig_s}{2}\left[\delta A_2\w\bG -\delta\bar A_2\w\G
\right]+\frac{ig_s}{4}\left[\delta A_2\w\delta\bar G_3 -\delta\bar A_2\w
\delta G_3\right]\ .\label{C4redefG3}\ee
It is important to note, however, that $\delta C'_4$ is \textit{not} the
fluctuation of $C_4$ as defined above but is shifted from that fluctuation
by a wedge product of $\delta A_2$ and the patched 2-form potential that
describes the background $\G$.  It is clear from \eqref{C4redefG3} that
$\delta C'_4$ is a globally-defined form; 
\cite{hep-th/0201029,0810.5768,1308.0323} demonstrated this fact using the
explicit field redefinition and the gauge transformations of the SUGRA fields.

Using the variables $\delta C'_4,\delta A_2,\delta\bar A_2$ rather than
$\delta C_4,\delta A_2,\delta\bar A_2$ serves two purposes: it removes the
background gauge transformations from the first-order potential and
clarifies the dependence of $\delta \t F_5$ on $\delta A_2,\delta\bar A_2$.
Since $\delta C'_4$ is globally-defined, it is the appropriate variable to
describe fluctuations in moduli (such as 4-form axions) or compensators.
However, because the explicit dependence of $\delta \t F_5$ on $\delta A_2$
changes, the field redefinition from $\delta C_4$ to $\delta C'_4$
also modifies the 10D EOM for $\delta A_2$
compared to the usual result from the SUGRA (while leaving the 4-form
EOM unchanged).  As can be determined either by direct variation (at linear
order) or by plugging the explicit field redefinition into \eqref{quadaction1},
the change to the linearized form of \eqref{E8} is to ensure that $A_2$
in the last term of that equation is 
the globally defined $\delta A_2$ (which may have a 
background value).  Henceforth, we will use this modified potential and
correspondingly modified EOM.% but drop the prime for notational convenience.

\subsubsection{Field Redefinition for D3-brane Source}

We can take a parallel approach for dynamical local sources; considering
only a single D3-brane, the Bianchi identity \eqref{BianchiF5} is
\be\label{bianchi-local1}
d \t F_5 = -2\kappa_{10}^2 T_3 \int d^4 \xi \sqrt{-\gamma} \, 
\star \epsilon_\| \, \delta^{10}(x,X(\xi)) \, ,
\ee
which we evaluate in the static gauge $\del_a X^\mu=\delta^\mu_a$.
With this gauge choice, the integral reduces to 
\be \int d^4 \xi \sqrt{-\gamma} \, 
\star \epsilon_\| \, \delta^{10}(x,X(\xi)) = -\left(\star_\perp - \star_\perp 
\h d Y_1 + \frac{1}{2} \, \star_\perp (\h d Y_1 \w \h d Y_1) +\cdots\right) 
\delta_\perp^6(y,Y(x))\, ,\ \ \ \ \ \label{bianchiexpand}\ee
following from the definition \eqref{epparallel}.  Remarkably, all factors
of the metric cancel on both sides of \eqref{bianchiexpand}, so we can
take $\star_\perp$ and $\delta_\perp^6$ to depend on an \textit{arbitrary}
6D metric $g_{\perp,mn}$ on the $y^m$.  For now we will leave this metric
arbitrary, making an advantageous choice later.  Because we can use an
arbitrary 6D metric, our procedure does not rely on factorizability of the
10D metric.  Note that explicit factors
of parallel propagators contracting $\h d Y$ have been suppressed since
they become Kronecker deltas at coincidence (as enforced by the delta
function).  Notationally, we continue to define $\h d =\del_\mu dx^\mu$, 
$\t d = \del_m dy^m$ rather than introduce $d_\|,d_\perp$.

Our approach, as when considering a fluctuating 3-form source in
\ref{sec:backgroundG3} above, is to demonstrate that terms containing
spacetime derivatives of the brane position are exact.  Then $\t F_5$
can be written as the exterior derivative of a (redefined) potential
plus delta-function-supported terms making explicit the entire dependence of
$\t F_5$ on the brane position $Y(x)$.  To separate out the the dynamics of
the brane position, we formally expand the right-hand side of
\eqref{bianchiexpand} around a fixed arbitrary point $Y_*^{\un m}$;
note that this is not necessarily the background value of the D3-brane
position.  The proper expansion quantity is Synge's
worldfunction $\sigma (Y_*,Y)$ (half the square geodesic
distance between $Y_*^{\un m}$ and $Y^{\sla m}$).\footnote{For additional 
details, see appendix \ref{sec:bitensors}.} 
The derivatives $\sigma_{\un m} \equiv \p_{\un m} \sigma(Y_*,Y)$
lie tangent to the geodesic from $Y^{\sla m}$ to $Y_*^{\un m}$
in $TM^*_{Y_*}$ as illustrated in figure \ref{fig:Synge}.  By taking 
further partial derivatives, we can see that
$\Lambda^{m}_{\sla m}\p_\mu Y^{\sla m} = - \Lambda^m_{\un m} \p_\mu \sigma^{\un m}$,
so we can replace $\h d Y_1\to -\h d\sigma_1$ in equation
\eqref{bianchiexpand} above.

In carrying out this formal expansion, we note that
\be\label{deltaexpand}
\delta_\perp^6(y, Y) = \delta_\perp^6(y, Y_*) - \sigma^{\un m} \p_{\un m} 
\delta_\perp^6(y, Y_*) + \frac{1}{2} \sigma^{\un m} \sigma^{\un n} \Del_{\un m} 
\p_{\un n} \delta_\perp^6(y, Y_*) + \mc O(\sigma^3) \, .
\ee
Order by order in $\sigma$, then, we find
\be
\label{F5 sigmas}
d \t F_5 &=& 2 \kappa_{10}^2 T_3 \left\{ \ep_\perp \dnot - \ep_\perp
\sigma^{\un m}\del_{\un m} \dnot - \h d \sperp (\sigma_1 \dnot) 
+\frac 12 \ep_\perp \sigma^{\un m}\sigma^{\un n} \Del_{\un m}\p_{\un n} \dnot
\right. \nonumber\\
&&\left. + (\h d \sperp\sigma_1) \sigma^{\un m}\del_{\un m} \dnot 
- \frac{1}{2} \sperp (\h d \sigma_1 \w \h d \sigma_1) \dnot \right\} \, .
\ee
Using \eqref{diracderivs}, and the fact that $\sigma^{\un m}$ is function of 
$Y_*,Y$ and not the position $y$ where $\t F_5$ is evaluated, 
we can rewrite
\be
\sigma^{\un m} \p_{\un m} \delta_\perp^6(y, Y_*) = -\Del_m \left(\sigma^{\un m}
\Lambda^m_{\un m}\delta_\perp^6(y, Y_*) \right)=
\star_\perp \t d \star_\perp \left( \sigma_1 \delta_\perp^6(y, Y_*) \right) \ .
\ee
We immediately see that the second and third terms of \eqref{F5 sigmas} 
combine into a total derivative $-d (\sperp \sigma_1 \dnot)$.

We wish to write the second order terms also as 10D total derivatives.  We
start by noting that similarly 
\be \sperp\sigma^{\un m}\sigma^{\un n} \Del_{\un m}\p_{\un n} \dnot =
\t d \left[(\sperp\sigma_1)\sigma^{\un m}\del_{\un m}\dnot\right]
\label{delta2deriv}\ee
and also 
\be \sperp (\h d \sigma_1 \w \h d \sigma_1) \dnot=\h d\left[\sperp(\sigma_1
\w\h d \sigma_1)\dnot\right]\ .\label{sigma2}\ee
The remaining term in \eqref{F5 sigmas} can be re-written in two ways; the
first is given by the simple differentiation
\be
\label{prop2}
\h d \left[(\sperp \sigma_1) \sigma^{\un m} \p_{\un m} \dnot\right] = 
(\h d \sperp \sigma_1) \sigma^{\un m} \p_{\sla m} \dnot 
- (\sperp \sigma_1) \h d \sigma^{\un m} \p_{\un m} \dnot \, .
\ee
While the left-hand side of \eqref{prop2} looks like the complement 
to \eqref{delta2deriv} that we desire, it enters \eqref{F5 sigmas} with an
incorrect coefficient, and the third term of \eqref{prop2} is not a 
derivative.  As it turns out, the complement of \eqref{sigma2}
is 
\be
\label{prop1}
\t d \left[\sperp (\sigma_1\w \h d \sigma_1) \dnot \right] = 
-\h d (\sperp \sigma_1) \sigma^{\un m} \p_{\un m} \dnot
-(\sperp \sigma_1)\h d \sigma^{\un m} \p_{\un m} \dnot  \, ,
\ee
where we have remembered to differentiate the parallel propagators 
$\Lambda^m_{\un m}$ in the definition of $\sigma_1$.
All told, 
\be
\h d (\sperp \sigma_1) \sigma^{\un m} \p_{\un m} \dnot = 
\frac{1}{2} \h d \left[(\sperp \sigma_1) \sigma^{\un m} \p_{\un m} \dnot\right]
-\frac{1}{2} \t d\left[ \sperp (\sigma_1\w \h d \sigma_1 \dnot\right]  \, ,
\ee
so, to second order in the formal expansion,
\be
d \t F_5 &=& 2\kappa_{10}^2 T_3 \left\{ \ep_\perp \dnot - d \left[\sperp\sigma_1 
\dnot\right] + \frac{1}{2} d\left[(\sperp\sigma_1) \sigma^{\un m} \p_{\un m} 
\dnot\right] \right. \nonumber \\
&& \left. - \frac{1}{2} d\left[\sperp (\sigma_1 \w \h d \sigma_1)\dnot
\right] \right\} \ .
\ee
While we do not carry out this calculation to higher order, 
we conjecture that all terms in the Bianchi identity at first or higher
order in the formal $\sigma$ expansion can be organized into total 
derivatives, as we have shown at first and second order.  It is important
to note that the source terms are actually independent of $Y_*$ when
all orders of the expansion are included, since they are simply a way of
re-writing a function of $Y(x)$.

Since the source for the Bianchi identity is a static delta function
plus a series of total derivatives according to our conjecture,
$\t F_5$ can be written in terms of a patched-together 
potential for a static magnetic monopole located at $Y_*$, 
%(plus perhaps a patched potential for a background flux), 
a globally-defined potential, and
additional terms that translate the Dirac string from $Y_*^{\un m}$ to
$Y^{\sla m}$ as
\begin{empheq}[box=\fbox]{align}
\label{mag F5 mono}
\t F_5 &= d S_4^*+d C'_4 -2\kappa_{10}^2 T_3 \left[\sperp \sigma_1 \dnot - 
\frac 12 (\sperp \sigma_1) \sigma^{\un m} \p_{\un m} \dnot\right. \nonumber \\
& \left. + \frac 12  \sperp (\sigma_1 \w \h d \sigma_1) \dnot
+\cdots \right]  \, .
\end{empheq}
Note that the static monopole potential $S_4^*$ satisfies 
$d^2 S^*_4 = 2\kappa_{10}^2 T_3 \sperp\dnot$, which is allowed since it is not 
globally defined and is singular at $y=Y_*$. Of course, $\t F_5$ and 
all physically meaningful quantities must be independent of the arbitrarily
chosen $Y_*$, so the equation $\del\t F_5/\del Y_*^{\un m}=0$ will
result in a system of relations among different orders of the expansion
similar to renormalization group flow.

We emphasize that \eqref{mag F5 mono} is \textit{not} our ansatz for 
D3-brane motion.  Rather, \textit{any} 5-form field strength with a monopole
source can be written in this form, as we have done for our ansatz in
\eqref{first-order cont}.  Specifically, $\delta\t F_5$ in that expression
is the first order term of $\t F_5$ in 
the fluctuation of the brane position; to make contact with that expression,
we should take $Y_*$ to be the fixed background brane position, $Y$ the
actual brane position including fluctuations, and $g_{\perp, mn}=\t g_{mn}$. 
Then $dS_4^*+dC'_4$ is the background 5-form, and $\sigma_1\to -\t Y_1$.  Then 
\eqref{mag F5 mono} is identically \eqref{first-order cont} to first order.

Just as in section \ref{sec:backgroundG3}, the field redefinition from
$C_4\to C'_4$ serves a dual purpose.  First, it is clear that 
$d^2C'_4=0$, so, unlike the original 4-form, the redefined potential is 
now globally defined and is suitable for dimensional reduction. Therefore, the
$C_4$ compensators and axion moduli appear in the redefined $C'_4$.
Further, the field redefinition cleanly separates $\t F_5$ into a contribution 
from the independent 4-form potential $C'_4$ and an explicit contribution from 
the D3-brane degrees of freedom (which is required by the Bianchi identity).  
As for other moduli, the explicit dependence of the field strength on $Y(x)$
must be supplemented by the appearance of compensators, which are necessary
to satisfy the constraints.  In addition, since $\t F_5$ explicitly depends
on the brane position when written in terms of $C'_4$, the D3-brane EOM
is not simply given by the DBI and WZ actions, as we discuss below.

\subsubsection{The Brane EOM in the Magnetic Picture}\label{sec:magneticEOM}

As we have noted previously, in the magnetic version of IIB SUGRA, 
D3-branes couple to $\t F_5$ only through the nontrivial Bianchi identity;
there is no WZ coupling between $C_4$ and a static D3-brane.  This
naively presents a puzzle, since the brane is mutually BPS with a GKP 
background (or the background of other static D3-branes) and should therefore
feel no force.  The DBI action provides a gravitational force, but there
is apparently no counter-balancing force from the 5-form, unlike in the
electric formulation of the theory!

The resolution of the puzzle lies in the redefinition of $\t F_5$ in
\eqref{mag F5 mono}.  While the redefinition leaves the $\t F_5$ EOM 
$E_6$ unchanged because $C'_4$ enters $\t F_5$ in the same way as $C_4$ does, 
the new explicit dependence of $\t F_5$ on the brane
position $Y^{\sla m}$ through $\sigma_{\un m}(Y_*,Y)$ modifies the D3-brane's
EOM, just as the SUGRA Chern-Simons terms in $\t F_5$ contribute to the EOM 
for $A_2$.  We show here that the delta-function-supported terms in equation
\eqref{mag F5 mono} introduce two new contributions to the brane EOM:
a force, which resolves the puzzle described above, and terms proportional
to the 5-form EOM $E_6$, which vanish on shell but will contribute to 
off-shell quantities including the effective action.

In the magnetic description, the D3-brane position degrees of freedom
appear in the DBI action \eqref{SDBI} and the $\t F_5$ kinetic terms,
which are 
\be
\label{SmagF5}
S_5 = -\frac{1}{2\kappa_{10}^2} \int d^{10} x \sqrt{-g} \, 
\left(\frac{1}{2}\right) \left(\frac{1}{5!}\right) \left. \t F_{MNPQR}
\t F^{MNPQR}\right|_{mag}  \ ,
\ee
where the subscript $mag$ indicates that the sum is over only the magnetic
components of $\t F_5$ (which are defined with indices lowered).
Note that the indices are raised with the full 10D metric $g^{MN}$.
In the static gauge, the pullback of the metric is
\be
P(g)_{\mu \nu} = g_{\mu \nu}(x, Y) + 2 g_{\mu \sla n}(x, Y) \h\p_\nu Y^{\sla n} 
+ g_{\sla m \sla n} (x, Y) \h\p_\mu Y^{\sla m} \h\p_\nu Y^{\sla n} \, ;
\ee
the EOM for $\gamma_{\mu\nu}$ also enforce $\gamma_{\mu\nu}=P(g)_{\mu\nu}$,
but we impose that constraint only after varying $S$ with respect to 
$Y^{\sla m}$.  The DBI part of the action, as previously, contributes
terms equal to the EOM \eqref{ED3_1} with $C_{\sla M\sla N\sla P\sla Q}=0$
and all indices $X^{\sla M}$ restricted to $Y^{\sla m}$ in static gauge.

We now determine the variation of the 5-form kinetic action with respect
to the brane position.  To first order in the $\sigma$ expansion, 
\be
\frac{\p \t F_{mnpqr}}{\p Y^{\sla m}} &=& 2\kappa^2_{10} T_3 
(\ep_\perp)_{mnpqrs}\left\{ \Lambda^s_{\sla m}\left[  \dnot - \frac 12 
\sigma^{\un m} \p_{\un m} \dnot\right] -\frac 12 \Lambda^s_{\un n}\sigma^{\un n} 
\Lambda^{\un m}_{\sla m} \p_{\un m} \dnot \right\} \nonumber \\
&=& 2\kappa^2_{10} T_3 (\ep_\perp)_{mnpqrs}\left\{ \Lambda^s_{\sla m}
\delta_\perp^6(y,Y)+\Lambda^s_{\un s}\Lambda^{[\un s}_{\sla m} \sigma^{\un m]}
\del_{\un m}\dnot\right\}\ ,\label{F05Ym}\\
\frac{\p \t F_{\mu mnpq}}{\p Y^{\sla m}} &=&  \kappa^2_{10} T_3 
(\ep_\perp)_{mnpqrs}\, \Lambda^r_{\sla m} \Lambda^s_{\sla n} \p_\mu Y^{\sla n} 
\dnot \, , \\
\frac{\p \t F_{\mu mnpq}}{\p ( \h\p_\nu Y^{\sla m})} &=& -\kappa_{10}^2 T_3 
\delta^\nu_\mu (\ep_\perp)_{mnpqrs}\,  \Lambda^r_{\un m} \sigma^{\un m}
\Lambda^s_{\sla m} \dnot \, .
\ee
We have used the relationship 
$\del\sigma^{\un m}/\del Y^{\sla m} = -\Lambda^{\un m}_{\sla m}$ as well 
as the expansion \eqref{deltaexpand} for the delta function. 

For the most part, the EOM can be evaluated using the typical Euler-Lagrange
formula.  However, the contribution from the second term of \eqref{F05Ym}
deserves special consideration.  In the variation of the action, we can
integrate by parts to remove the derivative from the delta function:
\be \delta S_5 = T_3\int d^{10}x\, \frac{1}{5!}\delta Y^{\sla m}\left\{
\Del^\perp_t\left[ \sqrt{-g}(\ep_\perp)_{mnpqrs}\t F^{mnpqr}\Lambda^s_{\un s}
\right]\Lambda^{[\un s}_{\sla m} \sigma^{\un m]}\Lambda^t_{\un m} \dnot+\cdots
\right\},\ \ \ \ \ \ \ \ee
where $\Del^\perp$ is the covariant derivative compatible with the 
as yet arbitrary metric $g_{\perp,mn}$.  With the antisymmetrization of the
$\un s,\un m$ indices, the derivative appears to be $\t d \sperp \t F_5$; 
however, the presence of the 10D metric requires a more delicate 
interpretation.  As we expect that the brane EOM will contain terms 
proportional to the $\t F_5$ EOM, we are motivated to re-write
\be \frac{1}{5!}\Del^\perp_{[t}\left[ \sqrt{-g}(\ep_\perp)_{|mnpqr|s]}\t F^{mnpqr}\right]
\Lambda^t_{\un n}\Lambda^s_{\sla m}= \frac 12 \sqrt{-\gamma}\sqrt{g_\perp}
(\star_\gamma \t d \star \t F)_{\un n\sla m}\ ,\ee
where $\star_\gamma$ is the 4D Hodge star for the induced metric; this
combination is ultimately independent of $\gamma_{\mu\nu}$.\footnote{Recall
that $g_{MN}$ and $\gamma_{\mu\nu}$ are independent variables until the
$\gamma_{\mu\nu}$ EOM is enforced.}  
Note that only certain components of $\t F$ appear in 
$(\star_\gamma \t d \star \t F)_{ts}$.
Similarly, the $\del_\mu(\del\mathcal L/\del(\del_\mu Y))$ term of the 
Euler-Lagrange equation contains
\be 
\frac{1}{4!}
\del_\mu \left[\sqrt{-g}(\ep_\perp)_{mnpqrs}\t F^{\mu mnpq}\Lambda^s_{\sla m}\right]
\Lambda^n_{\un n}
=\sqrt{-\gamma}\sqrt{g_\perp}\star_\gamma \h d\star \t F_{\un n\sla m}\ee
(with a slight abuse of notation).  These terms in fact add together 
to give a contribution proportional to the EOM $E_6$.

All told, the variation of the action with respect to the brane position
(in static gauge) is 
\be\delta S &=& -T_3\int d^{10}x\,\sqrt{-g}\,\delta Y^{\sla m}
\left\{ \left(\frac{1}{2}\right)
\int d^4\xi\sqrt{-\gamma} \left[ \gamma^{\mu\nu}\left(
\del_{\sla m} g_{\mu\nu}(x,Y)+2\del_{\sla m}g_{\sla n (\mu}\del_{\nu)}Y^{\sla n}
\right.\right.\right.\nonumber\\
&&\left.\left.\left.
+\del_{\sla m}g_{\sla n\sla p}\del_\mu Y^{\sla n}\del_\nu Y^{\sla p}\right)
-2\Del^\gamma_\mu\left(\gamma^{\mu\nu}g_{\sla m\sla n}\del_\nu Y^{\sla n}
+\gamma^{\mu\nu}g_{\nu\sla m}\right)\right]\delta^{10}(x,X)
\right.\nonumber\\
&&\left.-\left[\frac{1}{5!}(\ep_\perp)_{\sla m npqrs}\t F^{npqrs} 
\delta_\perp^6(y,Y)-\frac 12 
\frac{\sqrt{-\gamma}\sqrt{g_\perp}}{\sqrt{-g}} 
(\star_\gamma d\star \t F_5)_{\un n\sla m}\sigma^{\un n}
\dnot \right.\right.\nonumber\\
&&\left.\left.+\frac{1}{5!}
(\ep_\perp)_{\sla n \sla mpqrs}\t F^{\mu pqrs}\del_\mu Y^{\sla n}
\dnot\vphantom{\frac 12}\right]\right\}\ .\label{fullD3EOM1}
\ee
We need to make several comments.  First, we have worked only to the
first subleading order in the formal $\sigma$ expansion.  It is reasonable to
conjecture that the sole effect of the higher order terms
in $\sigma$ is to replace $\dnot\to\delta_\perp^6(y,Y)$ and perhaps to
add new contributions to the EOM which do not contribute at first order
in the D3-brane velocity (like the last term of \eqref{fullD3EOM1}).
We will primarily assume that this is the case in our discussion of the 
dimensionally reduced action below but also comment on the possibility that
the conjecture is false (we leave a check of the conjecture to \cite{cownden}).
Next, we note the appearance of several different metrics in
\eqref{fullD3EOM1} including the (as yet) arbitrary metric $g_{\perp,mn}$.
As we mentioned above, this formalism has allowed us to include geometries
in which $g_{MN}$ does not factorize into 4D and 6D metrics.
Finally, we recall from equations (\ref{quadaction1},\ref{ED3_1}) our 
convention that the EOM should be defined as (in static gauge)
\be \delta S = T_3\int d^{10}x\sqrt{-g}\int d^4\xi\sqrt{-\gamma}
\delta Y^{\sla m} E_{\sla m} \ \textnormal{with} \ E_{\sla m}\propto
\delta^{10}(x,X)\ .\ee
Some of the terms in \eqref{fullD3EOM1} manifestly take this form, but
others do not.  However, consider that
\be 
\int d^{10}x\sqrt{-g}\int d^4\xi\sqrt{-\gamma}\, f(x,y,Y)\delta^{10}(x,X)
=\int d^{10}x\sqrt{-\gamma}\sqrt{g_\perp}\, f(x,y,Y)\delta_\perp^6(y,Y)\ee
for any function $f(x,y,Y)$.  Furthermore, we have the identities
$\sqrt{-\gamma}\sqrt{g_\perp}\epsilon_\gamma^{\mu\nu\lambda\rho}
\epsilon_{\mu\nu\lambda\rho mnpqrs}=-4!\sqrt{-g}(\epsilon_{\perp})_{mnpqrs}$ and
$\sqrt{-\gamma}\sqrt{g_\perp}\epsilon_\gamma^{\mu\lambda\rho\sigma}
\epsilon_{\nu\lambda\rho\sigma mnpqrs}=-6\sqrt{-g}\delta^\mu_\nu
(\epsilon_{\perp})_{mnpqrs}$.
Therefore, we can re-write \eqref{fullD3EOM1} as the EOM
\begin{empheq}[box=\fbox]{align} \ \ 
E_{\sla m} &= \left\{\Del^\gamma_\mu\left(\gamma^{\mu\nu}g_{\sla m\sla n}
\del_\nu Y^{\sla n}+\gamma^{\mu\nu}g_{\nu\sla m}\right)-\frac 12 
\gamma^{\mu\nu}\left(\del_{\sla m} g_{\mu\nu}+2\del_{\sla m}g_{\sla n (\mu}
\del_{\nu)}Y^{\sla n}\right.\right.\nonumber\\
&\ \ \ \left.\left.+\del_{\sla m}g_{\sla n\sla p}\del_\mu Y^{\sla n}
\del_\nu Y^{\sla p}\right)-\frac 12 \left(\star_\gamma d\star\t F_5
\right)_{n\sla m}\Lambda^n_{\un n}\sigma^{\un n}\right.\nonumber\\
&\ \ \ \left.-\left( (\star_\gamma\star\t F_5)_{\sla m}+
(\star_\gamma\star\t F_5)^\mu{}_{\sla m\sla n}\del_\mu Y^{\sla n}\right)
\right\} \delta^{10}(x,X)\ .\label{fullD3EOM2}
\end{empheq}
Note that the final result depends only on the 10D metric and the induced
metric on the worldvolume, not the arbitrary metric $g_{\perp,mn}$; this
is physically necessary but occurs through nontrivial cancellations.
The field redefinition $C_4\to C'_4$ which showed the explicit dependence of 
$\t F_5$ on $Y^{\sla m}$ has introduced several new terms in the D3-brane EOM.
The last two terms represent the ``electromagnetic Lorentz'' 
force of $\t F_5$ on the brane and, at least to the order we have calculated 
in $\sigma$, they are independent of the arbitrary reference point $Y_*$, 
as any physical quantity should be.  Higher orders in the $\sigma$ expansion 
can contribute terms with higher powers of $\del_\mu Y^{\sla m}$ as necessary
for the 5-form version of the Lorentz force.
The remaining term, proportional to $E_6=d\star\t F_5$,
is somewhat more puzzling because it contains $\sigma^{\un n}$.  This term 
vanishes on shell, so it does not affect the
physically meaningful brane EOM, but it does contribute to the (off-shell) 
quadratic action that we wish to calculate.  We will discuss this contribution
in more detail when we calculate the quadratic action below.  For now,
we simply note that we can replace $(\star_\gamma E_6)_{n\sla m}\Lambda^n_{\un n}
\sigma^{\un n}=(\star_\gamma E_6)_{\sla m\sla n}\sigma^{\sla n}$ using properties
of the Synge world function (see appendix \ref{sec:bitensors}).

We can now return to the puzzle we raised earlier --- how does a static
magnetically-charged brane feel a BPS-like no-force condition even though 
it has no direct coupling to the 5-form?  We have recognized that the 
nontrivial Bianchi identity for $\t F_5$ modifies the D3-brane EOM
as above.  Consider a D3-brane in the background \eqref{back metric} 
at a constant position $Y^{\sla m}$. In this case, 
$\gamma_{\mu\nu}=g_{\mu\nu}=e^{2A}\hat\eta_{\mu\nu}$.  Then the brane EOM becomes
\be E_{\sla m} &=& \left[-\frac 12 \gamma^{\mu\nu}\del_{\sla m} g_{\mu\nu}
+(\star_\gamma\star \t F_5)_{\sla m}\right]\delta^{10}(x,X)
\nonumber\\
&=& \left[-2e^{-2A}\del_{\sla m}e^{2A}-e^{-6A}e^{10A}\del_{\sla m}e^{-4A}\right]
\delta^{10}(x,X)=0\ .
\ee
In other words, the new contributions to the brane EOM due to the $\t F_5$
field redefinition precisely restore the no-force condition for the brane.

It is also worth discussing the relationship of \eqref{fullD3EOM2} to the
D3-brane EOM in the electric formalism as given in \eqref{ED3_1}.  In
static gauge,\footnote{And excluding terms proportional to derivatives of
the delta function as in footnote \ref{deltaderivEOM}.} this is
\be E_{\sla m} &=& \left\{\Del^\gamma_\mu\left(\gamma^{\mu\nu}g_{\sla m\sla n}
\del_\nu Y^{\sla n}+\gamma^{\mu\nu}g_{\nu\sla m}\right)-\frac 12 
\gamma^{\mu\nu}\left(\del_{\sla m} g_{\mu\nu}+2\del_{\sla m}g_{\sla n (\mu}
\del_{\nu)}Y^{\sla n}\right.\right.\nonumber\\
&&\left.\left.+\del_{\sla m}g_{\sla n\sla p}\del_\mu Y^{\sla n}
\del_\nu Y^{\sla p}\right)+\epsilon_\gamma^{\mu\nu\lambda\rho}\Del^\gamma_\mu
\left(\frac 16C_{\sla m\nu\lambda\rho}+\frac 12C_{\sla m\sla n\lambda\rho}
\del_\nu Y^{\sla n}+\cdots\right)\right.\nonumber\\
&&\left.-\epsilon_\gamma^{\mu\nu\lambda\rho}\left(\frac{1}{24}\del_{\sla m}
C_{\mu\nu\lambda\rho}+\frac 16\del_{\sla m}C_{\sla n\nu\lambda\rho}\del_\mu Y^{\sla n}
+\cdots\right)\right\}\delta^{10}(x,X)\ ,\label{staticElecEOM}
\ee
where the $\cdots$ include higher powers of $\del_\mu Y^{\sla m}$, which we
did not calculate in the magnetic framework.  The first terms, which involve
the metric and its derivatives, are manifestly identical in the electric and
magnetic formalisms, so we are left to compare the terms involving the
potential/flux.  With some rearrangement,
\be 
\frac 16\epsilon_\gamma^{\mu\nu\lambda\rho}\Del^\gamma_\mu C_{\sla m\nu\lambda\rho}
=-\frac{1}{24}\epsilon_\gamma^{\mu\nu\lambda\rho}
(\hat dC_4)_{\mu\nu\lambda\rho\sla m}+\frac 16 \epsilon_\gamma^{\mu\nu\lambda\rho}
\del_{\sla n}C_{\sla m\nu\lambda\rho}\del_\mu Y^{\sla n}\ ;\ee
the first term combines with the first term of the third line of 
\eqref{staticElecEOM} to give $-(\star_\gamma\t F_5)_{\sla m}$ in terms of the
electric components.  We add the remainder to the other terms from
\eqref{staticElecEOM} and find
\be -\frac 13 \epsilon_\gamma^{\mu\nu\lambda\rho}\del_{[\sla m}C_{\sla n]\nu\lambda\rho}
\del_\mu Y^{\sla n}+\frac 12 \epsilon_\gamma^{\mu\nu\lambda\rho}\Del^\gamma_\mu
C_{\sla m\sla n\lambda\rho}\del_\nu Y^{\sla n}=
-(\star_\gamma\t F_5)^\mu{}_{\sla m\sla n}\del_\nu Y^{\sla n}+\cdots\ ,\ee
after remembering the no-torsion condition for derivatives of scalars.
As when deriving the D3-brane EOM originally, we have assumed that the
2-form potentials and 3-form fluxes do not contribute to the electric
components of $\t F_5$.  Now we see that the D3-brane EOM in the electric
and magnetic formalisms are equivalent (at least the terms we consider) if
\be\left[(\star_\gamma\t F_5)_{\sla m}+(\star_\gamma\t F_5)^\mu{}_{\sla m\sla n}
\del_\nu Y^{\sla n}\right]^{elec}=\left[(\star_\gamma\star\t F_5)_{\sla m}
+(\star_\gamma\star\t F_5)^\mu{}_{\sla m\sla n}\del_\nu Y^{\sla n}\right]^{mag}\ ,
\ee
which is simply the relation of the electric and magnetic components of
the fieldstrength to each other.

\subsubsection{Summary}

In the presence of a nontrivial Bianchi identity, the potential $C_4$ as 
usually defined contains both an independent degree of freedom in the 10D
SUGRA (which contributes both to the 4-form axions and compensators of
other moduli in dimensional reduction) and also a direct dependence on 
the D3-brane position.  This potential carries a Dirac string singularity
which moves with the brane (alternately described as gauge patching); 
because of the non-standard periodicity conditions, $C_4$ is not appropriate to
describe moduli or compensators in dimensional reduction, and it also
does not accurately reflect the contribution of the brane position to the
action.

As we describe in section \ref{sec:backgroundG3}, this situation is similar
to the case of background 3-form flux; we reviewed a field transformation
previously described by \cite{hep-th/0201029,0810.5768,1308.0323}
to a globally defined 4-form $C'_4$ which also makes explicit the dependence
of $\t F_5$ on fluctuations in $A_2$.  We then found that it is similarly
possible to write $\t F_5$ in the presence of a D3-brane monopole as
\begin{empheq}[box=\fbox]{align}
\t F_5 &= d S_4^*+d C'_4 -2\kappa_{10}^2 T_3 \left[\sperp \sigma_1 \dnot - 
\frac 12 (\sperp \sigma_1) \sigma^{\un m} \p_{\un m} \dnot\right. \nonumber \\
& \left. + \frac 12  \sperp (\sigma_1 \w \h d \sigma_1) \dnot
+\cdots \right]\tag{\ref{mag F5 mono}}
\end{empheq}
in terms of a new potential $C'_4$, which is globally defined and an 
independent degree of freedom.  

The explicit dependence of $\t F_5$ on the brane position contributes to the 
brane EOM, as derived in \eqref{fullD3EOM1}.  The modified EOM satisfies
the no-force condition on a static D3-brane in a GKP background, which
would otherwise be a mystery due to the lack of a WZ coupling to the magnetic
4-form.

%%%%%%%%%%%%%%%%%%%%%%%%%%%%

\subsection{4D Effective Action in Magnetic Formalism}

In this section, we calculate the 4D effective action in the magnetic
formulation of the SUGRA.  We begin by reviewing the ans\"atze for the
D3-brane positions, volume modulus, and axions, and then we compute the
quadratic action and integrate over the internal manifold.  Appendix
\ref{sec:MagneticEOM} contains details of the calculation.

\subsubsection{K\"ahler Moduli in Magnetic Components}

The 10D ans\"atze (and solutions to the constraints) for the volume
modulus and $C_4$ axions in terms of the magnetic components of $\t  F_5$ 
were presented in \cite{0810.5768,1308.0323}.  We review those results
here in concert with the ansatz for D3-brane motion as given in section
\ref{sec:magansatz}.

All these moduli are described by the same metric ansatz \eqref{metricansatz}
as in the electric formulation, including a compensator field
\be\label{magcompensator}
B_1 = -c(x)\t dK(y) +b_0^I(x)B_1^I(y)+B_1^Y(x,y)\ .\ee  
The 4-form perturbation is
\be\label{4formmagnetic}
\delta C'_4 = b_0^I(x)\t\star\omega_2^I(y) -\h db_0^I K_3^I(y)
-e^{2\Omega} \st \t d B^Y_1(x,y)\ ;
\ee
note that the volume modulus does not appear in the magnetic components of
$C'_4$, even through its compensator.  There is an additional compensator
for the axions in the 2-form potential, 
$\delta A_2=-\h db_0^I(x)\Lambda_1^I(y)$, which is nontrivial only in the 
presence of a background 3-form flux.  Altogether, the field strengths are
\be
\t F_5 &=& \st \t d e^{-4A}- e^{2\Omega} \h d(\st \t d B_1^Y)
+\h d b_0^I\w\left( \t\star\omega_2^I+\t d K_3^I-\frac{i g_s}{2}
\left(\Lambda_1^I\w \bG-\bar\Lambda_1^I\w G_3^{\0}\right)\right)
\nonumber\\ 
&&+\left[ e^{4\Omega} 
\h\epsilon \w \t d e^{4A}- e^{4\Omega}\sh \h d \t d (e^{4A} B_1)
+e^{2\Omega}\h\star\h db_0^I\w\gamma_2^I\right]\ \textnormal{and}\label{F5mag2}\\
G_3&=& \G + \h db_0^I\w\t d\Lambda_1^I\ .\label{G3mag}
\ee
The field strength $\t F_5$ is not the background plus $d\delta C'_4$ 
because $\t F_5$ includes extra terms as in (\ref{C4redefG3},\ref{mag F5 mono}).

The compensators and warp factor
are given by equations (\ref{warpshift1},\ref{KPoisson},\ref{BD3Solve}),
as in the electric formulation, along with 
\be\label{magaxcompensators}
\t\Del^2\Lambda^I_m=-\frac 12 G^{(0)}_{mnp}\gamma^{I,\widetilde{np}}\ ,\ \ 
\t\Del^2B^I_m=-e^{-2\Omega}\gamma^I_{mn}\del^{\t n}e^{-4A}-\frac{ig_s}{2}
\t\star\left(\t d\Lambda_1^I\wedge\bG-\textnormal{c.c.}\right)_m
\ee
for the axions.  The new form $\gamma_2^I$ in 
(\ref{F5mag2},\ref{magaxcompensators}) is harmonic (with the 
CY metric $\t g_{mn}$) satisfying
\be \boxed{\gamma_2^I\equiv e^{4A}\left[\omega_2^I+\t\star\left(\t dK_3^I-
\frac{ig_s}{2}\left(\t d\Lambda_1^I\wedge\bG-\textnormal{c.c.}
\right)\right)+e^{2\Omega}
\t dB^I_1\right]\equiv C^{IJ}\omega_2^J\ .}\label{gamma2def}\ee
The matrix $C^{IJ}$ is defined as in \eqref{Cinvdef}; see \cite{1308.0323}
for more details of the axion degrees of freedom.
The final constraint can be written conveniently as
\be\label{divB}
\wtn^{\t m} B_m = e^{-2\Omega}\delta e^{-4A}-e^{-4A}\delta e^{-2\Omega} \, ,
\ee
for the total compensator \eqref{magcompensator}
including the variations due to all the moduli; the compensator $B^I_1$ 
for the axions is divergenceless.

%%%%%%%%%%%%%%%%%%%%%%%%%%%%

\subsubsection{Effective Action}

The contribution to the quadratic action from the Einstein equations is only 
through the $(mn)$ component. By contracting the first-order parts of 
\be
\delta E_{mn} = \h\p^{2} \left[ 4 \del_{(m}A B_{n)} 
-2 \t g_{mn} \del^{\t p} A B_p + \wtn_{(m} B_{n)} -\frac 12 e^{-2\Omega}e^{-4A}
\t g_{mn}\right] \label{magEmn}
\ee
with
\be
\delta g^{mn} = \delta (e^{2A} \t g^{mn}) = -\frac 12 e^{6A}\delta e^{-4A} 
\t g^{mn} \, ,
\ee
we arrive at
\be
\label{mag SR}
S^R_{eff} = -\frac{1}{8\kappa_{10}^2}\int d^{4}x\int d^6y\sqrt{\t g}\, e^{4\Omega}
\left[ -6\delta e^{-4A}\h\del^{2}c+2e^{4A}\del^{\t m}\delta e^{-4A}\h\del^{2}
B_m\right]
\ee
after integration by parts and use of \eqref{divB}.  Note that 
$\delta e^{-2\Omega}=c(x)$, the volume modulus.  The first term of 
\eqref{mag SR} is proportional to 
\be \int d^6y\sqrt{\t g}\,\delta e^{-4A}=\t V c(x)\ee
because the brane motion does not change the warped volume per 
\eqref{YOmega}.  Therefore,
\be\label{mag SR2}
S^R_{eff} = \frac{1}{4\kappa_{10}^2}\int d^{4}x\, e^{4\Omega}\left[
3\t Vc\h\del^{2}c-
\int d^6y\sqrt{\t g}\,e^{4A}\del^{\t m}\delta e^{-4A}\h\del^{2}
B_m\right]\ .
\ee

Next, we examine the contribution from the 5-form EOM. The first-order parts 
of $E_6$ include the dynamical EOM for the axion as well as a contribution
from the compensator for all moduli:
\be
\delta E_6 = e^{2\Omega} \h d \sh \h d b_0^I \w \gamma_2^I
-e^{4\Omega} \h d \sh \h d \t d (e^{4A} B_1)  \, .
\ee
As in \eqref{quadaction1}, we wedge this with $\delta C'_4$ from 
\eqref{4formmagnetic} (which is globally defined and represents an independent
10D degree of freedom).  After integration by parts and some cancellation, 
we see that
\be
S^5_{eff} = \frac{1}{4\kappa^2_{10}} \left\{ \int d^{2\Omega} b_0^I\w\h d\h\star
\h db_0^J\int \t\star\omega_2^I\w\gamma_2^J-\int e^{6\Omega}e^{4A}
\t d\st \t dB_1^Y\w \h d\sh\h dB_1\right\}\ .
\ee
We can simplify this further using the 2-form inner product and
the constraint \eqref{F5offdiag} to find
\be\label{mag S5}
S^5_{eff} &=& \frac{1}{4\kappa^2_{10}}  \int d^4x \left[ 3\t V C^{IJ}e^{2\Omega}
b_0^I\h\del^{2}b_0^J + \int d^6y\sqrt{\t g}\, e^{4\Omega}e^{4A}\del^{\t m}
\delta e^{-4A}\h\del^{2}B_m\right]\nonumber\\
&&-\frac{T_3}{2}\int d^4x\int d^6y\sqrt{\t g}\,\t\delta^6(y,Y)e^{4\Omega}
e^{4A}\delta Y^{\sla m}\Lambda^m_{\sla m}\h\del^{2} B_m\ .
\ee

The linearized 3-form EOM $\delta E_8$ is non-trivial but does not contribute
to the quadratic action since $\delta A_2$ has the wrong legs to give a 
non-vanishing wedge product with it.  %In a different gauge, this sector
%does contribute, but the corresponding change in $\delta C_4$ introduces
%a compensating contribution from the 5-form sector.

Finally, the contribution to the action from the brane sector is determined by 
contracting the dynamical part of the brane EOM with the fluctuation in 
brane position. As there is no WZ term, there is no cancellation between 
terms involving $g_{\mu m}$ and $C'_4$.  Instead, given that the induced metric
is $g_{\mu\nu}$ to linear order, the linearized EOM is
\be\label{magEOM}
\delta E_{\sla m} = \left[ e^{-2\Omega}e^{-4A}\t g_{\sla m\sla n}\h\del^{2}
\delta Y^{\sla n}+\h\del^{2}B_{\sla m} -\frac 12 e^{-4\Omega}e^{-4A}
(\h\star\delta E_6)_{\sla m\sla n}\sigma^{\sla n}\right]\delta^{10}(x,X)\ .
\ee
Therefore, the quadratic action as in \eqref{quadaction1} is
\be
\label{mag SD3}
S^{D3}_{eff} &=& \frac{T_3}{2} \int d^4 x \int d^6 y \sqrt{\t g} 
\,\t\delta^{6}(y,Y)\delta Y^{\sla m} \left[ e^{2\Omega} \t g_{\sla m\sla n} 
\h\p^{2} \delta Y^{\sla n} +e^{4\Omega}e^{4A}\h\del^{2}B_{\sla m}\right.\nonumber\\
&&\left. -\frac 12 e^{2\Omega}\h\del^{2}b_0^I\gamma^I_{\sla m\sla n}\sigma^{\sla n}
+\frac 12 e^{4\Omega}\h\del^{2}\t d(e^{4A}B_1)_{\sla m\sla n}\sigma^{\sla n} \right]
\ .\ee
The second term will cancel with a similar term in $S^5_{eff}$ \eqref{mag S5}.
As in section \ref{ccs and dual}, we have
\(\t g_{\sla m\sla n}\del_\mu Y^{\sla m}\del^{\h\mu}Y^{\sla n}=2\t g_{i\bar\jmath}
\del_\mu Z^i\del^{\h\mu}\bar Z^{\bar\jmath}\);
here and in the following we drop slashes on complex indices for legibility
whenever the context is clear.

We are now forced to confront the terms that depend explicitly on the
arbitrary reference point through the appearance of $\sigma^{\sla n}$.
We begin with the term proportional to the harmonic (1,1)-form $\gamma_2$.  In
complex coordinates, we can write %(dropping slashes on indices for legibility)
\be \delta Y^{m}\gamma^I_{mn}\sigma^{n} =
C^{IJ} \left(\delta Z^{i}\omega^J_{i\bar{\jmath}}
\sigma^{\bar{\jmath}}+\delta\bar Z^{\bar{\jmath}}
\omega^J_{\bar{\jmath}i}\sigma^{i}\right)\ .
\ee
We know from our earlier calculation in the electric formulation that this
term should contain derivatives of the (locally-defined) K\"ahler potential
for the 2-form defined via 
$\omega^I_{i\bar\jmath}=i\del_i\bar\del_{\bar\jmath}k^I$, so we consider the 
derivative \(\bar\del_{\bar\jmath}(\omega^I_{i\bar k}\sigma^{\bar k})=
\omega_{i\bar\jmath}+(\bar\del_{\bar k}\omega_{i\bar\jmath})\sigma^{\bar k}\).  
Because their derivatives are the same, we
replace \(\omega^I_{i\bar\jmath}\sigma^{\bar\jmath}\to
-ik^I_i,\omega^I_{\bar{\jmath}i}\sigma^{i}\to ik^I_{\bar\jmath}\) to lowest order
in the formal $\sigma$ expansion.
\comment{\footnote{More precisely, a tensor expansion
gives $\omega^I_{i\bar\jmath}\sigma^{\bar\jmath}\equiv i\bar\del_{\bar\jmath} k^I_i
=k_i(Y)-\Lambda_i^{\un i}k_{\un i}(Y_0)+\Del_jk_i(Y)\sigma^i$; 
if we use a K\"ahler
transformation to set $k_{\un i}(Y_0)=0$, the error in the approximation is
given by the holomorphic derivative of $k_i$. \textbf{IS THERE A WAY TO 
PROVE THAT HOLOMORPHIC DERIVATIVE VANISHES B/C $\omega$ IS HARMONIC? CAN'T 
QUITE SEE IT.}} }
If, instead, our conjectured replacement $\dnot\to \delta_\perp^6(y,Y)$ is 
incorrect,  we should evaluate this term as 
$\gamma^I_{\un m\un n}\Lambda^{\un m}_{\sla m}\sigma^{\un n}(Y_*)$.
Since we can set both $k(Z,\bar Z)$ and $k_i(Z,\bar Z)$ to zero at $Y_*$
by a K\"ahler transformation, we se that
\(\gamma^I_{\un i\un{\bar\jmath}} \Lambda^{\un i}_{\sla i}\sigma^{\un{\bar\jmath}}
=C^{IJ}\omega^J_{\un i\un{\bar\jmath}} \Lambda^{\un i}_{\sla i}\sigma^{\un{\bar\jmath}}
\sim C^{IJ}k^J_{\sla i}(Y) \).
In fact, this is essentially the approximation used in \cite{Grana:2003ek} 
for these kinetic terms.  Presumably, higher order terms in the $\sigma$
expansion would make this approximation exact.  In either case, we are led
to replace
\be 
e^{2\Omega} \h\del^2b_0^I \gamma^I_{\sla m\sla n}\delta Y^{\sla m}\sigma^{\sla n}
\to -ie^{2\Omega}C^{IJ}\h\del^2b_0^I\left(k_i^J\delta Z^i-k^J_{\bar\imath}
\delta \bar Z^{\bar\imath}\right)\ ,\label{crossterms}
\ee
in precise agreement with \eqref{ElectricEffActionTot}.  

The other term, proportional to $\t d(e^{4A}B_1)$, is somewhat more subtle
because both $e^{-4A}$ and $B_1$ contain Green's functions evaluated at the
singular coincidence limit of their arguments.  Specifically, $e^{4A}\to 0$
at at D3-brane, so the volume modulus and axion compensators do not contribute
to \eqref{mag SD3}, while the divergence of $B_1^Y$ can lead a priori to a 
finite contribution for the D3-brane.\footnote{The compensators $K$ and 
$B_1^I$ do in fact
solve Poisson equations with singular sources (given by $\t G(y,Y)$), but 
the subsequent convolution against another Green's function ensures the 
compensators themselves are smooth.}  
It is possible to show for any 1-form $v_1$ that
\be \del_{[\sla p}\left((\t dv_1)_{\sla m]\sla n}\sigma^{\sla n}\right)
=(\t dv_1)_{\sla p\sla m}-\frac 12 \sigma^{\sla n}\t\Del_{\sla n}
\left(\t dv_{\sla p\sla m}\right)\ ,\ee
so it seems reasonable 
at lowest order in the formal $\sigma$ expansion to replace
\be 
\delta Y^{\sla m} e^{4\Omega}\del^{\h 2}\t d(e^{4A}B_1)_{\sla m\sla n}\sigma^{\sla n}
\to  \delta Y^{\sla m} e^{4\Omega}\del^{\h 2} e^{4A}B_{\sla m}\to
-e^{2\Omega}\delta Y^{\sla m}\del^{\h 2}\delta Y^{\sla n}\lim_{y\to Y}
\left(\Lambda^m_{\sla m}\t g_{mp}\frac{\t G^p_{\sla n}(y,Y)}{\t G(y,Y)}\right)
\hspace{1cm}\label{greenkinetic}\ee
in the coincidence limit for the warp factor \eqref{warpD3promote} and
compensator \eqref{BD3Solve}.
This term has the same general structure as the final term in 
\eqref{ElectricEffActionTot}, as it provides a not-necessarily-Hermitian 
contribution to the field space metric for the brane positions.
However, the precise form is puzzling.  Consistency with the calculation
in the electric formalism suggests that we can identify the Green's function
form of \eqref{greenkinetic} with derivatives of K\"ahler potentials in
combinations proportional to $C^{IJ}k^I_ik^J_j$, etc.  However, while it would
be interesting to conjecture that the Green's functions may be related to the
K\"ahler potentials in a similar way, the Green's functions know only about the
unwarped CY metric and not the global 
warp factor or flux information contained in
$C^{IJ}$.  Possibly, higher order terms in the formal $\sigma$ expansion
contain this information; we leave it to future work \cite{cownden}
to determine if that is true.

Alternately, we can recall that we assumed that the $\delta$ function in
the last term of \eqref{mag SD3} should be evaluated at $Y$ rather than $Y_*$.
Evaluating it at $Y_*$ regulates the Green's functions in $e^{4A}B_1$.  Since
the effective action cannot depend on the arbitrary point $Y_*$, one way to 
remove the $Y_*$ dependence is to average $Y_*$ over the CY.  This certainly
contains global information about the warp factor and flux.  Ultimately,
a resolution of these issues will require carrying the $\sigma$ expansion
to higher (or all) orders.  One approach, which we will not pursue here,
may be to note that physical quantities like $\t F_5$ are independent of $Y_*$,
leading to relations between terms at different orders in the $\sigma$
expansion.  These relations may help resum the series in a manner similar
to renormalization group flow.

Collecting the effective actions from each sector, we can construct the 
total effective action. Once again, there is significant cancellation between 
the different sectors. The total remaining action is
\begin{empheq}[box=\fbox]{align}
S_{eff} &= -\frac{3 \t V}{4\kappa^2_{10}} \int d^4 x \, 
\left( e^{4\Omega}\p_\mu c(x) \p^{\h \mu} c(x) + e^{2\Omega}C^{IJ} 
\p_\mu b^I_0 \p^{\h \mu} b^J_0 \right) \nonumber \\
& -T_3 \int d^4 x 
\left[e^{2\Omega} \t g_{i,\bar\jmath}(Y) \p_\mu Z^{i} \p^{\h \mu}
\bar Z^{\bar\jmath} -\frac i4 C^{IJ}\del_\mu b_0^I\left(k^J_{\sla i}
\del^{\h\mu} Z^{\sla i} -k^J_{\bar{\sla\imath}}\del^{\h\mu}
\bar Z^{\bar{\sla\imath}}\right)
\right.\nonumber\\ 
&\left. -\frac 12 e^{2\Omega}\t g_{mp}\Lambda^m_{\sla m}\left(
\frac{\t G^p_{\sla n}(y,Y)}{\t G(y,Y)}\right)\p_\mu \delta Y^{\sla m} 
\p^{\h \mu}\delta Y^{\sla n}\right] \, ,\label{MagneticEffActionTot}
\end{empheq}
assuming we use the replacements (\ref{crossterms},\ref{greenkinetic}).

Our final result \eqref{MagneticEffActionTot} is not manifestly equal to
the effective action \eqref{ElectricEffActionTot} found using the electric
formalism for IIB SUGRA, and the two actions may not be equal at all.  
While a proper treatment must yield the same effective action whether we
take electric or magnetic components for $\t F_5$, \eqref{MagneticEffActionTot}
relies on two conjectured replacements (\ref{crossterms},\ref{greenkinetic})
as well as a D3-brane EOM truncated at second order in a formal expansion; 
we have not yet been able to verify all our assumptions.
As a result, it is unsurprising that we do not have precise agreement with
the complete derivation of \eqref{ElectricEffActionTot}.  However, the 
agreement of the general form of the action is striking: in the magnetic
formalism, backreaction of the D3-brane on $\t F_5$ induces a 
$\del b_0\del Z$ cross term of the correct form (and correct coefficient, 
if (\ref{crossterms}) is correct) as well as non-Hermitian $\del Y \del Y$
kinetic terms.  The possibility that higher orders in the formal $\sigma$
expansion will lead to a relation between the derivatives of the CY
K\"ahler potential and Green's functions is intriguing, but we leave that
to the future.

\subsubsection{Summary}

We provided a unified ansatz for fluctuations of the volume modulus, $C_4$
axions, and D3-brane position in the magnetic formalism and performed a
dimensional reduction on this ansatz.  Interestingly, 
the D3-brane EOM as modified according to the results of the previous 
subsection contains terms of the form $i\del b k_i\del Z^I+\textnormal{c.c.}$ 
and additional $\del Y\del Y$ kinetic terms. In the electric formalism,
these terms had appeared due to the WZ action, which does not exist in the
magnetic formalism.  Although we have not been able to give a full
interpretation of these terms in the magnetic formalism --- there are 
possibly more contributions from higher orders in the formal $\sigma$
expansion --- it is important to note that these crucial ``cross terms'' 
in the kinetic action arise due to the backreaction of the brane motion
on the 5-form.

%%%%%%%%%%%%%%%%%%%%%%%%%%%%
%%%%%%%%%%%%%%%%%%%%%%%%%%%%
%%%%%%%%%%%%%%%%%%%%%%%%%%%%

\section{Discussion}
\label{sec:discussion}

D-branes are important ingredients in flux compactifications, and 
their dynamics are essential for
understanding the structure of the K\"ahler moduli of the low energy 
effective theory as well as applications in string phenomenology and cosmology.
For instance, several models of inflation in string theory explicitly 
use the motion of D-branes in warped regions in their construction
(see \cite{Kachru:2003sx,Baumann:2007ah}), so a correct description 
of D-brane dynamics is not of idle interest.

We have shown that a consistent dimensional reduction of 10D supergravity 
in the presence of a D3-brane requires the inclusion of fluctuations in the 
10D metric and 5-form gauge potential, in addition to the degrees of freedom 
of the transverse motion of the D3-brane.
The D3-brane can couple to the 4-form as either an electric or 
magnetic source, and we presented for both
cases the first consistent set of fluctuations that solve the 10D constraint 
equations. For a D3-brane coupling as a magnetic source, we find a novel 
field redefinition of the magnetic 4-form potential
that allows the 5-form field strength to be written in terms of a 
globally-defined 4-form plus delta-function-supported terms that make
the dependence of $\t F_5$ on the D3-brane position explicit. 
The field redefinition leads to additional terms 
in the D3-brane equation of motion from $\t F_5$,
which resolve a puzzle involving the no-force condition on a D3-brane 
in the magnetic description, as well as
contributing important terms to the effective action.  

Combining our consistent 10D description of transverse D3-brane degrees of 
freedom with existing
descriptions for the volume modulus \cite{0810.5768} and $C_4$ 
axions \cite{1308.0323}, we performed a careful dimensional
reduction to obtain the 4D effective action. The resulting effective action 
contains important contributions
due to flux and warping, as previously seen in the axion sector in 
\cite{1308.0323}.  The calculation involves a remarkable set of cancellations 
between the compensators and Green's functions, demonstrating
the importance of a consistent 10D solution of the constraint equations.
We also explicitly demonstrated that there is no ``self-energy'' problem for 
the dynamical effective action which might arise from inserting the 
backreacted brane solution into the brane effective action.
The kinetic action includes the expected brane-axion cross-terms as well as
kinetic terms for transverse brane motion in addition to the manifest
kinetic term in the DBI action.  When treating the brane as an electric
source, these terms arise from the Hodge dualization of the axions from 2-forms
to scalars in the presence of the brane WZ action.  When the underlying
CY manifold has only a single K\"ahler modulus, the K\"ahler potential
agrees with the proposal of \cite{hep-th/0208123,Kachru:2003sx}.  
In the dimensional 
reduction treating the brane as a magnetic source, these additional terms 
arise from the explicit dependence of $\t F_5$ on the brane position along
with the backreaction of the brane on the 5-form.  While we have not
reproduced the precise form of the additional D3-brane kinetic terms, 
there are intriguing hints that these terms could be related to Green's 
functions on the internal space.

Many cancellations in the dimensional reduction likely occur 
because of the high degree of structure of the background, including
supersymmetry and no-scale structure. 
An effective action for D3-branes in a more general warped 
background with interesting applications will likely be more complex, and
the techniques developed here can play an important role in the necessary 
calculation.  Moving
beyond the probe approximation, which we have argued is necessary, it would be 
interesting to see if there
are modifications to the kinetic part of the effective action arising from 
the interaction of the D3-brane with the 10D fields.
As another example, the dynamics of $\overline{\textnormal{D3}}$-branes 
in warped flux backgrounds \cite{Kachru:2002gs} beyond
the probe approximation should also include perturbations to the 10D fields 
in the effective description.
We leave a detailed investigation of these and other applications to future 
work. Nevertheless, we have seen the importance of a consistent 10D 
description, solving the constraint equations,
for constructing a 4D effective action from dimensional reduction.

\acknowledgments
We would like to thank L.~McAllister for helpful discussions during the 
early stages of this project and D.~Andriot for comments on an earlier version
of the manuscript.
B.~U.~would like to acknowledge support by NSERC, an Institute of 
Particle Physics Postdoctoral Fellowship, and a Lorne
Trottier Fellowship from McGill University for support during the early stages 
of this project.  The work of A.~F.~and B.~C.~has also been supported by 
NSERC.

%%%%%%%%%%%%%%%%%%%%%%%%%%%%
%%%%%%%%%%%%%%%%%%%%%%%%%%%%
%%%%%%%%%%%%%%%%%%%%%%%%%%%%

\appendix

\section{Conventions}\label{a:conventions}

Here we summarize our conventions and notational shorthand.  
External, noncompact
spacetime coordinates denoted $x^\mu$, while internal, compact dimension 
coordinates are $y^m$; when used, $x^M$ include all coordinates.  
Brane worldvolume coordinates are $\xi^a$, and the embedding of the
worldvolume into spacetime is denoted $X^{\sla M}(\xi)$ (with $\xi$
dependence sometimes suppressed).  The slashed index indicates that the 
coordinate transforms under diffeomorphisms as the position of the brane
as opposed to the spacetime point $x^M$ where SUGRA fields are evaluated
(see appendix \ref{sec:bitensors} on bitensors).

Quantities with a hat $\hat{}$ are associated
with the 4D metric $\h\eta_{\mu\nu}$, such as raised or lowered indices, the
antisymmetric tensor $\h\epsilon_{\mu\nu\lambda\rho}$ 
(or volume form $\h\epsilon$).  Similarly, any quantity with a tilde $\t{}$ 
is associated with the unwarped CY metric $\t g_{mn}$.  However, as 
partial derivatives are metric-independent, we do not accent them (ie,
we write $\del_\mu,\del_m$) except for raised indices.  However, for appearance,
we accent the derivatives rather than the square in Laplacians/d'Alembertians,
writing $\h\del^2$ and $\t\Del^2$ rather than $\del^{\h 2}$ and $\t\Del^{\t 2}$.
Ten-dimensional quantities have capital indices but no accents.  A
superscript $(0)$ with parentheses indicates a background value.

Two other metrics appear in this paper, the worldvolume metric $\gamma_{ab}$
and an arbitrary metric $g^\perp_{mn}$.  Indices $a,b$ denote quantities 
associated with $\gamma_{ab}$; in static gauge $\Del^\gamma_\mu$ is also 
the covariant derivative associated with the worldvolume metric.  Quantities
associated with $g^\perp_{mn}$ are denoted with a $\perp$ sub- or superscript.

We work with a mostly + metric and define the antisymmetric symbol 
$\epsilon$ as a tensor.
Similarly, delta functions are defined as scalars, so $\epsilon$ and $\delta$
implicitly carry $\sqrt{|g|}$ factors.
Combinatorial factors for differential forms are defined as in appendix B of 
\cite{Polchinski:1998rr}.  The wedge symbol in a wedge product 
may be omitted in in-line mathematics.

Harmonic 2-forms are written in terms of a basis
$\{\omega_2^I\}$ on the CY manifold, so any harmonic form can be written
as $e^I\omega_2^I$ for constant coefficients $e^I$.  This is not the basis of 
harmonic forms at a single point; if $h_{1,1}$ is greater than the 
2nd Betti number of $T^6$, the $\omega_2^I$ are not all linearly independent at
any given point, only as functions. We orthonormalize the basis with respect
to the inner product
\be \int \omega_2^I\w\t\star\omega_2^J =3\t V \delta^{IJ}\ .\ee
This normalization allows us to choose the K\"ahler form as 
$\omega_2^1=\t J$, since $\t J^3=6\t\epsilon$ and $\t\star\t J =\t J^2/2$.  
(Strictly speaking, in this paper we only consider 2-forms with positive
parity under the orientifold involution, but similar considerations would
apply for those with negative parity.)

%%%%%%%%%%%%%%%%%%%%%%%%%%%%%%%%%%%%%%%%%%%%%%%
%%%%%%%%%%%%%%%%%%%%%%%%%%%%%%%%%%%%%%%%%%%%%%%

\section{Bitensors, Expansions, and Green's Functions}
\label{sec:bitensors}

As we have noted previously, any attempt to describe the influence of D-branes
on SUGRA fields necessarily involves (at least) two points in spacetime:
the position where the SUGRA fields are evaluated and the position of the
localized brane source (respectively $y^m$ and $Y^{\sla m}$ in static gauge).  
The SUGRA fields are generally functions of 
\emph{both} of these positions.  We also must consider both positions when
evaluating the 10D EOM, as there are bulk equations evaluated at $y^m$ and
brane equations involving fields evaluated at $Y^{\sla m}$.
At some points, we also introduce a fixed reference point $Y_*^{\un m}$.
As diffeomorphisms in general act differently at
different points, we use distinct markings to indicate which transformation
acts on a given tensor index (ie, unmarked, slashed, or underlined).
Here we give a very brief review of the properties of tensor functions of
two spacetime points, known as \emph{bitensors}, following \cite{1102.0529}.

A bitensor is a tensorial function of two points in spacetime, which may
have indices marked for either of the two points.  As an example, the 
5-form $\t F_5$ has 5 indices associated with the evaluation point $y$ but 
also depends on the brane position $Y$, ie, $\t F_{mnpqr}(y,Y)$.
A key example for us is the 6-dimensional biscalar Dirac distribution
(delta function) $\delta^6(y,Y)$ (for some metric $g_{mn}$) defined by
\be
\int d^6y \sqrt{g(y)} f(y) \delta^6(y,Y) = f(Y)\ ,\ \ 
\int d^6Y \sqrt{g(Y)} f(Y) \delta^6(y,Y) = f(y)\ .
\ee
Note that the Dirac distribution integrates as a scalar in both coordinates,
so it implicitly includes a factor of $1/\sqrt{g}$.  (In the main text,
we will consider Dirac distributions for metrics $\t g_{mn}$ and $g^\perp_{mn}$.)

\begin{figure}[t]
\centering \includegraphics[scale=0.55]{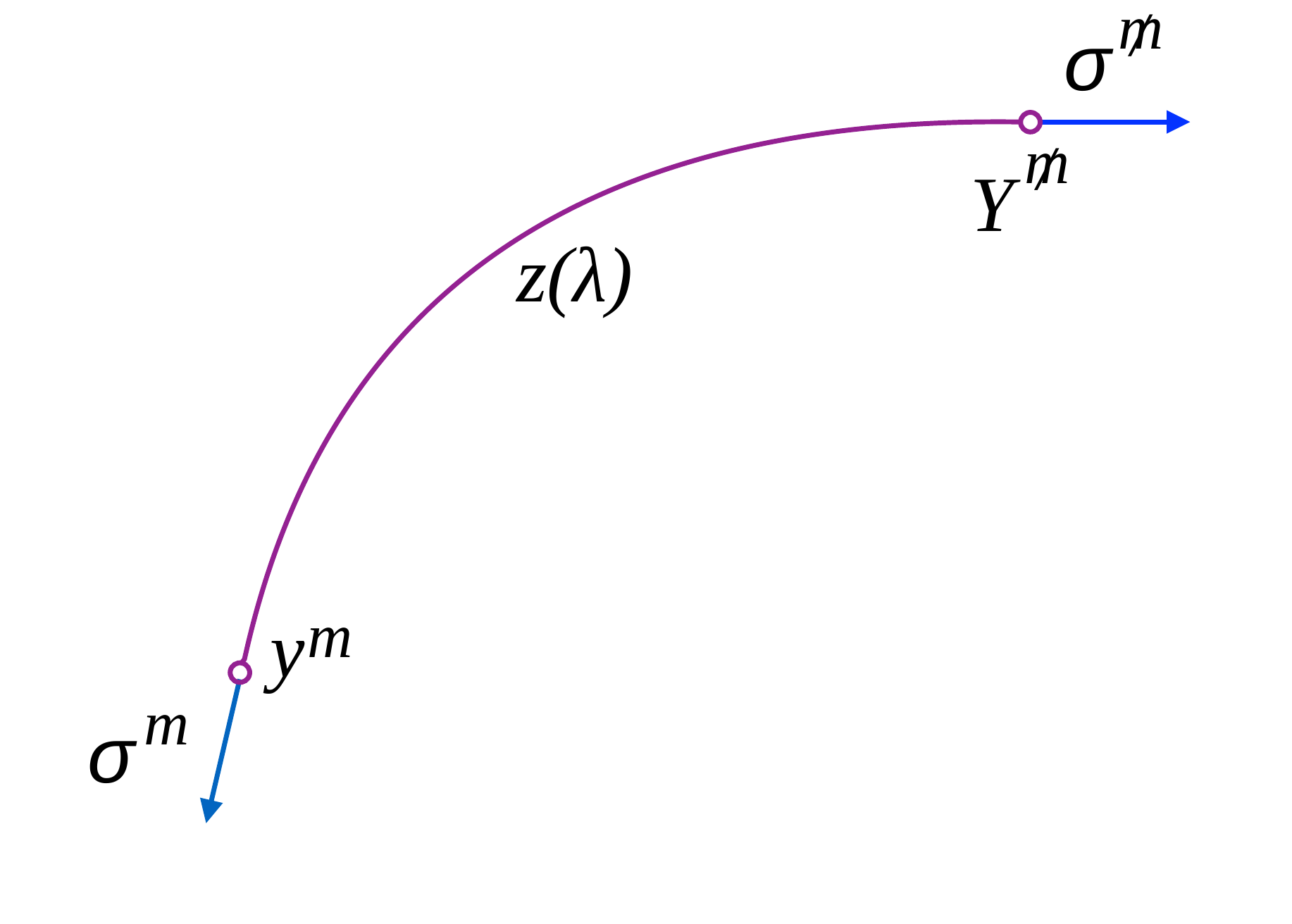}
\caption{The geodesic $z(\lambda)$ connects $y^m$ to $Y^{\sla m}$. 
$\sigma^m,\sigma^{\sla m}$ are outgoing tangents to the geodesic at the 
endpoints with length equal to the geodesic distance.}
\label{fig:Synge}
\end{figure}

To understand the coincidence limit of a bitensor as well as the expansion of
a tensor around a fixed point, we consider the Synge world function.
This is a biscalar function of $y,Y$ defined by
\be
\label{Synge}
\sigma(y,Y) = \frac{1}{2} (\lambda_1 - \lambda_0) \int_{\lambda_0}^{\lambda_1} 
g_{mn}(z) t^m t^n d\lambda \ ,
\ee
where $z^m(\lambda)$ describes a geodesic with $z(\lambda_0) = y$ and 
$z(\lambda_1) = Y$ as depicted in figure \ref{fig:Synge}.
For affine parameter $\lambda$, $t^m$ is tangent to the geodesic, and
$\sigma(y,Y)$ is half the squared geodesic distance between $y$ and $Y$. 
The derivatives $\sigma_m\equiv \del_m\sigma$ and 
$\sigma_{\sla m}\equiv\del_{\sla m}\sigma$ are tangent to the geodesic at 
the respective endpoint and directed outward, as in the figure.  It can
be shown that $\sigma_m\sigma^m=\sigma_{\sla m}\sigma^{\sla m}=2\sigma$.
A covariantly constant vector $A^{m}$ can be parallel transported from $y$ 
to $Y$ along the geodesic via the parallel propagator $\Lambda^{\sla m}_n$
as $A^{\sla m}(Y) = \Lambda^{\sla m}_{n}(y,Y) A^{n}(y)$, with the corresponding
generalization for covariantly constant tensors.  At coincidence
$y=Y$, $\Lambda^{\sla m}_{n}=\delta^{\sla m}_n$, so \emph{any} bitensor
(with any distribution of indices) satisfies
\be T_{m_1\cdots m_a \sla{n}_1\cdots\sla{n}_b}{}^{p_1\cdots p_j \sla{q}_1\cdots\sla{q}_k}
(y,Y)\left(\Lambda^{\sla{n}_1}_{n_1}\cdots \right)
\left(\Lambda^{q_1}_{\sla{q}_1}\cdots \right)\delta(y,Y)=
T_{m_1\cdots m_a n_1\cdots n_b}{}^{p_1\cdots p_jq_1\cdots q_k}(y,y)\delta(y,Y),
\ \ \ \ \ 
\ee
etc.

%%%
%%%

There are several useful identities among the Dirac distribution, parallel
propagator, and derivatives of the worldfunction.  
First, because $\sigma^m\propto t^m$, the tangent to the geodesic, we have
\be
\sigma_{\sla n} = - \Lambda^{m}_{\sla n} \sigma_{m}  \qquad \text{and} \qquad 
\sigma_{n} = - \Lambda^{\sla m}_{n} \sigma_{\sla b}\, .
\ee
Next, in the coincidence limit, parallel propagators are covariantly
constant with respect to either endpoint.  Finally,
the Dirac distribution and the parallel propagator satisfy the identities
\be\label{diracderivs}
\nabla_m\left( \Lambda^m_{\sla n}(y,Y) \delta(y,Y)\right) = - \partial_{\sla n} 
\delta(y,Y) \ ,\ 
\nabla_{\sla m} \left( \Lambda_n^{\sla m}(y,Y) \delta(y,Y)\right) = 
- \partial_{n} \delta(y,Y)
\ee
in any dimensionality.

Of course, it is often useful to evaluate a tensor as a series expansion
around a fixed reference point, preferably in a manifestly covariant manner.
One possible application is to consider the expansion of a bitensor
$T_{m_1\cdots \sla{n}_1\cdots}(y,Y)$ around coincidence $y=Y$; in covariant 
form, this is an expansion in powers of $\sigma^m$ (or alternately
$\sigma^{\sla m}$) \cite{1102.0529}.  Another application is the expansion
of a tensor as a function of $Y^{\sla m}$ near a reference point 
$Y_\star^{\un m}$.  As an example, the expansion of a scalar is
\be
\label{SyngeExpand}
A(Y) = A(Y_*) - \del_{\un n} A(Y_*) \sigma^{\un n} + \frac{1}{2} 
\nabla_{\un m} \del_{\un n} A(Y_*) \sigma^{\un m} \sigma^{\un n} + \cdots \, ,
\ee
where the dots represent terms higher order in $\sigma^{\un m}$.  In the main
text, we use this to expand $\t \delta^6(y,Y)$ around $Y=Y_\star$, taking
$y$ as a constant, so the delta function is just a scalar function of $Y$.

Finally, let us discuss the behavior of Green's functions on curved space.
Consider a minimally coupled massless scalar $\Phi$ and a vector field 
$A^m$ which satisfy the Poisson equations
\be
\nabla^{2} \Phi = - \mu(y)\ , \ \ \nabla^{2} A^m = - j^m(y)\, ,
\ee
where $\mu,j^m$ are sources.  We can write the solutions in terms of the 
biscalar and bitensor Green's functions $G(y,y')$ and $G^m_{m'}(y,y')$ as
\be
\Phi(y) = \int dY \sqrt{g(Y)}\, G(y,Y)\ \mu(Y) \ , \ \ 
A^m(y) = \int dY \sqrt{g(Y)}\, G^m_{\sla m}(y,Y)\ j^{\sla m}(Y) 
\ee
(in any dimensionality). The Green's functions are defined to satisfy
\be
\label{G relations}
\nabla^{2} G(y,Y) = - \delta(y,Y) \ ,\ \ 
\nabla^{2} G^m_{\sla m}(y,Y) = -\Lambda^m_{\sla m} \delta(y,Y)\ .
\ee
The scalar and tensor Green's functions are related by
\be
\label{Gs}
\nabla_m G^m_{\sla m} (y,Y) = - \p_{\sla m} G(y,Y)
\ee
and vice-versa.  We are mostly concerned with the 6D metric $\t g_{mn}$
and corresponding Green's functions $\t G(y,Y)$ and $\t G^m_{\sla m}(y,Y)$.

%%%%%%%%%%%%%%%%%%%%%%%%%%%%%%%%%%%%%%%%%%%%%%%
%%%%%%%%%%%%%%%%%%%%%%%%%%%%%%%%%%%%%%%%%%%%%%

\section{Linearized Equations of Motion}

In this appendix, we assemble all the linearized 10D EOM for the 10D SUGRA 
fields; these are the Einstein equations, 5-form and 3-form flux equations, 
and D3-brane position equations.  We will identify how the EOM divide into
constraints and dynamical EOM; the electric form of the equations are listed
first, followed by the magnetic form.

%%%%%%%%%%%%%%%%%%%%%%%%%%%%

\subsection{Electric Formalism Equations of Motion}
\label{sec:ElectricEOM}

Here we provide the EOM for the electric formalism for D3-brane motion.
For reference, our ansatz for the SUGRA fields is
\begin{align}
ds^2 &= e^{2\Omega}e^{2A} \h\eta_{\mu\nu}dx^\mu dx^\nu + 2e^{2\Omega}e^{2A} 
\p_\mu B_m (x,y) dx^\mu dy^m + e^{-2A} \til g_{mn} dy^m dy^n \, ,
\tag{\ref{metricansatz}} \\
\t F_5&= e^{4\Omega}\h\epsilon\w\t de^{4A}+d\left(e^{4\Omega}e^{4A}\h\star\h dB_1
\right)+\h d b_2^I\w \omega_2^I \nonumber \\
& +\left[\t\star\t d e^{-4A}-e^{2\Omega}e^{-4A} \t\star\left(
\h dB_1\w\t de^{4A}\right) +\star d\left(e^{4\Omega}e^{4A}\h\star\h dB_1\right) 
+ e^{-2\Omega}\h\star\h d b_2^I\w e^{-4A}\t\star \omega_2^I\right]
 ,\tag{\ref{F5ansatz2}} \\
G_3 &= G_3^{(0)} - e^{-2\Omega} \hat d \hat \star \hat d b_2^I \w \Lambda_1^I 
+ e^{-2\Omega} \hat \star \hat d b_2^I \w \t d \Lambda_1^I\, ,
\tag{\ref{G3ansatz}}
\end{align}
along with $Y^{\sla m}(x)=Y^{(0)\sla m}+\delta Y^{\sla m}(x)$.  As in the main
text, the components of $\t F_5$ in square brackets are the magnetic 
components provided for reference as $\star\t F_5$ of the electric components.
Above, and throughout, the metric compensator $B_1(x,y)$ is the 
\emph{total} compensator for all the moduli
\begin{align}
\h d B_1(x,y)\equiv -\h dc(x)\wedge \t d K(y)
+e^{-4\Omega}\h\star\h db_2^I(x)\wedge B_1^{b,I}(y)+\h dB_1^Y(x,y)\ . 
\tag{\ref{B1 tot}}
\end{align}
This is clearly an abuse of notation; for reference,
$\h d\h\star\h dB_1=-\h d\h\star\h dc\t dK+\h d\h\star\h dB_1^Y$ and
$\h d^2B_1=e^{-4\Omega}\h d\h\star\h db_2^IB_1^{b,I}$ to first order.
Furthermore, the warp and Weyl factors include first-order contributions, 
ie $A=A(x,y)$ and $\Omega=\Omega(x)$.  Finally, to allow for the presence
of a compensator $\delta C_4\sim -\hat d b_2^I K_1^I$, we do not yet require 
that $\omega_2^I$ be harmonic (though it must be closed to avoid terms in 
$\t F_5\sim b_2^I\t d\omega_2^I$).
A number of the results in this
appendix follow from calculations in \cite{1308.0323}.

\subsubsection{Einstein Equations}

The components of the Ricci tensor to first order are
\be
\label{app:Rmunu}
 R_{\mu \nu} &=& \p_\mu \p_\nu (4A - 2\Omega) - \h\eta_{\mu \nu} \h\p^{2} 
(A+\Omega) + e^{4A}e^{2\Omega} \left(\p^{\t \ell} A \h\p^{2} B_\ell 
\h\eta_{\mu \nu} - \wtn^{2} A \h\eta_{\mu \nu} + \p_\mu \p_\nu \wtn^{\t \ell} 
B_\ell \right) \, , \hspace{.4in}\\
\label{app:Rmum}
R_{\mu m} &=& 2 \p_\mu \p_m A - 8\p_\mu A \p_m A + e^{4A}e^{2\Omega} 
\left( \p_\mu \wtn^{\t \ell} (\wtn_{[m}B_{\ell]})  - \wtn^{2} A \p_\mu B_m + 
4 \del^{\t \ell} A \p_\mu \til\p_{[m}B_{\ell]} \right) \, , \\
\label{app:Rmn}
R_{mn} &=& \h\p^{2} \wtn_{(m}B_{n)} + 4 \del_{(m}A \h\p^{2} B_{n)} - 
\del^{\t\ell}A\h\p^{2} B_{\ell} \til g_{mn} + \wtn^{2} A \til g_{mn} + 
e^{-4A}e^{-2\Omega} \h\p^{2} A \til g_{mn} \nonumber \\
&& - 8 \p_m A \p_n A + \til R_{mn} \, .
\ee
We note for later that we did not need to use $\h d^2 B_1=0$, which is not
true off shell for \eqref{B1 tot}.  Here, $\t R_{mn}$ is the Ricci tensor of
$\t g_{mn}$; for the CY metrics we consider here, it vanishes, so we set
$\t R_{mn}=0$ henceforth.

Using these, we can calculate the Ricci curvature, $\mc R$, to first order
\be
\mc R = 6 e^{-2A}e^{-2\Omega} \h\p^{2} (A-\Omega) + 2 e^{2A} \left( \h\p^{2} 
\wtn^{\t \ell} B_\ell + \del^{\t \ell} A \h\p^{2} B_\ell + \wtn^{2} A - 
4 \del^{\t \ell} A \del_\ell A \right) \, ,
\ee
and the Einstein tensor, whose components are
\be
\label{app:Gmunu}
G_{\mu \nu} &=& (\p_\mu \p_\nu - \h\eta_{\mu \nu} \h\p^{2} )\left(4A - 2\Omega + 
e^{4A}e^{2\Omega} \wtn^{\t \ell} B_\ell\right) + 2 e^{4A}e^{2\Omega} \h\eta_{\mu \nu}
\left(2 \del^{\t \ell} A \del_\ell A - \wtn^{2} A\right) \, , \\
\label{app:Gmum}
G_{\mu m} &=& 2\p_\mu \p_m A - 8 \p_\mu A \p_m A + e^{4A}e^{2\Omega} \p_\mu
\left( \wtn^{\t \ell} \wtn_{[m}B_{\ell ]} + 4 \del^{\t \ell} A \wtn_{[m}B_{\ell]} 
\right. \nonumber \\
&& \left. + 2 B_m (2 \del^{\t \ell} A \del_\ell A - \wtn^{2} A) \right) \, , \\
\label{app:Gmn}
G_{mn} &=& \h\p^{2} \wtn_{(m}B_{n)} - \h\p^{2} \wtn^{\t \ell} B_\ell\t g_{mn}+ 
4 \del_{(m}A \h\p^{2} B_{n)} - 2  \del^{\t \ell} A \h\p^{2} B_\ell\t g_{mn}
- 8 \p_m A \p_n A +  4 \del^{\t \ell} A \del_\ell A\til g_{mn} \nonumber \\
&& + e^{-4A}e^{-2\Omega} \til g_{mn} \h\p^{2} (3\Omega - 2A) \, .
\ee

Next, we determine the stress-energy tensor. We remind the reader that 
the contributions of the Ramond-Ramond fluxes to the energy-momentum are
\begin{align}
\tag{\ref{T5}}
T^5_{MN} &= \frac{1}{4 \cdot 4!} \til F_{MPQRS} \til F_N{}^{PQRS} \, ,\quad
T^3_{MN}=\frac{g_s}{4} \left(G_{(M}{}^{PQ} \bar G_{N)PQ}-g_{MN} |G|^2\right)\ .
\end{align}
%Our ansatz for the electric form of the self-dual 5-form $\t F_5$ is
The resulting energy-momentum tensor for the 5-form \eqref{F5ansatz2}, 
including terms up to first-order, has components
\be
T^5_{\mu\nu} &=& 2 e^{4A} e^{2\Omega} \hat \eta_{\mu\nu}\left(\del^{\t \ell} A 
\hat \partial^{2} B_\ell - 2 \del^{\t \ell} A \del_\ell A\right)\, , \\
T^5_{\mu m} &=& 4 e^{4A} e^{2\Omega} \left(\del^{\t \ell} A \partial_\mu 
\t \nabla_{[m} B_{\ell]} - \del^{\t \ell} A \del_\ell A \partial_\mu B_m\right)
- 2 e^{-2\Omega} (\h \star \h d b_2^I)_\mu \omega_{mn} \partial^{\t n} A \, ,\\
T^5_{mn} &=& 4 \del^{\t \ell} A\del_\ell A \ \t g_{mn} - 8 \del_m A \del_n A 
+ 4 \del_{(m}A \h\partial^{2} B_{n)} - 2 \del^{\t \ell}A \h \del^{2} B_\ell \t g_{mn}
\, .\ee
The resulting energy-momentum tensor for the 3-form \eqref{G3ansatz}, 
including terms up to first-order, has components
\be
T^3_{\mu\nu} &=& -\frac{g_s}{4}e^{2\Omega} e^{8A} \left|G_3^{(0)}\right|^{\t 2} 
\h \eta_{\mu\nu}\, ,\\
T^3_{\mu m} &=& -\frac{g_s}{4} \left[ie^{-2\Omega} e^{4A} 
(\h \star \h d b_2^I)_\mu 
\t\star\left(\t d \Lambda_1^{I}\wedge \bar G^{(0)} - \textnormal{c.c.}\right)_m
+e^{2\Omega} e^{8A}\partial_\mu B_m \left|G_3^{(0)}\right|^{\t 2}\right] \, ,\ \  
\label{app:TmumG3}\\
T^3_{mn} &=& \frac{g_s}{4} \left(e^{4A} (G^{(0)})_m {}^{\t p\t q} 
(\bar G^{(0)})_{npq} - \t g_{mn} e^{4A} \left|G_3^{(0)}\right|^{\t 2}\right)=0
\, .\label{app:T3mnG3}\ee
We have used the imaginary self-duality of the background flux, 
ie $\t \star G_3^{(0)} = i G_3^{(0)}$, to simplify 
(\ref{app:TmumG3},\ref{app:T3mnG3}).
%, since
%$(\t d \Lambda_1)^{\t p \t q} \bar G_{mpq}^{(0)} = -i (\t d \Lambda_1 \w \bar G_3^{(0)})_m$.

The energy-momentum tensor for our mobile D3-brane comes from
\begin{align}
T_{MN}^{D3} &= - \kappa_{10}^2T_3 \int d^4 \xi \sqrt{-\gamma} \, \gamma^{ab} 
\Lambda_M^{\sla M}\Lambda_N^{\sla N}g_{\sla M\sla P} g_{\sla N\sla Q} \p_a X^{\sla P} 
\p_b X^{\sla Q} \delta^{10}(x,X(\xi)) \, . \tag{\ref{TD3}}
\end{align}
We will use slashed indicies $X^{\sla M}$ to refer to the embedding 
coordinates of the D3-brane 
and will work in static gauge $\xi^a = \delta^a_{\sla \mu} X^{\sla \mu}(\xi)$.
The D3-brane energy-momentum tensor, up to first order, has components
\be
\label{app:TmunuD3} T^{D3}_{\mu\nu} &=& -\kappa_{10}^2 T_3 e^{8A} e^{2\Omega} 
\hat \eta_{\mu\nu} \t \delta^6(y,Y)\, ,\\
\label{app:TmumD3} T^{D3}_{\mu m} &=& -\kappa_{10}^2T_3  e^{8A} e^{2\Omega} 
\partial_\mu B_m \t \delta^6(y,Y) -T_3 e^{4A} \t g_{m\sla n} 
\partial_\mu Y^{\sla n}(x) \t \delta^6(y,Y)\, ,\\
\label{app:TmnD3} T^{D3}_{mn} &=& 0 \, .
\ee
Other localized sources (such as other D3-branes and O3-planes) also contribute
an energy momentum tensor $T^{loc}_{MN}$ identical in form to 
(\ref{app:TmunuD3},\ref{app:TmumD3},\ref{app:TmnD3}), with the exception
that $T^{loc}_{\mu m}$ does not contain the term with explicit dependence on 
the single mobile brane's position, $\partial_\mu Y^{\sla n}$.

At the end of the day, we obtain the 10D Einstein equations  
$E_{MN} = G_{MN} -(T^5_{MN}+T^3_{MN}+T^{D3}_{MN}+T^{loc}_{MN})$ explicitly in terms
of our ansatz (through first order in fluctuations):
\be
\label{app:Emunu}
E_{\mu\nu}&=& e^{2\Omega}e^{4A}\left[2(4 \del^{\t \ell} A \del_\ell A - 
\t \nabla^{2} A)+T_3 e^{4A}\t\delta^6(y,Y)+\frac{g_s}{4} e^{4A} 
\left|G_3^{(0)}\right|^{\t 2} + \cdots \right] \h \eta_{\mu\nu} \nonumber \\
&&+(\p_\mu\p_\nu - \h\eta_{\mu \nu} \h\p^{2} )(4A - 2\Omega + e^{4A}e^{2\Omega} 
\wtn^{\t \ell} B_\ell) - 2 e^{4A} e^{2\Omega} \h\eta_{\mu\nu} \del^{\t \ell} A 
\h \partial^{2} B_\ell\, ,\\
E_{\mu m} &=& e^{2\Omega}e^{4A}\partial_\mu B_m \left[ 2(4 \del^{\t \ell} A 
\del_\ell A - \t \nabla^{2} A)+\kappa_{10}^2 T_3 e^{4A} \t\delta^6(y,Y)
+ \frac{g_s}{4} e^{4A} \left|G_3^{(0)}\right|^{\t 2}+ \cdots \right] \nonumber\\
&& +2\p_\mu \p_m A - 8 \p_\mu A \p_m A + e^{4A}e^{2\Omega} \p_\mu \wtn^{\t \ell} 
\wtn_{[m}B_{\ell]}+T_3 e^{4A} \t g_{m\sla n}\p_\mu Y^{\sla n}(x) \t \delta^6(y,Y) 
\nonumber \\
&& +2 e^{-2\Omega} e^{4A} (\h \star \h d b_2^I)_\mu \left[e^{-4A}\omega^I_{mn} 
\partial^{\t n} A + \frac{i g_s}{8} \t \star \left(\t d \Lambda_1^I\w 
\bar G^{(0)}_3 - \textnormal{c.c.}\right)\right]\, ,\label{app:Emum} \\
\label{app:Emn}
E_{mn} &=& \hat \partial^{2} \left( \t \nabla_{(m} B_{n)}
-\t g_{mn} \t \nabla^{\t \ell} B_\ell \right)	
+ e^{-4A} e^{-2\Omega} \t g_{mn} \hat \partial^{2} \left(3\Omega -2 A\right)\, ,
\ee
where $\cdots$ denotes the contributions due to local sources other than our 
mobile D3-brane, whose precise forms are unimportant.  The Einstein equations
contain both constraints and dynamical EOM  For example, the $(\mu\nu)$ and 
$(\mu m)$ components both contain the Poisson equation
\eqref{coulomb2} that determines the warp factor to be \eqref{warpshift1},
which now must be satisfied to first order point-by-point
on the external spacetime (in this way, our choice of coordinates is similar
to the Coulomb gauge of Maxwell theory).  The $(\mu\nu)$ component 
\eqref{app:Emunu} also contains a constraint proportional to 
$\del_\mu\del_\nu-\h\eta_{\mu\nu}\h\del^2$ (yielding \eqref{Bdiv2})
and a dynamical EOM in the last term, while the remainder of the $(\mu m)$
component \eqref{app:Emum} is a constraint \eqref{offdiagconstraint}
determining the contribution of each modulus to $B_m$ (this gives both
(\ref{BD3Solve},\ref{axeinsteinconstraint})).
Finally, $E_{mn}$ is entirely second-order 
in external spacetime derivatives and contributes a dynamical EOM.

\subsubsection{Form Flux EOM}

In addition to the 10D Einstein equations, we also have 10D EOM 
from the 5-form and 3-form fluxes, given by
\begin{align}
E_6 &= d\star \tilde{F}_5 -\frac{ig_s}{2} G_3\w\bar G_3
+ 2\kappa_{10}^2 T_3 \int d^4 \xi \sqrt{-\gamma} \, \star 
\epsilon_\| \, \delta^{10}(x,X(\xi)) \, ,\tag{\ref{E6}} \\
E_8&=d\star G_3+iG_3\w\left(\t F_5+\star\t F_5\right) +\frac i2 A_2\w E_6\, .
\tag{\ref{E8}}
\end{align}
Using the ansatz (\ref{metricansatz},\ref{F5ansatz2},\ref{G3ansatz}), 
the 5-form EOM becomes
\be
E_6 &=& \t d \t \star \t d e^{-4A} - \frac{i g_s}{2} G_3^{(0)} \w \bar G_3^{(0)} - 2 \kappa_{10}^2 T_3 \t \epsilon\, \t \delta^6(y,Y) 
+ \t d \left(e^{-4A} \t \star \h \partial^{2} B_1\right) \nonumber \\
&&+\hat d \left[ \t\star\t de^{-4A} +e^{2\Omega}\t d\t\star\t dB_1
-2\kappa_{10}^2T_3\t\star \t Y_1 \t\delta^6(y,Y) +e^{-4A}\hat\del^2\t\star B_1
\right]\nonumber\\
&&-e^{-2\Omega}\hat\star\hat d b_2^I\w\t d\left[e^{-4A}\t\star
\omega_2^I + \frac{i g_s}{2} \left(\Lambda_1^I\w \bar G_3^{(0)} - 
\textnormal{c.c.}\right)\right]  \nonumber \\
&&+ e^{-2\Omega} \h d \h \star \h d b_2^I \w \left[e^{-4A} \t \star \omega_2^I 
-\t\star\t d B_1^{b,I}
+ \frac{i g_s}{2} (\Lambda_1^I \w \bar G_3^{(0)} - \textnormal{c.c.})\right]
\label{app:E6} \, .
\ee
Each component of $E_6$ leads to a distinct constraint or dynamical EOM.
The constraints largely repeat those from the Einstein equations: the first
three terms of \eqref{app:E6} are once again the instantaneous Poisson
equation for the warp factor.  For reasons explained in footnote \ref{higher},
we ignore the $\h d\h\del^2\t\star B_1$ term, so the (1,5) components of
$E_6$ nearly reproduce \eqref{app:Emum} (up to a Hodge star).  They differ
only by a term proportional to $\t d\t\star \omega_2^I$; therefore, we see
that $\omega_2^I$ must be harmonic or alternately that any compensator
$\h db_2 K_1$ in $C_4$ must vanish.
The dynamical EOM include (0,6) and (2,4) components and yield
\eqref{E6eom} once the definition \eqref{gamma4def} is taken into account.

Meanwhile, the 3-form EOM becomes
\be
E_8 &=& -\h db^I_2 \w \left[\t d \t \star \t d \Lambda_1^I + 
i \omega^I_2 \w G_3^{(0)}\right] +e^{-2\Omega}e^{-4A}(\h d \h\star\h d\h\star
\h db_2^I)\w\t\star\Lambda_1^I\nonumber\\
&&+e^{-2\Omega}(\h\star\h d\h\star\h db_2^I)\w\t d(e^{-4A}\t\star\Lambda_1^I)
+ie^{-2\Omega}\h d\h\star\h db_2^I\w(\t\star\t d e^{-4A})\w \Lambda_1^I
\ .\label{app:E8} \ee
The first term (in square brackets)
gives \eqref{G3constraint}, the constraint determining $\Lambda_1^I$.  
The remainder of the terms give the dynamical EOM $\delta E_8$, but do not
contribute to the 2-derivative quadratic action because they either have the
wrong legs to wedge with $\delta A_2$, contribute only at higher derivative
order, or both.

\subsubsection{Brane EOM}

Finally, the D3-brane EOM is
\begin{align}
E_{\sla M} &= \Del_a\left[\left(g_{\sla M\sla N}\del^a X^{\sla N}
+\frac 16 \frac{\mu_3}{T_3}\epsilon^{abcd}C_{\sla M\sla N\sla P\sla Q}
\del_b X^{\sla N}\del_c X^{\sla P}\del_d X^{\sla Q}\right)
\delta^{10}(x,X)\right]\tag{\ref{ED3_1}}\\
&-\left[\frac 12 \del_{\sla M}g_{\sla N\sla P}\del_a X^{\sla N}\del^a X^{\sla P}
+\frac{1}{4!}\frac{\mu_3}{T_3} \epsilon^{abcd}\del_{\sla M}
C_{\sla N\sla P\sla Q\sla R}\del_a X^{\sla N}\del_b X^{\sla P}\del_c X^{\sla Q}
\del_d X^{\sla R}\right]\delta^{10}(x,X)\ .\nonumber
\end{align}
As noted in the main text, we have ignored terms proportional to the 
derivative of the delta function, as they vanish in the variation of the
action upon integration by parts.

We evaluate these EOM in the static gauge
using the 4-form background corresponding to (\ref{F5ansatz2}), ie
\be
C_4 = e^{4\Omega} e^{4A} \hat \epsilon + e^{4\Omega} 
e^{4A} \hat \star \hat d B_1 + b_2^I \wedge \omega_2^I\, .\label{app:electricC4}
\ee
Then the $\sla M = \sla \mu$ component of the D3-brane EOM becomes
(again, dropping terms proportional to the derivative of the delta function)
\be
E_{\sla \mu} =-\left[2 e^{-2A-2\Omega} \partial_{\sla \mu} \left(e^{2A+2\Omega}
\right) - e^{-4A-4\Omega} \partial_{\sla \mu} \left(e^{4A+4\Omega}\right)\right] 
\delta^{10}(x,X)= 0\, ,\label{app:ED3mu}
\ee
which is a trivial constraint.
The $\sla M = \sla m$ component gives a dynamical EOM
\begin{align}
E_{\sla m} = e^{-4A} e^{-2\Omega} \t g_{\sla m \sla n} 
\left(\h \partial^{2} Y^{\sla n}\right) \delta^{10}(x,X)\, .
\tag{\ref{D3EOMdynamical}}
\end{align}
The second line of \eqref{ED3_1} vanishes identically, and the terms in the
first line containing $g_{\sla m\sla\mu}$ and $C_{\sla m\sla\mu\sla\nu\sla\rho}$,
both of which include the $B_1$ compensator, cancel each other.

%%%%%%%%%%%%%%%%%%%%%%%%%%%%

\subsection{Magnetic Equations of Motion}\label{sec:MagneticEOM}
In this section, we will compute the 10D EOM for the gravity, 5-form flux, 
3-form flux, and local sources for the magnetic form.
The ansatz for all moduli is
\begin{align}
ds^2 &= e^{2\Omega}e^{2A} \h\eta_{\mu\nu}dx^\mu dx^\nu + 2e^{2\Omega}e^{2A} 
\p_\mu B_m (x,y) dx^\mu dy^m + e^{-2A} \til g_{mn} dy^m dy^n \, ,
\tag{\ref{metricansatz}} \\
\t F_5 &= \st \t d e^{-4A}- e^{2\Omega} \h d(\st \t d B_1^Y)
+\h d b_0^I\w\left( \t\star\omega_2^I+\t d K_3^I-\frac{i g_s}{2}
\left(\Lambda_1^I\w \bG-\bar\Lambda_1^I\w G_3^{\0}\right)\right)
\nonumber\\ 
&+\left[ e^{4\Omega} 
\h\epsilon \w \t d e^{4A}- e^{4\Omega}\sh \h d \t d (e^{4A} B_1)
+e^{2\Omega}\h\star\h db_0^I\w\gamma_2^I\right]\ ,\tag{\ref{F5mag2}}\\
G_3&= \G + \h db_0^I\w\t d\Lambda_1^I\ ,\tag{\ref{G3mag}}
\end{align}
plus $Y^{\sla m}(x) =Y^{(0)\sla m}+\delta Y^{\sla m}(x)$.  In the main text,
we also consider the relation of $Y^{\sla m}(x)$ to reference point 
$Y^{\un m}_*$; they are connected by a geodesic which has outward-pointing 
tangents $\sigma^{\un m},\sigma^{\sla m}$ at the endpoints. The metric
compensator and (redefined) 4-form perturbation are
\begin{align}
B_1 &= -c(x)\t dK(y) +b_0^I(x)B_1^I(y)+B_1^Y(x,y)\ ,\tag{\ref{magcompensator}}\\
\delta C'_4 &= b_0^I(x)\t\star\omega_2^I(y) -\h db_0^I K_3^I(y)
-e^{2\Omega} \st \t d B^Y_1(x,y)\ .\tag{\ref{4formmagnetic}}
\end{align}
As before, $\Omega(x),A(x,y)$ contain both background and first-order 
parts.  Here, $\omega_2^I$ is harmonic, and there is an explicit compensator
$K_3^I$ for the axions.  The form $\gamma_2^I$ is shorthand for
\begin{align} 
\gamma_2^I&\equiv e^{4A}\left[\omega_2^I+\t\star\left(\t dK_3^I-
\frac{ig_s}{2}\left(\t d\Lambda_1^I\wedge\bG-\textnormal{c.c.}
\right)\right)+e^{2\Omega}
\t dB^I_1\right]\equiv C^{IJ}\omega_2^J\ ,\tag{\ref{gamma2def}}
\end{align}
which we will motivate from the constraints below; as we will also see that
$\gamma_2^I$ must be harmonic, $C^{IJ}$ is a change-of-basis matrix
defined as in \eqref{Cinvdef}, which can depend on the background values of
the moduli in general.  Again, many of the following results are adapted 
from \cite{1308.0323}.

\subsubsection{Einstein Equations}

The metric is the same in the magnetic formalism as the electric formalism,
with the exception that the $B_m$ compensator takes a somewhat different
form.  However, the Einstein tensor is independent of the particular form 
of $B_m$, so the Einstein tensor is still given by equations 
(\ref{app:Gmunu},\ref{app:Gmum},\ref{app:Gmn}).

The energy-momentum tensors are similar to the electric formalism but 
not quite identical.  We find
\be
T^5_{\mu\nu}&=& -4e^{2\Omega}e^{4A}\left(\del^{\t\ell}A\del_\ell A\right)
\h\eta_{\mu\nu}\ ,\label{app:magT5munu}\\
T^5_{\mu m} &=& -4e^{2\Omega}e^{4A}\left(\del^{\t\ell}A\del_\ell A\right)
\del_\mu B_m+2e^{2\Omega}e^{4A}\del_\mu(\t dB)_{mn}\del^{\t n}A
-2\del_\mu b_0^I\gamma^I_{mn}\del^{\t n}A\ ,\label{app:magT5mum}\\
T^5_{mn}&=& 4\del^{\t\ell}A\del_\ell A\t g_{mn}-8\del_m A\del_n A
\label{app:magT5mn}
\ee
for the 5-form contribution based on \cite{1308.0323}.  Since the only
compensator in $G_3$ is for the axions, its contribution to the 
energy-momentum tensor is identical to \cite{1308.0323}:
\be
T^3_{\mu\nu}&=& -\frac{g_s}{4}e^{2\Omega} e^{8A} \left|G_3^{(0)}\right|^{\t 2} 
\h\eta_{\mu\nu}\ ,\label{app:magT3munu}\\
T^3_{\mu m} &=& -\frac{g_s}{4}\left[-ie^{4A}\del_\mu b_0^I\t\star\left(
\t d\Lambda_1^I\w\bG-\textnormal{c.c.}\right)_m
+e^{2\Omega} e^{8A}\partial_\mu B_m \left|G_3^{(0)}\right|^{\t 2}\right]
\ ,\label{app:magT3mum}\\
T^3_{mn}&=& 0\ .\label{app:magT3mn}
\ee
In static gauge, the energy-momentum tensor for the D3-brane is still
given by (\ref{app:TmunuD3},\ref{app:TmumD3},\ref{app:TmnD3}) since it is
unaffected by the ansatz for the flux.

In the end, the Einstein equations through first order are
\be
\label{app:magEmunu}
E_{\mu\nu}&=& e^{2\Omega}e^{4A}\left[2(4 \del^{\t \ell} A \del_\ell A - 
\t \nabla^{2} A)+T_3 e^{4A}\t\delta^6(y,Y)+\frac{g_s}{4} e^{4A} 
\left|G_3^{(0)}\right|^{\t 2} + \cdots \right] \h \eta_{\mu\nu} \nonumber \\
&&+(\p_\mu\p_\nu - \h\eta_{\mu \nu} \h\p^{2} )(4A - 2\Omega + e^{4A}e^{2\Omega} 
\wtn^{\t \ell} B_\ell) \, ,\\
E_{\mu m} &=& e^{2\Omega}e^{4A}\partial_\mu B_m \left[ 2(4 \del^{\t \ell} A 
\del_\ell A - \t \nabla^{2} A)+\kappa_{10}^2 T_3 e^{4A} \t\delta^6(y,Y)
+ \frac{g_s}{4} e^{4A} \left|G_3^{(0)}\right|^{\t 2}+ \cdots \right] \nonumber\\
&& +2\p_\mu \p_m A - 8 \p_\mu A \p_m A + e^{4A}e^{2\Omega} \p_\mu \wtn^{\t \ell} 
\wtn_{[m}B_{\ell]}+T_3 e^{4A} \t g_{m\sla n}\p_\mu Y^{\sla n}(x) \t \delta^6(y,Y) 
\nonumber \\
&& +2 e^{4A} \del_\mu b_0^I \left[e^{-4A}\gamma^I_{mn} 
\partial^{\t n} A + \frac{i g_s}{8} \t \star \left(\t d \Lambda_1^I\w 
\bar G^{(0)}_3 - \textnormal{c.c.}\right)\right]\, ,\label{app:magEmum} \\
\label{app:magEmn}
E_{mn} &=& \hat \partial^{2} \left( \t \nabla_{(m} B_{n)}+\del_{(m}AB_{n)}
-\t\Del^{\t\ell}B_\ell\t g_{mn}
-2\t g_{mn} \t \nabla^{\t \ell} A B_\ell \right)	
+ e^{-4A} e^{-2\Omega} \t g_{mn} \hat \partial^{2} \left(3\Omega - 2 A\right)\, .
\hspace{0.5in}\ee
As before, \eqref{app:magEmunu} contains the instantaneous version of the 
Poisson equation determining the warp factor, which yields \eqref{warpshift1}, 
along with the constraint \eqref{Bdiv2} for $\t\Del^{\t n}B_n$ (including
\eqref{KPoisson} for $K$, the volume modulus compensator).  
The off-diagonal Einstein equation
\eqref{app:magEmum} also includes the Poisson equation for the warp factor,
along with a Poisson equation \eqref{BYGreen} for $B_m^Y$ (which is satisfied 
by \eqref{BD3Solve}), and the Poisson equation 
\be \t\Del^2B^I_m=-e^{-2\Omega}\gamma^I_{mn}\del^{\t n}e^{-4A}-\frac{ig_s}{2}
\t\star\left(\t d\Lambda_1^I\wedge\bG-\textnormal{c.c.}
\right)_m\ .\label{app:BIpoisson}\ee
The internal component \eqref{app:Emn} is once again a dynamical EOM only,
and we can simplify it to the form \eqref{magEmn} using \eqref{Bdiv2}.

\subsubsection{Form Flux EOM}

Because the magnetic ansatz \eqref{F5mag2} for $\t F_5$ differs considerably
from the electric case, the EOM for $\t F_5$ and $G_3$ also differ
significantly from the electric formalism.

The first thing to note is that neither the 3-form or D3-brane source
terms have the correct components to contribute to $E_6$ in the magnetic
formalism; $E_6=d\star\t F_5$, as explained in section \ref{sec:SUGRAwD3}.  
Therefore, the EOM becomes
\be E_6 = -e^{4\Omega}\h d\h\star\h d\t d(e^{4A}B_1)+ e^{2\Omega}
 \h d\h\star\h db_0^I\w\gamma_2^I+e^{2\Omega}\h\star\h db_0^I\w\t d\gamma_2^I
\ .\ee
The first two terms contribute to the dynamical EOM, while the last term
is a constraint requiring that $\gamma_2^I$, defined in terms of the
moduli and compensators as \eqref{gamma2def}, be closed.  Meanwhile,
the Bianchi identity is now the constraint
\begin{align}
d \t F_5 -\frac{ig_s}{2} G_3\w\bar G_3
+ 2\kappa_{10}^2 T_3 \int d^4 \xi \sqrt{-\gamma} \, \star \epsilon_\| 
\, \delta^{10}(x,X(\xi))&=0 \, .\tag{\ref{BianchiF5}}
\end{align}
Written as in \eqref{F5mag2}, $\t F_5$ automatically satisfies the Bianchi
identity as long as $B_1^Y$ is given by \eqref{BD3Solve}.  However, in terms
of the shorthand variable $\gamma_2^I$, the magnetic components of $\t F_5$
are $\t F_5 = \t d e^{-4A}- e^{2\Omega} \h d(\st \t d B_1)+e^{2\Omega}e^{-4A}\h 
db_0^I\t\star\gamma_2^I$. Like the ($\mu m$) component of the Einstein 
equation, the constraint from the Bianchi identity in these variables leads to 
\eqref{BYGreen} for $B_1^Y$ and \eqref{app:BIpoisson} if and only if
$\t d\t\star\gamma_2^I=0$, which implies that $\gamma_2^I$ is harmonic.

The 3-form EOM is 
\be
E_8 &=& e^{2\Omega}\h d\h\star\h db_0^I\w \t\star\t d\Lambda_1^I -
e^{2\Omega}\h\star\h db_0^I\w \t d\t\star\t d\Lambda_1^I+i
e^{4\Omega}e^{4A}\h d\h\star\h dB_1\w\G -i e^{2\Omega}\h\star\h db_0^I\w
\gamma_2^I\w\G\, .\hspace{0.5in}
\ee
The first and third terms contribute to the dynamical EOM, while 
the second and fourth, when acted on by $\t\star$, give the Poisson equation
\eqref{magaxcompensators} for $\Lambda_1^I$.

\subsubsection{Brane EOM}

As noted in the main text, the D3-brane action has no WZ term through 
second order in $\h dY$ in the magnetic formalism because $C_4$ has the wrong 
legs.  However, as we discussed extensively, we should properly think of
$\t F_5$ as depending explicitly on the brane position because of the 
nontrivial Bianchi identity, much as it depends on $A_2$ and $G_3$.
As a result, the $\t F_5$ kinetic action contributes to the D3-brane EOM,
which we derived in section \ref{sec:magneticEOM}.  With some conjectures
about higher-order terms in a formal expansion, the EOM in static gauge 
comes out to
\begin{align} 
E_{\sla m} &= \left\{\Del^\gamma_\mu\left(\gamma^{\mu\nu}g_{\sla m\sla n}
\del_\nu Y^{\sla n}+\gamma^{\mu\nu}g_{\nu\sla m}\right)-\frac 12 
\gamma^{\mu\nu}\left(\del_{\sla m} g_{\mu\nu}+2\del_{\sla m}g_{\sla n (\mu}
\del_{\nu)}Y^{\sla n}+\del_{\sla m}g_{\sla n\sla p}\del_\mu Y^{\sla n}
\del_\nu Y^{\sla p}\right)\right.\nonumber\\
&\left. -\frac 12 \left(\star_\gamma d\star\t F_5\right)_{n\sla m}
\Lambda^n_{\un n}\sigma^{\un n}+\frac{\sqrt{-g}}{\sqrt{-\gamma}\sqrt{g_\perp}}
\frac{1}{5!}
\left( (\ep_\perp)_{\sla m npqrs}\t F^{npqrs}+(\ep_\perp)_{\sla m \sla n pqrs}
\t F^{\mu pqrs}\del_\mu Y^{\sla n}\right)\right\}\nonumber\\
&\times \delta^{10}(x,X)\ .\tag{\ref{fullD3EOM2}}
\end{align}
We recall that $\gamma_{\mu\nu}$ is the (independent) worldvolume metric,
$g_{\perp, mn}$ is an arbitrary metric on the $y^m$ coordinates, and 
$\sigma^{\un m}$ is the tangent to a geodesic from the brane position
$Y^{\sla m}$ to an arbitrary reference point $Y_*^{\un m}$ at $Y_*^{\un m}$.
As usual, the EOM for $\gamma_{\mu\nu}$ sets it equal to the pullback of
$g_{MN}$ to the brane worldvolume, but $\gamma_{\mu\nu}$ has no first-order
fluctuation.  For convenience, we take $g_{\perp, mn}=\t g_{mn}$.

In \eqref{fullD3EOM2}, the background terms involving $\del_{\sla m}g_{\mu\nu}$
and $(\epsilon_\perp\cdot \t F_5)_{\sla m}$ cancel each other; this is 
the no-force condition on the D3-brane in the magnetic formalism.
A number of terms enter at second order, leaving a first-order dynamical
EOM of 
\begin{align}
\delta E_{\sla m} &= \left[ e^{-2\Omega}e^{-4A}\t g_{\sla m\sla n}\h\del^{2}
\delta Y^{\sla n}+\h\del^{2}B_{\sla m} -\frac 12 e^{-4\Omega}e^{-4A}
(\h\star\delta E_6)_{\sla m\sla n}\sigma^{\sla n}\right]\delta^{10}(x,X)\ .
\tag{\ref{magEOM}}\end{align}

%%%%%%%%%%%%%%%%%%%%%%%%%%%%
%%%%%%%%%%%%%%%%%%%%%%%%%%%%
%%%%%%%%%%%%%%%%%%%%%%%%%%%%
%%%%%%%%%%%%%%%%%%%%%%%%%%%%

\section{K\"ahler Potential and Kinetic Action}
\label{sec:kahler}

As is well-known, GKP compactifications have $\mathcal{N}=1$ SUGRA (possibly
with spontaneously broken supersymmetry) as their 4D effective theory.
As a result, the metric on moduli space must be K\"ahler, meaning that
moduli space is complex and that the hermitean metric on moduli space is
K\"ahler $G_{a \bar b} = \p_a \p_{\bar b} \mc K$.
Here we consider the kinetic Lagrangian 
$\mc L=G_{a\bar b}\del_\mu\phi^a\del^\mu\bar\phi^{\bar b}$ for the K\"ahler
potential
\be
\label{Kpot}
\mc K = -3\log\left[ -i(\rho - \bar \rho) - \gamma k(Z, \bar Z) 
\right] \, ,
\ee
which is appropriate for the case that $h^{1,1}=1$.
In relation to the variables of the 10D SUGRA fields,
$\rho = b + i ( c + \gamma k(Z, \bar Z)/2)$, 
and $k(Z,\bar Z)$ is the K\"{a}hler potential of the underlying
CY manifold. 

The kinetic Lagrangian for these moduli takes the form
\be
\mc L = \p_{\rho}\p_{\bar \rho} \mc K \, \p\rho \p\bar\rho + \p_\rho \p_{\bar Z} 
\mc K \, \p\rho \p\bar Z + \p_{\bar \rho}\p_Z \mc K \, \p\bar\rho \p Z + 
\p_Z \p_{\bar Z} \mc K \, \p Z \p\bar Z \, .
\ee
Using \eqref{Kpot}, each term is:
\be
\p_{\rho}\p_{\bar \rho} \mc K \, \p\rho \p\bar\rho &=& \frac{3}{4c^2}\left( 
(\p b)^2 + (\p c)^2 +\gamma\p c (\p_Z k \, \p Z + \p_{\bar Z} k \, 
\p\bar Z)\right. \nonumber \\
&& \left.+ \frac{\gamma^2}{4}(\p_Z k \, \p Z + \p_{\bar Z}k \, \p\bar Z)^2 
\right)\ , \\
\p_\rho \p_{\bar Z} \mc K \, \p\rho \p\bar Z &=& i 
\frac{3 \gamma \p_{\bar Z}k}{4c^2} \left( \p b + i\p c + i 
\frac{\gamma}{2}(\p_Z k \, \p Z + \p_{\bar Z} k \, \p\bar Z) \right) \p\bar Z 
\ , \\
\p_{\bar\rho}\p_Z \mc K \, \p\bar\rho \p Z &=& -i  
\frac{3 \gamma \p_{Z}k}{4c^2} \left( \p b - i\p c - 
i \frac{\gamma}{2}(\p_{\bar Z}k \, \p\bar Z + \p_Z k \, \p Z)\right) \p Z\ , \\
\p_Z \p_{\bar Z} \mc K \, \p Z \p\bar Z &=& \frac{3\gamma}{2c} 
\left( \p_Z\p_{\bar Z} k + \frac{\gamma \p_Z k \p_{\bar Z} k}{2c} \right) 
\p Z \p\bar Z \, .
\ee
Adding everything together gives
\be
\mc L &=& \frac{3}{4c^2}\left( (\p b)^2 + (\p c)^2 \right) + 
\frac{3\gamma}{2 c} \p_Z\p_{\bar Z} k \, \p Z \p\bar Z - 
\frac{3 i \gamma}{4 c^2}\p b (\p_Z k \, \p Z - \p_{\bar Z} k \, \p\bar Z) 
\nonumber \\
&&  - \frac{3\gamma^2}{16 c^2} \left( \vphantom{\frac 12}\p_Z k \p_Z k \, 
\p Z \p Z - 2 \p_Z k \p_{\bar Z} k \, \p Z \p\bar Z + \p_{\bar Z} k \p_{\bar Z} k 
\, \p\bar{Z} \p\bar{Z}\right) \, .\label{LK}
\ee

Based on the form of \eqref{LK}, what we anticipate seeing in the
quadratic action are separate quadratic terms for the scalar axion,
volume modulus, and the D3-brane; a second-derivative of the internal
K\"ahler potential; sets of derivatives (holomorphic, antiholomorphic, 
and mixed) acting on $k$; and a coupling between the axion and 
D3-brane moduli. 

%%%%%%%%%%%%%%%%%%%%%%%%%%%%
%%%%%%%%%%%%%%%%%%%%%%%%%%%%
%%%%%%%%%%%%%%%%%%%%%%%%%%%%

\bibliographystyle{JHEP}
\bibliography{D3Branes}

%%%%%%%%%%%%%%%%%%%%%%%%%%%%
%%%%%%%%%%%%%%%%%%%%%%%%%%%%
%%%%%%%%%%%%%%%%%%%%%%%%%%%%

\end{document}